\documentclass[superscriptaddress,preprint,prd,tightenlines,nofootinbib,longbibliography,eqsecnum]{revtex4-2}

\pdfoutput=1

\usepackage{amsmath, amssymb, amsfonts, amsthm, latexsym, mathrsfs, bbm, slashed, bm}
\usepackage{booktabs}
\usepackage{array}
\usepackage{makecell}
\usepackage{multirow}
\usepackage{mathtools}
\usepackage{stmaryrd}

\usepackage[usenames,dvipsnames]{xcolor}
\usepackage[title,titletoc]{appendix}
\usepackage{graphicx}
\usepackage{physics}

\usepackage{xifthen}
\usepackage{tensor}
\usepackage{dsfont}

\usepackage[inline]{enumitem}

\usepackage{setspace}
\usepackage[marginal, multiple]{footmisc}

\usepackage[T1]{fontenc}
\usepackage[utf8]{inputenc}
\usepackage{lmodern}
\usepackage[normalem]{ulem}

\definecolor{DullDarkRed}{rgb}{0.6,0.1,0.1}
\definecolor{RefColor}{rgb}{0,0.5,0}

\definecolor{darkgreen}{rgb}{0,0.5,0}
\usepackage[unicode, colorlinks, allcolors=blue!70!black, linktocpage, pdfusetitle]{hyperref}
\hypersetup{
    bookmarks=true,
    unicode=false,
    pdftoolbar=true,
    pdfmenubar=true,
    pdffitwindow=false,
    pdfstartview={FitH},
    pdfnewwindow=true,
    colorlinks=true,
    linkcolor=darkgreen,
    citecolor=DullDarkRed,
    filecolor=magenta,
    urlcolor=darkgreen,
    linktocpage=true
}
\usepackage[all]{hypcap}

\usepackage{orcidlink}

\usepackage{verbatim}

\usepackage{cancel}

\usepackage[capitalize]{cleveref}

\usepackage{microtype}

\usepackage{standalone}
\usepackage{tikz}
\usetikzlibrary{arrows.meta, decorations.markings}

\ifdefined\myext
\usetikzlibrary{external}
\fi

\newcommand{\pd}{\partial}

\newcommand{\ty}[1]{\text{#1}}

\DeclareMathOperator{\sn}{sn}

\DeclareMathOperator{\am}{am}

\allowdisplaybreaks

\newcommand{\MeudonText}{Laboratoire d'étude de l'Univers et des phénomènes eXtrêmes (LUX), Observatoire de Paris, Université PSL, Sorbonne Université, CNRS, 92190 Meudon, France}

\newcommand{\UofTrujilloText}
{Escuela Profesional de F\'{\i}sica, Facultad de Ciencias F\'{\i}sicas y Matem\'aticas, Universidad Nacional de Trujillo, Av. Juan Pablo II s/n, Trujillo, Per\'u}

\newcommand{\utecText}{Universidad de Ingenieria y Tecnologia – UTEC, Jr. Medrano Silva 165 - Barranco 15063, Lima, Peru}

\makeatletter
\usepackage{etoolbox}
\patchcmd{\frontmatter@abstract@produce}
  {\vskip200\p@\@plus1fil
   \penalty-200\relax
   \vskip-200\p@\@plus-1fil}
  {}{}{}
\makeatother

\newcommand{\lnmiit}{The LNM Institute of Information Technology (LNMIIT), Jaipur 302031, India}

\begin{document}
	
\title{
Orbit-averaged spinning and eccentric binary black holes: Hamiltonian and integrability
}	

\author{Tom Colin\,\texorpdfstring{\orcidlink{0009-0004-8392-2808}}{}}
\email{tom.colin@obspm.fr}
\affiliation{\MeudonText}

\author{Sashwat Tanay\,\texorpdfstring{\orcidlink{0000-0002-2964-7102}}{}}
\email{stanay@utec.edu.pe}
\affiliation{\MeudonText}
\affiliation{\utecText}

\author{Manuel Alva-Morales\,\texorpdfstring{\orcidlink{0000-0001-8925-6042}}{}}
\email{malvam@unitru.edu.pe}
\affiliation{\UofTrujilloText}

\author{Prateek Anand{}}
\affiliation{\lnmiit}

\begin{abstract}
\noindent
For over two decades, the orbit-averaged precession equations of spinning
binary black holes at second post-Newtonian order have been a standard
model for their spin dynamics. One of their constants of motion, the
projected effective spin, underlies the stability analysis of aligned-spin
configurations and the construction of analytical solutions. The system is
known to be Hamiltonian and integrable, its Hamiltonian being the orbit
average of the full post-Newtonian Hamiltonian. Independently, we derive
this Hamiltonian from the equations of motion alone, show why the orbit
average gives the correct equations, and relate its constants of motion to
those of the full system. We then construct the action-angle variables and
frequencies with the recently introduced method of fictitious variables,
and propose a criterion for the breakdown of the integrability conditions.
This criterion recovers the separatrices of simple classical systems and
the up-down instability, whose thresholds we give explicitly for eccentric
orbits. We also show that, when all binaries of an ensemble share the same
parameters and radiation reaction does not depend on the spins, the
Hamiltonian structure preserves an isotropic distribution of spin
directions. We confirm numerically that isotropy can be lost when the 
parameters are correlated with the spin directions. Finally, we use 
the orbit-averaged Hamiltonian to generalize the projected effective spin 
to binary neutron stars, for which it is not conserved, and obtain a closed 
form valid for eccentric orbits.
\end{abstract}

\maketitle

\section{Introduction}
\label{sec:intro}

In a binary black hole (BBH) with spins misaligned with the orbital
angular momentum, spin-orbit and spin-spin couplings make the orbital
angular momentum $\mathbf{L}$ and the two spins $\mathbf{S}_1$ and
$\mathbf{S}_2$ precess and nutate, which modulates the emitted gravitational
waves (GWs)~\cite{kidder1995}. Spins are also of astrophysical interest,
since their orientations depend on how the binary was formed, and measuring
them with GWs helps discriminate between formation
channels~\cite{Rodriguez2016, Gerosa:2013laa}.

In the post-Newtonian (PN) regime, the orbital period of a spinning BBH is
much shorter than its precession timescale, which is itself much shorter
than the radiation-reaction (RR) timescale~\cite{gerosa2015multi,
gerosa2023fastprecession}. This hierarchy allows one to average the
precession equations over a Newtonian orbit. The resulting 2PN
orbit-averaged (OA) precession equations were introduced by
Schnittman~\cite{schnittman} and extended by Racine to include the
quadrupole-monopole couplings~\cite{Racine2008}. Racine also found a new
constant of motion (COM), later called the projected effective spin
$\chi_{\rm eff}$~\cite{gerosa2015multi}.

The OA equations have since been one of the main frameworks for studying
spin evolution. Schnittman found that they admit spin-orbit resonances, in
which $\mathbf{L}$, $\mathbf{S}_1$, and $\mathbf{S}_2$ are coplanar and
precess jointly about the total angular momentum
$\mathbf{J}$~\cite{schnittman}. Bogdanovi\'{c} et al.~\cite{Bogdanovic_2007}
found that, for circular BBHs with identical masses, spin magnitudes, and
initial separations, evolved with spin-independent RR, an initially
isotropic distribution of spin directions remains isotropic during the
inspiral. Gerosa et al.~\cite{Gerosa:2013laa} showed that anisotropic
distributions can instead become locked near resonant planes.

On the analytical side, the conservation of $\chi_{\rm eff}$ has been the
key to understanding the OA system. It reduces the precession to the motion
of a single variable in an effective potential, which led to closed-form
solutions~\cite{Kesden2015, gerosa2015multi, Klein2017b, colin-tanay}. The
stability of the aligned-spin configurations relies on $\chi_{\rm eff}$ as
well. The up-down configuration was shown to become unstable in a finite
range of orbital separations, first from the shape of the effective
potentials~\cite{gerosa2015updown}, and later by describing small
perturbations around it as a harmonic oscillator~\cite{PhysRevD.101.124037}.
Secular precession equations for eccentric orbits and arbitrary spin-induced
quadrupole moments were derived in Ref.~\cite{PhysRevD.103.084024}, and the
linear stability of the aligned configurations was analyzed in
Ref.~\cite{PhysRevD.103.084025}.

The OA system has also been studied with the tools of celestial mechanics.
Correia~\cite{correia:hal-02319812}\footnote{We were not aware of
Ref.~\cite{correia:hal-02319812} during our work, and learned of 
it only after completing the results presented here. We thank Paul 
Ramond for pointing it out to us.} orbit averaged the full
Hamiltonian, obtained the OA equations of motion (EOMs) from it through the
Poisson brackets of the angular momenta, noted that the system is
integrable, and argued that the spin-orbit resonances are Cassini states
rather than resonances. An analogous problem, the precession of a planet
with a satellite, involves three precessing angular momenta and a
Hamiltonian quadratic in their mutual cosines. Bou\'e and
Laskar~\cite{boue:hal-00335321} showed that it is integrable and obtained
its nutation and precession frequencies by quadratures.

In this article, we revisit the 2PN OA system from the point of view of
Hamiltonian mechanics. We show why the orbit average of the full PN
Hamiltonian generates the OA EOMs, which is not immediate for $\mathbf{L}$,
and we derive the same Hamiltonian independently from the OA EOMs alone.
We relate the three COMs in involution of the OA system to those of the
full 2PN system, compute the action variables with the fictitious variables
introduced in Ref.~\cite{Tanay2023}, and then obtain the frequencies and
angle variables. The frequencies confirm that the spin-orbit resonances are
not resonances in the usual sense.

We then propose a criterion for the breakdown of the integrability
conditions, expressed through the Jacobian between the COMs and the action
variables. Like the method of fictitious variables, it is a general tool of
Hamiltonian mechanics, not tied to PN theory. We illustrate it on simple
systems. For the OA system, it recovers the up-down instability and the
separations at which it starts and ends. Since the criterion depends on the
orbit only through the magnitude of $\mathbf{L}$, it gives these separations 
explicitly for eccentric orbits, for which the onset is set by $a(1-e^2)$
rather than by the semi-major axis $a$ alone. The linear stability
analysis of Ref.~\cite{PhysRevD.103.084025}, which holds for eccentric
orbits, also depends on the orbit only through $L$, but does not give the
thresholds $L_\pm$ explicitly.

The Hamiltonian structure also has astrophysical consequences. We show that
it explains the preservation of isotropy observed in
Ref.~\cite{Bogdanovic_2007}, through an argument close to Liouville's
volume theorem. The argument holds when all members of the ensemble share
the same parameters and RR does not depend on the spins. It also predicts
when isotropy can be lost, for instance when the masses of the members are
correlated with the direction of $\mathbf{L}$, which we confirm
numerically.

The methods developed here also apply to OA binary neutron stars (BNSs).
Their quadrupole-monopole terms differ from the black-hole
values~\cite{Schafer2024}, and $\chi_{\rm eff}$ is no longer
conserved~\cite{PhysRevD.108.043018}. 
The OA BNS system nevertheless remains integrable, and in 
Sec.~\ref{sec:chi_eff_BNS} we use its OA Hamiltonian to generalize 
$\chi_{\rm eff}$ to BNSs, in closed form and for eccentric orbits.

The paper is organized as follows. Section~\ref{sec:setup} introduces the
system and its OA EOMs. Section~\ref{sec:Hamiltonian} discusses its
Hamiltonian structure and derives the OA Hamiltonian.
Section~\ref{sec:Integrability_AAV} discusses its integrability and
computes its action variables, and Sec.~\ref{sec:frequencies_angles} its
frequencies and angle variables. Section~\ref{sec:separatrix} presents the
criterion for the breakdown of the integrability conditions and its
applications. Section~\ref{sec:isotropic_spins} studies the consequences
of the Hamiltonian structure for ensembles of BBHs,
Sec.~\ref{sec:chi_eff_BNS} extends part of these results to BNSs, and
Sec.~\ref{sec:conclusions} concludes.

\section{Setup}   \label{sec:setup}

\subsection{Notation and conventions}\label{sec:Notation}
We consider a spinning and eccentric BBH with masses $m_1>m_2$. 
The system has a total mass $m = m_1 + m_2$, a reduced mass $\mu = m_1 m_2 / m$, and a 
symmetric mass ratio $\nu = \mu / m$.  We set $G = 1$ and 
keep $c$ explicit as a PN bookkeeping parameter.

Throughout this work, 
three-dimensional vectors are denoted by boldface letters 
(e.g., $\mathbf{V}$), their Euclidean norms  by the corresponding 
unbolded letter, $V \equiv \|\mathbf{V}\|$, and unit vectors by 
a hat, $\hat{\mathbf{V}} \equiv \mathbf{V}/V$. $t$ denotes the 
coordinate time and overdots denote $d/dt$.

We work in the center-of-mass frame, where 
$\mathbf{P}_1 = -\mathbf{P}_2 \equiv \mathbf{P}$, with relative 
separation vector $\mathbf{R} = \mathbf{R}_1-\mathbf{R}_2$, and 
unit vector $\mathbf{n} = \mathbf{R}/R$. The orbital angular 
momentum is $\mathbf{L}=\mathbf{R} \times \mathbf{P}$, the spins 
of the two black holes are $\mathbf{S}_1$ and $\mathbf{S}_2$, and 
the total angular momentum is $\mathbf{J} = \mathbf{L} + \mathbf{S}_1 + \mathbf{S}_2$. 
We refer to the three vectors $(\mathbf{L},\mathbf{S}_1,\mathbf{S}_2)$ 
as the triad, and to a given set of their values, i.e., a point of the 
orbit-averaged phase space, as a configuration.

We use two mass-weighted spin combinations 
$\mathbf{S}_{\text{eff}} \equiv \sigma_1 \mathbf{S}_1 + \sigma_2 \mathbf{S}_2$ 
and $\mathbf{S}_0 \equiv \Sigma_1 \mathbf{S}_1 + \Sigma_2 \mathbf{S}_2$, 
where $\sigma_a \equiv 1 + {3 m_b}/{4 m_a}$ and 
$\Sigma_a \equiv {m}/{m_a}$, with $(a,b) \in \{(1,2),(2,1)\}$. 
$\mathbf{S}_{\text{eff}}$ features in the spin-orbit coupling terms while 
$\mathbf{S}_0$  features in the spin-spin coupling terms. 
$\kappa_1, \kappa_2$, and $\gamma$ denote 
the angles between $(\vb{L}, \vb{S}_1)$, $(\vb{L}, \vb{S}_2)$,
and $(\vb{S}_1, \vb{S}_2)$, whereas  $\Delta \Phi$ is the angle between 
the projections of the two spins on the plane perpendicular to $\vb{L}$.

\subsection{The system} \label{sec:system}
The two BHs move on a bound orbit, and their dynamics is governed by a
Hamiltonian expressed in the Arnowitt-Deser-Misner (ADM) 
coordinates~\cite{Arnowitt2008} with the Newton-Wigner-Pryce 
(NWP) spin supplementary condition (SSC)~\cite{Pryce1948, Newton1949}, 
which fixes the center-of-mass worldline of each spinning body. 
The Hamiltonian reads~\cite{Damour2001, Hartl:2004xr, Barker:1966zz, PhysRevD.12.329}
\begin{equation} \label{eq:hamiltonian}
   H = H_{\text{N}} + \frac{1}{c^2} H_{1\text{PN}} + \frac{1}{c^2} H_{\text{SO}} + \frac{1}{c^4} H_{2\text{PN}}+ \frac{1}{c^2} H_{\text{SS}},
\end{equation}
where we display the spin-dependent terms
\begin{subequations}
\label{eq:spin_components}
\begin{align}
H_{\text{SO}} &= \frac{2}{R^3}\, \mathbf{L} \cdot \mathbf{S}_{\text{eff}} \,, \\
H_{\text{SS}} &= \frac{\nu}{2 R^3}\left[3(\mathbf{n} \cdot \mathbf{S}_0)^2 - S_0^2\right] .
\end{align}
\end{subequations}
With $\mathbf{S}_a = m_a^2 \boldsymbol{\chi}_a / c$, $H_{\text{SO}}$ 
and $H_{\text{SS}}$ enter at 1.5PN and 2PN order, respectively. 
The purely orbital terms $H_{\ty{N}}$, $H_{1\text{PN}}$, and 
$H_{2\text{PN}}$ can be found in Eqs.~(2.2) of Ref.~\cite{colin-tanay}. 
The form of $H_{\text{SS}}$ in terms of the single combination 
$\mathbf{S}_0$ includes the $S_1 S_2$ couplings as well as 
the self-spin ($S_1^2$, $S_2^2$) couplings. The evolution of any 
function $f$ of the phase-space variables is given by
\begin{equation} \label{eq:evolution_eq}
    \frac{df}{dt} = \lbrace f, H \rbrace,
\end{equation}
with the fundamental canonical Poisson brackets (PBs)
\begin{subequations}\label{eq:PB-rules-1}
\begin{align} 
    \{ R_i, P_j \} &= \delta_{ij}, \\
    \{ S_{ai}, S_{bj} \} &= \delta_{ab} \epsilon_{ij}{}^k S_{ak}, \quad a, b \in \{1, 2\}, 
\end{align}
\end{subequations}
combined with the usual PB rules of anti-symmetry and chain 
differentiation~\cite{Samanta2022} (with $v_l$ representing 
the phase space variables),
\begin{subequations}    \label{eq:PB-rules-2}
\begin{align}
    \{ f, g \} &= -\{ g, f \}, \\
    \{ f, g(v_l) \} &= \{ f, v_l \} \frac{\partial g}{\partial v_l}, 
\end{align}
\end{subequations}
and any PB that is not covered by Eqs.~\eqref{eq:PB-rules-1}--\eqref{eq:PB-rules-2} 
simply vanishes. If $\pb{f}{g} = 0$, then we say that $f$ and $g$ commute,
or they are in involution.

One can obtain the EOMs for the three angular 
momenta directly in terms of physical variables\footnote{
The  terms $H_{\text{N}}$, $H_{\text{1PN}}$ and $H_{\text{2PN}}$ 
on the right-hand side of Eq.~\eqref{eq:hamiltonian} are pure orbital 
terms that do not involve spin variables. Hence, they commute with 
the spins and the orbital angular momenta under PBs and do not 
affect the angular momenta evolution.}
\footnote{The corresponding equations in Ref.~\cite{Racine2008} 
have typos. See Footnote~[9] of Ref.~\cite{colin-tanay} 
for more details.\label{fn:shared_note}}
\begin{subequations}\label{eq:eom}
\begin{align}
\frac{d\mathbf{S}_a}{dt} &= \frac{1}{c^2 R^3}\left\{2\sigma_a \mathbf{L} + \nu\Sigma_a\left[3(\mathbf{n}\cdot \mathbf{S}_0)\mathbf{n}-\mathbf{S}_0\right] \right\} \times \mathbf{S}_a\,, \\
\frac{d\mathbf{L}}{dt} &= -\frac{1}{c^2 R^3}\left\{ 2\,\mathbf{L}\times\mathbf{S}_{\text{eff}} + 3\nu (\mathbf{n}\cdot\mathbf{S}_0)(\mathbf{n} \times \mathbf{S}_0)\right\}\,.
\end{align}
\end{subequations}
The so-called 2PN orbit-averaged (OA) EOMs for the three 
angular momenta $\vb{S}_1, \vb{S}_2$, and $\vb{L}$ were 
introduced by Schnittman~\cite{schnittman} and have been 
the subject of many analyses~\cite{gerosa2015multi, Racine2008}. 
They are obtained by averaging~\cref{eq:eom} over a Newtonian 
orbit, keeping the angular momenta ($\vb{L}, \vb{S}_a$) fixed, 
using~\cite{Racine2008}
\begin{align}       \label{eq:how_to_average}
\left\langle\frac{1}{R^3}\right\rangle \equiv  \frac{1}{D^3} = \frac{1}{(a\sqrt{1-e^2})^3} , \qquad
\left\langle\frac{n^i n^j}{R^3}\right\rangle=\frac{1}{2 D^3}\left(\delta^{i j}-\hat{L}^i \hat{L}^j\right),
\end{align}
where $D$, $a$ and $e$ are respectively the semi-minor axis, 
the semi-major axis and the eccentricity of the associated 
Newtonian ellipse. The OA spin dynamics therefore depend on the shape 
of the ellipse only through $D$. The averaging procedure is 
justified because the spins precess on a timescale $\sim (c/v)^2\, T_{\text{orb}}$ 
and the orbital angular momentum on $\sim (c/v)^3\, T_{\text{orb}}$, 
both much longer than the orbital period. The resulting OA EOMs can 
be written as\footref{fn:shared_note}
\footnote{\label{footnote-schnittman}Eqs.~(2.2), (A6), and (A7) 
of Ref.~\cite{schnittman} do not have all the terms of our 
Eqs.~\eqref{eq:eom} or \eqref{eq:OA-eom}  because 
the EOMs of Ref.~\cite{schnittman} correspond to a Hamiltonian 
where the 2PN self-spin-squared terms (involving two copies of 
either only $\vb{S}_1$ or $\vb{S}_2$; also known as 
monopole-quadrupole terms) have been dropped.}
\begin{subequations}
\label{eq:OA-eom}
\begin{align}
\left\langle \frac{d\mathbf{S}_a}{dt} \right\rangle &= \frac{1}{c^2 D^3} \left\{ 2\sigma_a\,\mathbf{L} + \frac{\nu\Sigma_a}{2}\left[\mathbf{S}_0- 3 (\hat{\mathbf{L}}\cdot \mathbf{S}_0)\hat{\mathbf{L}}\right]\right\} \times \mathbf{S}_a, \\
\left\langle \frac{d\mathbf{L}}{dt} \right\rangle &= \frac{1}{c^2 D^3} \left\{ 2\,\mathbf{S}_{\mathrm{eff}} - \frac{3\nu}{2}\frac{(\mathbf{L}\cdot \mathbf{S}_0)}{L^2} \mathbf{S}_0 \right\} \times \mathbf{L}. \label{eq:dldt_avg}
\end{align}
\end{subequations}
One can check directly that Eqs.~\eqref{eq:OA-eom} 
conserve $\mathbf{J}$, $L$, $S_1$, and $S_2$. In what follows, 
we omit the angular brackets on the OA time derivatives for brevity.

\section{Hamiltonian nature of the system}
\label{sec:Hamiltonian}

\subsection{Hamiltonian of the orbit-averaged system}   \label{sec:SPS_PB}

We begin this section with the statement that the 2PN OA spinning BBH
system is Hamiltonian, i.e., there exists a Hamiltonian $\bar{H}$ with a set
of fundamental PBs such that the EOMs~\eqref{eq:OA-eom} follow from
$df/dt = \pb{f}{\bar{H}}$. This Hamiltonian is
\begin{equation} \label{eq:H_bar}
\bar{H} = \frac{1}{c^2 D^3}\left[ 2\,\mathbf{L}\cdot\mathbf{S}_{\text{eff}} + \frac{1}{2}\,\mathbf{S}_1\cdot\mathbf{S}_2 - \frac{3\nu}{4}(\mathbf{\hat{L}}\cdot\mathbf{S}_0)^2 \right],
\end{equation}
together with the fundamental SO(3) PBs
\begin{align} \label{eq:SO3-PBs}
\{S_{ai}, S_{bj}\} = \delta_{ab}\,\epsilon_{ijk} S_{ak}, \qquad
\{L_i, L_j\} = \epsilon_{ijk} L_k, \qquad
\{L_i, S_{aj}\} = 0 .
\end{align}
For any $\mathbf{V}\in\{\mathbf{L},\mathbf{S}_1,\mathbf{S}_2\}$,  
Eqs.~\eqref{eq:PB-rules-2} and~\eqref{eq:SO3-PBs} let us  write
\begin{equation} \label{eq:LP-eom}
\frac{d\mathbf{V}}{dt} = \{\mathbf{V},\bar H\} = \frac{\partial \bar H}{\partial \mathbf{V}}\times\mathbf{V}.
\end{equation}
This directly reproduces Eqs.~\eqref{eq:OA-eom}.

For the full non-OA system, $\mathbf{L} = \mathbf{R}\times\mathbf{P}$ 
is built from the canonical position and momenta variables, and the 
canonical PBs~\eqref{eq:PB-rules-1} imply $\{L_i, L_j\} = \epsilon_{ijk} L_k$, 
the same algebra as the spins. In the OA system, $\mathbf{R}$ and 
$\mathbf{P}$ have been averaged out and only $\mathbf{L}$ remains; 
we therefore take Eqs.~\eqref{eq:SO3-PBs}, inherited from the full system, 
as the fundamental brackets. $L^2$, $S_1^2$, and $S_2^2$ commute with every 
function of $(\mathbf{L},\mathbf{S}_1,\mathbf{S}_2)$, and they are called 
Casimir invariants. Thus they are conserved and the OA phase space is the 
product of the spherical manifolds for $\vb{S}_1, \vb{S}_2$ and 
$\vb{L}$: $S^2(\mathbf{S}_1) \times S^2(\mathbf{S}_2) \times S^2(\mathbf{{L}})$, 
with $L$, $S_1$, and $S_2$ as fixed parameters.

However, we note that the variables $(\mathbf{L},\mathbf{S}_1,\mathbf{S}_2)$ 
are not canonical. The canonical pairs are $(\phi_{S_1}, S_{1z})$, $(\phi_{S_2}, S_{2z})$, 
and $(\phi_{L}, L_{z})$, where $\phi_V$ is the azimuthal angle corresponding to 
$\mathbf V$ about a fixed axis $\hat{\mathbf z}$, chosen not parallel to 
$\mathbf J$ so that $J^2$ and $J_z$ are independent. 
One can verify that Eqs.~\eqref{eq:SO3-PBs} imply~\cite{Tanay2023}
\begin{equation} \label{eq:angular_momenta_canonical_brackets}
\{\phi_{S_1}, S_{1z}\} = \{\phi_{S_2}, S_{2z}\} = \{\phi_L, L_z\} = 1,
\end{equation}
and all other brackets among them vanish (apart from the ones connected via 
anti-symmetry), thereby supporting the canonical nature of these coordinates.

This Hamiltonian and these brackets were obtained in
Ref.~\cite{correia:hal-02319812} by orbit averaging the full
Hamiltonian. In the rest of this section, we check that the OA EOMs are
Hamiltonian without assuming prior knowledge of $\bar{H}$ (Sec.~\ref{sec:discovering}), 
then show why orbit averaging the full Hamiltonian gives the OA EOMs, and
finally derive $\bar{H}$ from the OA EOMs alone (Sec.~\ref{sec:derivation_OA}).

\subsection{Testing the Hamiltonian nature of the system} \label{sec:discovering}
 
As per Ref.~\cite{jose1998classical}, a system is Hamiltonian if 
and only if for any two functions $f$ and $g$ of the phase space 
variables $\vb{V}$,
 \begin{align}
     \frac{d}{dt} \pb{f}{g} - \pb{\dot{f}}{g} - \pb{f}{\dot{g}} = 0,
 \end{align}
holds true.~\cref{eq:PB-rules-2} lets us write
\begin{equation}
\frac{d}{dt}\{f,g\} - \{\dot{f},g\} - \{f,\dot{g}\}
= \partial_i f\,\partial_j g
\left[\frac{d}{dt}\{V^i,V^j\} - \{\dot{V}^i,V^j\} - \{V^i,\dot{V}^j\}\right],
\end{equation}
which implies that the condition
\begin{align}      \label{eq:leibniz-defect}
   \frac{d}{dt}\{V^i,V^j\} - \{\dot{V}^i,V^j\} - \{V^i,\dot{V}^j\}  = 0,
\end{align}
alone guarantees the Hamiltonian nature of 
a system. It can be checked that the system 
of Eq.~\eqref{eq:OA-eom} with PBs
given by Eq.~\eqref{eq:SO3-PBs} indeed 
satisfies~\cref{eq:leibniz-defect}.
This confirms, without using the explicit form of 
$\bar{H}$, that the OA EOMs~\eqref{eq:OA-eom} 
correspond to a Hamiltonian. We now show how this Hamiltonian 
is obtained.

\subsection{Derivation of the orbit-averaged Hamiltonian} \label{sec:derivation_OA}
We obtain $\bar H$ in two ways, (a) by orbit averaging the full
Hamiltonian $H$, and (b) from an ansatz built from the OA EOMs alone.

\subsubsection{Derivation via direct orbit averaging}
If the full non-OA Hamiltonian (presented in \cref{eq:hamiltonian}) 
is known, one can derive $\bar{H}$ directly. We first remark that 
the purely orbital terms $H_{\text{N}}$, $H_{\text{1PN}}$, and $H_{\text{2PN}}$ 
commute with $(\mathbf{L},\mathbf{S}_1,\mathbf{S}_2)$ and do not 
contribute to Eqs.~\eqref{eq:eom}. Averaging the remaining terms 
$H_{\text{SO}}$ and $H_{\text{SS}}$ over $\mathbf{R}$ (including $\mathbf{n}$) 
with Eqs.~\eqref{eq:how_to_average}, while holding $(\mathbf{L},\mathbf{S}_1,\mathbf{S}_2)$ 
fixed, gives
\begin{equation} \label{eq:H_averaged}
\langle H \rangle = \bar{H} + \frac{1}{4c^2D^3}\left[\frac{m_2}{m_1}S_1^2 + \frac{m_1}{m_2}S_2^2\right].
\end{equation}
Holding the angular momenta fixed is justified  because they evolve 
on timescales much longer than $T_{\text{orb}}$ (Sec.~\ref{sec:system}). 
The Casimir terms in square brackets in Eq.~\eqref{eq:H_averaged} 
commute with all functions in phase space of the OA system and therefore 
do not contribute to the dynamics. They can thus be omitted. 
One can check that $\langle H \rangle $ generates the OA 
EOMs~\eqref{eq:OA-eom} using Eq.~\eqref{eq:evolution_eq}. 
However it is instructive to see how it comes about using simple arguments, 
as we do below, first for $\vb{S}_a$ and then for $\vb{L}$.

\textit{Spins.} The right-hand sides of Eqs.~\eqref{eq:how_to_average} depend 
on $\hat{\mathbf L}$ but not on $\mathbf S_a$, and $\mathbf S_a$ is fixed 
while orbit averaging. Averaging therefore commutes with $\partial/\partial\mathbf S_a$, 
and since $\langle H\rangle=\bar H$ up to the omitted Casimir terms, 
\begin{equation}       \label{eq:spin-OA-EOM-4}
\left\langle\frac{d\mathbf S_a}{dt}\right\rangle
=\left\langle\frac{\partial H}{\partial\mathbf S_a}\right\rangle\times\mathbf S_a
=\frac{\partial\bar H}{\partial\mathbf S_a}\times\mathbf S_a=\{\mathbf S_a,\bar H\}.
\end{equation}

\textit{Orbital angular momentum.} The above argument does not apply 
to $\mathbf L$, because the averages \eqref{eq:how_to_average} depend on 
$\mathbf L$, so averaging does not commute with $\partial/\partial\mathbf L$. 
We use instead the conservation of the total angular momentum 
$\mathbf J=\mathbf L+\mathbf S_1+\mathbf S_2$ in both systems. Only the status 
of $\mathbf L$ differs. In the full system, $\mathbf L=\mathbf R\times\mathbf P$ 
is a function of the canonical variables, with the PBs~\eqref{eq:PB-rules-1}, 
from which $\{L_i,L_j\}=\epsilon_{ijk}L_k$ follows. In the OA system, $\mathbf L$ 
is itself a phase-space variable, with the PBs~\eqref{eq:SO3-PBs}. In both cases, 
for any fixed unit vector $\hat{\mathbf{e}}$, the flow generated by $\hat{\mathbf{e}}\cdot\mathbf{J}$,
\begin{equation}\label{eq:rotation_generator}
\frac{d\mathbf V}{d\lambda_{\hat{\mathbf e}\cdot\mathbf J}}=\{\mathbf V,\hat{\mathbf e}\cdot\mathbf J\}=\hat{\mathbf e}\times\mathbf V,
\qquad
\mathbf V\in\{\mathbf R,\mathbf P,\mathbf S_a\}\ \text{(full)},\quad
\mathbf V\in\{\mathbf L,\mathbf S_a\}\ \text{(OA)},
\end{equation}
rigidly rotates all vectors about $\hat{\mathbf{e}}$~\cite{jose1998classical, arnold2013mathematical}. 
$H$ and $\bar H$ depend on these vectors only through scalar products and magnitudes, 
with no fixed external direction. They are therefore invariant under these rotations, 
$\{H,\mathbf J\}=\{\bar H,\mathbf J\}=0$, and $\mathbf J$ is conserved in both systems. 
It follows that
\begin{equation}
\left\langle\frac{d\mathbf L}{dt}\right\rangle
=-\sum_a\left\langle\frac{d\mathbf S_a}{dt}\right\rangle
=-\sum_a\{\mathbf S_a,\bar H\}
=\{\mathbf L,\bar H\},
\end{equation}
where the first equality uses the conservation of $\mathbf J$ in the full non-OA system, 
the second uses the averaged spin EOM~\cref{eq:spin-OA-EOM-4}, and the third uses the 
conservation of $\mathbf J$ in the OA system.

\subsubsection{Derivation via ansatz matching}
Alternatively, $\bar H$ can be reconstructed from the OA EOMs~\eqref{eq:OA-eom} 
alone without the knowledge of the full Hamiltonian. We take the most
general scalar built from $(\mathbf L,\mathbf S_1,\mathbf S_2)$ at
spin-orbit and spin-spin order that is invariant under rotations and time
reversal. Since time reversal flips all angular momenta, it excludes the
triple product $\mathbf L\cdot(\mathbf S_1\times\mathbf S_2)$. The ansatz
reads
\begin{align}\label{eq:H_ansatz}
\bar H    =\frac{1}{c^2D^3}\Big[ & k_1\,\mathbf L\cdot\mathbf S_1+k_2\,\mathbf S_1\cdot\mathbf S_2+k_3\,\mathbf L\cdot\mathbf S_2 
+k_{4}\frac{(\mathbf L\cdot\mathbf S_1)^2}{L^2}+k_5\frac{(\mathbf L\cdot\mathbf S_1)(\mathbf L\cdot\mathbf S_2)}{L^2}
+k_6\frac{(\mathbf L\cdot\mathbf S_2)^2}{L^2}\Big].
\end{align}
Being built from scalar products only, it is invariant under the rotations 
generated by $\mathbf J$, so $\{\mathbf J,\bar H\}=0$. Requiring that 
Eq.~\eqref{eq:LP-eom} reproduces Eqs.~\eqref{eq:OA-eom} gives the unique solution
\begin{equation}\label{eq:ansatz_solution}
k_1=2\sigma_1,\quad k_2=\frac12,\quad k_3=2\sigma_2,\quad
k_4=-\frac34\,\frac{m_2}{m_1},\quad k_5=-\frac32,\quad k_6=-\frac34\,\frac{m_1}{m_2},
\end{equation}
i.e., Eq.~\eqref{eq:H_bar}. $\bar H$ is fixed up to an additive term which 
could be a function of the Casimirs $L^2$, $S_1^2$ and $S_2^2$; this term 
has vanishing PBs with all variables and does not affect the EOMs.

\section {Integrability and action variables}
\label{sec:Integrability_AAV}

\subsection{The setup}

On the 6-dimensional standard phase space (SPS)
$S^2(\mathbf{S}_1) \times S^2(\mathbf{S}_2) \times S^2(\mathbf{L})$, 
where the Casimirs $L$, $S_1$, and $S_2$ are fixed, the OA EOMs 
admit the three COMs
\begin{align} \label{eq:COMs}
    \vb{C} = \left( J^2, J_z, \Lambda \right) ,
\end{align}
with $J \equiv \norm{\vb{L} + \vb{S}_1 + \vb{S}_2}$ and 
$\Lambda = \vb{L} \cdot \vb{S}_0$. We use $J^2$ rather than $J$ 
as a COM, since $J^2$ is a smooth function on the whole phase space, 
whereas $J$ is not differentiable at $J=0$ (see Sec.~\ref{sec:separatrix}).
$J^2$ and $J_z$ are conserved since $\{\bar H, \mathbf{J}\} = 0$ 
(Sec.~\ref{sec:derivation_OA}). For $\Lambda$, we use $J^2 = L^2 + S_1^2 + S_2^2 + 2\,
\mathbf{L}\cdot(\mathbf{S}_1+\mathbf{S}_2) + 2\,\mathbf{S}_1\cdot\mathbf{S}_2$ 
to rewrite $\bar{H}$ of Eq.~\eqref{eq:H_bar} in terms of $J^2$, $\Lambda$, and the 
Casimirs only
\begin{equation} \label{eq:hamiltonian_dependent}
    \bar{H} = \frac{1}{4 c^2 D^3} \left[ J^2 - L^2 - S_1^2 - S_2^2 + 6 \Lambda - \frac{3 \nu \Lambda^2}{L^2} \right] .
\end{equation}
Since $\Lambda$ and $J^2$ are scalars, they commute with every component of $\mathbf J$, the
generator of rotations [Eq.~\eqref{eq:rotation_generator}], so that $J^2$, $J_z$ and $\Lambda$ are in
involution. Eq.~\eqref{eq:hamiltonian_dependent} then gives
$\{\Lambda,\bar H\}=\{\Lambda,J^2\}\,\partial\bar H/\partial J^2=0$, so $\Lambda$ is conserved.
Hence $\bar H$ is not an independent COM and is not included in $\vb{C}$.
The three COMs in $\vb{C}$ are also independent, $dC_1 \wedge dC_2 \wedge dC_3 \neq 0$, 
except at some special configurations like the (anti-)aligned configurations, spin–orbit 
resonances and certain other points (as discussed in Sec.~\ref{sec:separatrix}).\footnote{At 
the so-called spin-orbit resonances, the $(\vb{L}, \vb{S}_1, \vb{S}_2)$ triad becomes coplanar 
and rotates about the fixed $\vb{J}$ rigidly. Hence, $d \Lambda \wedge d(J^2) = 0$. 
The triad becoming coplanar signifies a reduction in the dimension of the ``invariant torus'' 
(to be introduced later) on which the dynamics evolve, which is due to
$dC_1 \wedge dC_2 \wedge dC_3 =0$.
}
The 2PN OA system is therefore integrable~\cite{arnold2013mathematical,
jose1998classical}, as noted in Ref.~\cite{correia:hal-02319812}, where an
equivalent form of Eq.~\eqref{eq:hamiltonian_dependent} appears.

Integrability also extends to the case $m_1 = m_2$, where $\Lambda = \Sigma\, \vb{L}\cdot(\vb{S}_1 + \vb{S}_2)$, 
with $\Sigma_1 = \Sigma_2 \equiv \Sigma$, remains independent of $J^2$ and $J_z$. Only the explicit 
expression of the third action variable (Sec.~\ref{sec:third_action}) degenerates in this limit.

\subsection{Connection between the orbit-averaged and the full 2PN constants of motion}
A natural question is how the OA COMs relate to those of the full (non-OA) 2PN system.
We intuitively expect that orbit averaging maps 
the COMs of the full system to those of the OA system.

Besides $H$ and $\mathbf{J}$, the full 2PN system possesses two more COMs~\cite{Tanay2020}
\begin{align}
\widetilde{L^2} &= \underbrace{L^2}_{\text{0PN}} + \underbrace{\delta L^2}_{\text{2PN}}, \\
\widetilde{\mathbf{S}_{\text{eff}} \cdot \mathbf{L}} &= \underbrace{\mathbf{S}_{\text{eff}} \cdot \mathbf{L}}_{\text{0PN}} + \underbrace{\frac{1}{4}\,\mathbf{S}_1 \cdot \mathbf{S}_2}_{\text{0.5PN}} + \underbrace{\delta_2(\mathbf{S}_{\text{eff}} \cdot \mathbf{L})}_{\text{1.5PN}},
\end{align}
where the PN orders are indicated relative to the leading term. The OA 
Hamiltonian~\eqref{eq:H_bar} contains only the leading spin-orbit term 
and the spin-spin term, the latter being smaller by a relative 0.5PN factor. 
The OA COMs can therefore match the full COMs only up to relative 0.5PN order, 
and $\delta L^2$ and $\delta_2$ must be dropped. Doing so gives:
\begin{itemize}
    \item $\widetilde{L^2} \to L^2$, which is a Casimir of the OA system and is therefore conserved;
    \item $\widetilde{\mathbf{S}_{\text{eff}} \cdot \mathbf{L}} \to \Lambda_{\text{eff}} \equiv \mathbf{S}_{\text{eff}} \cdot \mathbf{L} + \mathbf{S}_1 \cdot \mathbf{S}_2/4$, which is a COM of the OA system, since
    \begin{equation} \label{eq:Lambda_eff_OA}
    \Lambda_{\text{eff}} = \frac{1}{2}\, c^2 D^3\, \bar H + \frac{3\nu}{8}\frac{\Lambda^2}{L^2} .
    \end{equation}
\end{itemize}
Each COM of the full system thus corresponds to a COM of the OA system:
\begin{equation}
\big(H,\ \mathbf{J},\ \widetilde{L^2},\ \widetilde{\mathbf{S}_{\text{eff}} \cdot \mathbf{L}}\big) \;\longrightarrow\; \big(\bar H,\ \mathbf{J},\ L^2,\ \Lambda_{\text{eff}}\big) .
\end{equation}
Moreover, $k = 1/4$ is the only coefficient for which $\mathbf{S}_{\text{eff}} 
\cdot \mathbf{L} + k\,\mathbf{S}_1 \cdot \mathbf{S}_2$ is conserved by the OA 
EOMs~\eqref{eq:OA-eom}. This provides an independent consistency check between the 
full COM of Ref.~\cite{Tanay2020} and the OA Hamiltonian~\eqref{eq:H_bar}. 

Finally, $\Lambda$ and $\Lambda_{\text{eff}}$ are related by
\begin{equation} \label{eq:linear_comb}
\Lambda = \frac{4}{3}\, \Lambda_{\text{eff}} - \frac{1}{6}\left(J^2 - L^2 - S_1^2 - S_2^2\right) .
\end{equation}
Both are valid choices for the third COM for the OA system. We use $\Lambda$ 
because its flow equation is simpler, which simplifies the computation of the 
action variables in Sec.~\ref{sec:third_action}.

\subsection{Defining canonical coordinates and action variables}\label{sec:canonical_actions}

The OA system is integrable, and its 6-dimensional phase space
$S^2(\mathbf{S}_1)\times S^2(\mathbf{S}_2)\times S^2(\mathbf{L})$ 
is compact. By the Liouville--Arnold theorem~\cite{arnold2013mathematical}, 
at fixed $L$, $S_1$, and $S_2$, each connected component of a regular 
level set, on which the COMs are independent, 
\begin{equation}\label{eq:torus}
    \vb{C} = \text{const},
\end{equation}
is then a 3-torus. 

The torus is built from the flows of the COMs. The flow generated by a 
phase-space function $h$ is encapsulated by the following evolution 
equation~\cite{arnold2013mathematical, jose1998classical}
\begin{equation}\label{eq:flow_def}
    \frac{dg}{d\lambda_h} = \{g, h\},
\end{equation}
for any phase-space function $g$, and $\lambda_h$ is the flow parameter.
When $g$ and $h$ are in involution, $dg/d\lambda_h = 0$. Thus both
$g$ and $h$ remain constant along the flow of the other, and these 
flows stay on the level set $\vb{C}=$ const. Since $\bar H$ is a
function of the COMs,  which are in mutual involution, 
[Eq.~\eqref{eq:hamiltonian_dependent}], the same is true for the time evolution flow,
\begin{equation}
   \frac{dg}{dt}  = \frac{dg}{d\lambda_{\bar{H}
    }} = \{g, \bar{H}\},
\end{equation}
and we say that the torus is invariant. Since our COMs are in involution, 
their flows commute (Sec.~39 of~\cite{arnold2013mathematical}), and they 
can be applied in any order without leaving the torus.

For an integrable system with $n$ canonical pairs $(q_i,p_i)$, the action 
variables are defined as
\begin{equation}\label{eq:action-general}
\mathcal{J}_k = \frac{1}{2\pi}\oint_{\mathcal{C}_k}\sum_{i=1}^{n}p_i\,dq_i ,
\qquad k=1,\dots,n ,
\end{equation}
where the homotopy classes of the loops $\mathcal{C}_k$ 
generate the fundamental group of the torus. $\mathcal{J}_k$ 
does not depend on the choice of loop within its class 
(Proposition~11.2 of Ref.~\cite{Fasano2006}).
Here $n=3$, and with the canonical pairs $(\phi_V,V_z)$ of
Eq.~\eqref{eq:angular_momenta_canonical_brackets},
\begin{equation}\label{eq:action-defined}
\mathcal{J}_k = \frac{1}{2\pi}\oint_{\mathcal{C}_k}\sum_{\mathbf V}V_z\,d\phi_V,
\qquad \mathbf V\in\{\mathbf L,\mathbf S_1,\mathbf S_2\} .
\end{equation}
Two of these loops come directly from the flows of $J_z$ and 
$J^2$, which rotate the whole system rigidly about 
$\hat{\mathbf z}$ and $\hat{\mathbf J}$ 
[Eq.~\eqref{eq:rotation_generator}]. The third
one is more complicated and involves the flow of $\Lambda$ (Sec.~\ref{sec:third_action}). 

Since $\phi_V$ is undefined at the poles 
$\mathbf V\parallel\pm\hat{\mathbf z}$, a deformation
of a loop that crosses a pole shifts $\mathcal{J}_k$ by 
$\pm V$ (App.~\ref{app:invariant_action_stokes_rotation}). 
The actions are thus defined up to $n_L L+n_{S_1}S_1+n_{S_2}S_2$, with 
integer $n$'s, which gives again action-angle variables
(Proposition~11.3 of Ref.~\cite{Fasano2006}).

\subsection{First two action variables}
\label{sec:first_two_actions}

The flow generated by $J^2$ follows from Eq.~\eqref{eq:rotation_generator} with
$\hat{\vb{e}} = \hat{\vb{J}}$ and the chain rule 
\begin{equation} \label{eq:V_flow}
    \frac{d\vb{V}}{d\lambda_{J^2}} = \{\vb{V}, J^2\} = 2J \,\hat{\vb{J}} \times \vb{V} , \qquad \vb{V} \in \{\vb{L}, \vb{S}_1, \vb{S}_2\} .
\end{equation}
All three vectors thus rotate rigidly about the fixed $\hat{\vb{J}}$ 
at angular velocity $2J$, and the configuration returns to itself after 
$\lambda_{J^2} = \pi/J$. This closed loop gives the first action
\begin{equation} \label{eq:first_action_expansion}
    \mathcal{J}_1 = \frac{1}{2\pi} \oint_{J^2\text{-flow}} \left( L_z\, d\phi_L + S_{1z}\, d\phi_{S_1} + S_{2z}\, d\phi_{S_2} \right) .
\end{equation}

\begin{figure*}[t] 
    \centering
    \includegraphics[width=0.9\textwidth]{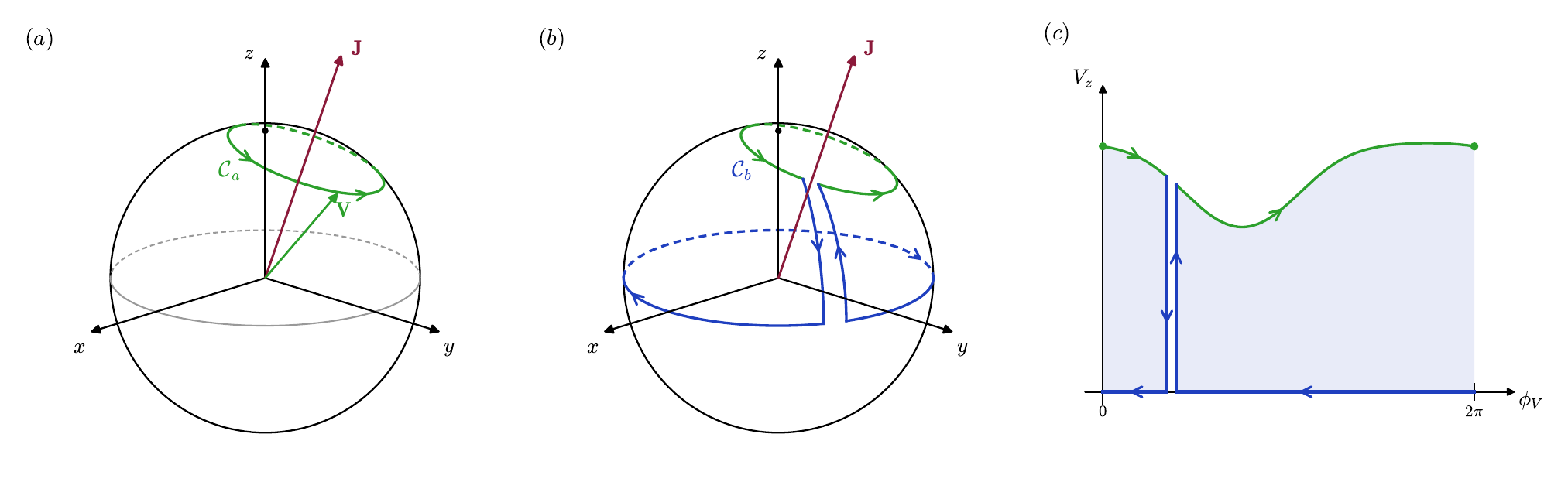}
    \hfill
    \caption{
    Computation of $\oint V_z\, d\phi_V$ for $\vb{V}\in \{\vb{L},\vb{S}_1,\vb{S}_2\}$, 
    when the north pole lies inside the cap bounded by the loop and the south pole outside (case shown in all three panels).
 (a).~Under the flow generated by $J^2$, the tip of $\vb{S}_1$ traces a circle (green) about $\vb{J}$.
 (b).~Closed contour: the loop (green), completed by a meridian arc, the equator traversed backward
and the same meridian arc (blue). The blue arcs do not contribute, so the loop and the contour give the same integral.
 (c).~The same contour in the $(\phi_{V}, V_{z})$ plane. Since the loop goes once around the
$z$-axis, $\phi_{V}$ runs from $0$ to $2\pi$ along it. The shaded area equals
$\oint V_{z}\, d\phi_{V}$. In  (b) and  (c), the two meridian arcs are drawn slightly apart for
clarity.}
    \label{fig:stokes_theorem}
\end{figure*}

Along this loop, each vector $\vb{V} \in \{\vb{L},\vb{S}_1,\vb{S}_2\}$ 
rotates counterclockwise (when seen from outside the sphere) about $\hat{\vb{J}}$ at 
fixed magnitude, so its tip traces a circle on the sphere of radius $V$ 
[Fig.~\ref{fig:stokes_theorem}(a)].
This circle divides the sphere into two parts.
We call the part that is towards $\hat{\vb{J}}$, i.e., the part containing the 
point $V\hat{\vb{J}}$, the \emph{cap}.
The area of the cap is $2\pi V(V-\vb{V}\cdot\hat{\vb{J}})$.
Therefore a cap defined as such can be bigger than a hemisphere.
The evaluation of $\oint V_z\, d\phi_V$ depends on the position 
of the poles relative to this cap, because $\phi_V$ is measured about 
the $z$-axis and is undefined at the poles. If the north pole lies inside 
the cap and the south pole outside, i.e., if 
$-\hat{\vb{z}}\cdot\hat{\vb{J}} < \hat{\vb{V}}\cdot\hat{\vb{J}} < \hat{\vb{z}}\cdot\hat{\vb{J}}$, 
the loop goes once around the $z$-axis and $\phi_V$ increases by $2\pi$ along it. 
We treat this case here, as in all panels of Fig.~\ref{fig:stokes_theorem}, 
and show in App.~\ref{app:invariant_action_stokes_rotation} that the three 
other cases lead to the same action, up to $n_L L+n_{S_1}S_1+n_{S_2}S_2$, 
with $n_V \in \mathbb{Z}$.

To evaluate $\oint V_z\, d\phi_V$ as an area $\int d\phi_V\, dV_z$ with Stokes'
theorem, the loop must bound a region on which $\phi_V$ is defined everywhere.
Neither region bounded by the loop qualifies, since the cap contains the north
pole and its complement the south pole. We therefore close the loop
$\mathcal{C}_a$ of Fig.~\ref{fig:stokes_theorem}(a) into the contour
$\mathcal{C}_b$ of Fig.~\ref{fig:stokes_theorem}(b), adding a meridian arc down
to the equator, the equator traversed backward, and the same meridian arc back
up, with the two meridians infinitesimally close. The new contour encloses no
pole, and the added arcs do not contribute, since $d\phi_V = 0$ on a meridian
and $V_z = 0$ on the equator. Therefore
\begin{equation} \label{eq:loop_equals_contour}
    \oint_{\mathcal{C}_a} V_z\, d\phi_V = \oint_{\mathcal{C}_b} V_z\, d\phi_V .
\end{equation}

In the $(\phi_V,V_z)$ plane, the loop of Fig.~\ref{fig:stokes_theorem}(a)
becomes a curve $V_z(\phi_V)$ with $\phi_V$ running from $0$ to $2\pi$, and
$\oint V_z\,d\phi_V$ is the area under this curve
[Fig.~\ref{fig:stokes_theorem}(c)]. Since the area element on the sphere of
radius $V$ is $V\,d\phi_V\,dV_z$, this area is $1/V$ times the corresponding
area on the sphere, the region between the equator and the loop. It is
therefore the area of the upper hemisphere minus that of the cap, divided by
$V$\footnote{If the loop crosses the equator, the parts below it count
negatively, and the result still holds.}
\begin{equation} \label{eq:loop_integral}
    \oint_{\mathcal{C}_a} V_z\, d\phi_V
    = \frac{2\pi V^2 - 2\pi V\,(V-\vb{V}\cdot\hat{\vb{J}})}{V}
    = 2\pi\, \vb{V}\cdot\hat{\vb{J}} .
\end{equation}
Inserting Eq.~\eqref{eq:loop_integral} into Eq.~\eqref{eq:first_action_expansion} gives
\begin{equation} \label{eq:action_1_final}
    \mathcal{J}_1 = \vb{L}\cdot\hat{\vb{J}} + \vb{S}_1\cdot\hat{\vb{J}} + \vb{S}_2\cdot\hat{\vb{J}} = J .
\end{equation}

The other cases occur in practice. For instance, when $L \gg S_1, S_2$, $\vb{L}$ is close to
$\vb{J}$ and its cap is small, so that it generally contains neither pole. 
Going from the case when the loop contains a pole to the one where it does not
requires a pole to cross the loop, which we discuss in
Sec.~\ref{sec:canonical_actions}. In App.~\ref{app:invariant_action_stokes_rotation}, we show that,
for any position of the poles (with respect to the loop of Fig.~\ref{fig:stokes_theorem}(a)),
$\oint V_z\, d\phi_V = 2\pi[\vb{V}\cdot\hat{\vb{J}} + (n_N + n_S - 1)V]$, where $n_N$ ($n_S$) is $1$
if the north (south) pole lies inside the cap and $0$ otherwise. The other cases thus shift
$\mathcal{J}_1$ by an integer combination of $L$, $S_1$, and $S_2$. 
So  Eq.~\eqref{eq:action_1_final} holds for any
orientation of the $z$-axis, up to integer multiples of 
$L$, $S_1$ and $S_2$.

The second action follows from the flow of $J_z$, which by Eq.~\eqref{eq:rotation_generator} with
$\hat{\vb{e}}=\hat{\vb{z}}$ rotates all vectors rigidly about the $z$-axis,
$d\vb{V}/d\lambda_{J_z} = \hat{\vb{z}}\times\vb{V}$. Along this loop, each tip traces a circle
about the $z$-axis itself, so $V_z$ is constant and $\phi_V$ increases by $2\pi$. 
Here we have $(n_N, n_S) = (1, 0)$. Each term then contributes $2\pi V_z$, and
\begin{equation} \label{eq:action_2_final}
    \mathcal{J}_2 = L_z + S_{1z} + S_{2z} = J_z .
\end{equation}

\subsection{The third action variable}

\subsubsection{Introducing fictitious variables}
The first two actions follow from rigid rotations about fixed axes, $\hat{\vb{J}}$ and
$\hat{\vb{z}}$, along which each integral reduces to an area on a sphere. The loop defining the
third action is instead built from the $\Lambda$-flow, which is not
a rigid rotation. As we will see in Sec.~\ref{sec:third_action}, $\vb{L}$ precesses about $\vb{S}_0$, 
and each $\vb{S}_a$ about $\vb{L}$, so each vector rotates about an axis that itself moves. 
Along such a motion, $V_z$ and $\phi_V$ vary in a complicated way, and we have not found a direct 
way to compute $\oint V_z\, d\phi_V$, as for the previous two actions.
The same difficulty arises in Ref.~\cite{Tanay2023},\footnote{In Ref.~\cite{Tanay2023}, the action variables 
were computed for the non-OA BBH system at 1.5PN order. The analog of $\Lambda$ was $\vb{S}_{\text{eff}}\cdot\vb{L}$.}
where it is bypassed by introducing fictitious variables. We follow this method and refer to
Ref.~\cite{Tanay2023} for details.

The method rests on the following observation. When $\vb{V}$ is rotated by a small angle $d\alpha$
about $\hat{\vb{z}}$, $\phi_V$ increases by $d\alpha$, and the integrand is
$V_z\, d\phi_V = (\hat{\vb{z}}\cdot\vb{V})\, d\alpha$. In words, the integrand is the component of
$\vb{V}$ along the rotation axis times the rotation angle. For a rotation about any other axis,
$V_z\, d\phi_V$ no longer takes this form.
If instead $\vb{V}$ is written 
as $\vb{V}=\vb{r}\times\vb{p}$, with canonical pairs $(\vb r,\vb p)$, the integrand 
of Eq.~\eqref{eq:action-general} becomes $\vb{p}\cdot d\vb{r}$, which takes this form for 
every axis.
Indeed, a rotation by $d\alpha$ about an arbitrary axis $\hat{\vb{n}}$ gives
$d\vb{r} = d\alpha\, \hat{\vb{n}}\times\vb{r}$, so that
\begin{equation} \label{eq:pdr_rotation}
    \vb{p}\cdot d\vb{r} = (\hat{\vb{n}}\cdot\vb{V})\, d\alpha .
\end{equation}
This form makes the action computation feasible; see Eqs.~(28)--(33) of Ref.~\cite{Tanay2023}.

For each of the angular momenta $\vb{V}\in\{\vb{L},\vb{S}_1,\vb{S}_2\}$, which we call 
the physical variables in contrast to the fictitious ones, we introduce a fictitious position 
$\vb{r}_V$ and a fictitious momentum $\vb{p}_V$ such that
\begin{equation} \label{eq:V_fictitious}
    \vb{V} = \vb{r}_V \times \vb{p}_V ,
\end{equation}
with the canonical PBs
\begin{equation} \label{eq:PB_fictitious}
    \{ r_{V i}, p_{V' j} \} = \delta_{VV'}\,\delta_{ij},
\end{equation}
and all other PBs between these variables being zero apart from the ones connected 
via anti-symmetry.\footnote{Fictitious variables can be seen as instances of symplectic 
realizations in the literature concerning Poisson manifolds~\cite{vaisman2012lectures, crainic2021lectures}. 
We thank Leo Stein for pointing this out.} These nine canonical pairs span an
18-dimensional extended phase space (EPS), as opposed to the standard phase space (SPS) 
of Sec.~\ref{sec:SPS_PB}. With Eq.~\eqref{eq:V_fictitious}, the PBs~\eqref{eq:PB_fictitious}
of the fictitious variables imply the SO(3) PBs~\eqref{eq:SO3-PBs} of the physical variables.
Any function of the physical variables, in particular $\bar H$, therefore has 
the same dynamics in the SPS and in the EPS. In the EPS, the action integrand 
is $\sum_{\vb{V}} \vb{p}_V\cdot d\vb{r}_V$.

Along the flow of any function $F(\vb{L},\vb{S}_1,\vb{S}_2)$, the
PBs~\eqref{eq:PB_fictitious} and the chain rule applied to
Eq.~\eqref{eq:V_fictitious} give
\begin{equation} \label{eq:fictitious_rotation}
    \frac{d\vb{r}_V}{d\lambda_F} = \frac{\partial F}{\partial \vb{V}}\times\vb{r}_V, \qquad
    \frac{d\vb{p}_V}{d\lambda_F} = \frac{\partial F}{\partial \vb{V}}\times\vb{p}_V .
\end{equation}
The fictitious vectors $\vb{r}_V$ and $\vb{p}_V$ thus obey the same equation as $\vb{V}$
[Eq.~\eqref{eq:LP-eom}]: at each instant, $\vb{r}_V$, $\vb{p}_V$, and $\vb{V}$ undergo the same
rotation, with angular velocity $\partial F/\partial\vb{V}$. Their lengths and mutual angles are
therefore preserved, and $\vb{r}_V$ and $\vb{p}_V$ remain in the plane orthogonal to $\vb{V}$.
Applying Eq.~\eqref{eq:pdr_rotation} to each pair, with
$\hat{\vb{n}}\, d\alpha = (\partial F/\partial\vb{V})\, d\lambda_F$, gives
\begin{equation} \label{eq:EPS_integrand}
    \sum_{\vb{V}} \vb{p}_V\cdot d\vb{r}_V
    = \sum_{\vb{V}} \vb{V}\cdot\frac{\partial F}{\partial \vb{V}}\, d\lambda_F .
\end{equation}
For $F=J^2$, $\partial F/\partial\vb{V} = 2\vb{J}$ for each $\vb{V}$, and the
sum in Eq.~\eqref{eq:EPS_integrand} is $\sum_{\vb{V}}\vb{V}\cdot 2\vb{J} = 2J^2$.
Since the $J^2$-loop closes after $\lambda_{J^2} = \pi/J$, we obtain
$2\pi\mathcal{J}_1 = 2J^2\,\pi/J$, i.e., $\mathcal{J}_1 = J$ [Eq.~\eqref{eq:action_1_final}], which provides a check of the method. 

Three points are worth mentioning. First, the fictitious variables are not unique. 
Indeed, for a given $\vb{V}$, any pair $(\vb{r}_V,\vb{p}_V)$ lying in the plane 
orthogonal to $\vb{V}$ and satisfying $\vb{r}_V\times\vb{p}_V=\vb{V}$ is admissible.
This freedom affects the actions only through an integer combination of the Casimirs, 
$n_L L + n_{S_1} S_1 + n_{S_2} S_2$, as shown below [Eq.~\eqref{eq:EPS_SPS_loop}].

Second, the EPS and SPS integrands differ, because the EPS integrand $\vb{p}_V\cdot d\vb{r}_V$ 
is sensitive to a motion that the physical variables cannot see. 
Indeed, by the cross product structure of Eq.~\eqref{eq:LP-eom}, 
the component of $\partial F/\partial\vb{V}$ along $\vb{V}$ leaves 
$\vb{V}$ fixed. By Eq.~\eqref{eq:fictitious_rotation}, however, it 
rotates $\vb{r}_V$ and $\vb{p}_V$ (which are orthogonal to $\vb{V}$) 
about $\vb{V}$. Using Eq.~\eqref{eq:pdr_rotation}, such a rotation by 
$d\alpha$ gives $\vb{p}_V\cdot d\vb{r}_V = V\, d\alpha$, whereas 
$V_z\, d\phi_V = 0$. To separate this rotation from the motion of $\vb{V}$, 
we measure the orientation of the pair by the angle $\psi_V$ such that 
\begin{equation} 
    \hat{\vb{r}}_V = \cos\psi_V\,\hat{\vb{e}}_\theta + \sin\psi_V\,\hat{\vb{e}}_\phi ,
\end{equation}
where $\hat{\vb{e}}_\theta$ and $\hat{\vb{e}}_\phi$ are unit polar 
and azimuthal vectors corresponding to the polar and azimuthal angles 
of $\vb{V}$. Along any flow of the form of
Eq.~\eqref{eq:fictitious_rotation}, we show in 
App.~\ref{app:invariant_action_stokes_rotation} that
\begin{equation} \label{eq:EPS_SPS}
    \vb{p}_V\cdot d\vb{r}_V = V_z\, d\phi_V + V\, d\psi_V .
\end{equation}
Along a loop that is closed in the EPS, $\vb{V}$, $\vb{r}_V$, 
and $\vb{p}_V$ all return to their initial values, so $\psi_V$ 
changes by $2\pi n_V$, with $n_V$ an integer, and
\begin{equation} \label{eq:EPS_SPS_loop}
    \oint \vb{p}_V\cdot d\vb{r}_V = \oint V_z\, d\phi_V + 2\pi n_V V .
\end{equation}
Summing over $\vb{V}$, the action computed in the EPS 
equals the SPS action up to $n_L L + n_{S_1} S_1 +n_{S_2} S_2$, 
as mentioned above in Sec.~\ref{sec:canonical_actions}.

Third, the EPS is integrable as well, so the loops used there lie on tori, as in
the SPS. Since the EPS is 18-dimensional, this requires nine independent COMs in
involution. Six of them are $J^2$, $J_z$, $\Lambda$, $L^2$, $S_1^2$, and $S_2^2$,
where the last three are Casimirs in the SPS but ordinary COMs in the
EPS.\footnote{In the EPS, they no longer commute with every variable; e.g.,
$\{r_{Li}, L^2\} = 2(\vb{L}\times\vb{r}_L)_i$. They still commute with every
function of the physical variables, including $\bar{H}$.}
These six COMs are functions of the physical variables only, and they leave the
shape of each pair $(\vb{r}_V,\vb{p}_V)$ free, e.g., the length $r_V$ and the
product $\vb{r}_V\cdot\vb{p}_V$, so one more COM per pair is needed. Along the
flow of any function of the physical variables, $\vb{r}_V$ and $\vb{p}_V$ only
rotate rigidly [Eq.~\eqref{eq:fictitious_rotation}], so their lengths are
unchanged. The three quantities $r_V^2+p_V^2$ therefore commute with the six
COMs above, and with each other since they involve different pairs, and they
are independent of them. Fixing $r_V^2+p_V^2$ bounds $\vb{r}_V$ and $\vb{p}_V$,
so the level sets are compact and, by the Liouville--Arnold theorem, are tori.\footnote{
Ref.~\cite{Tanay2023} used $\vb r_V\cdot\vb p_V$ instead of $r_V^2+p_V^2$. 
Its flow, $\vb r_V\to e^{\lambda}\vb r_V$, $\vb p_V\to e^{-\lambda}\vb p_V$, 
is unbounded, so the corresponding level sets are not compact and are not tori.
The actions obtained there are unaffected, since their loops involve only rotations.
}
\subsubsection{The third action variable}
\label{sec:third_action}
To compute $\mathcal{J}_3$, we need a third closed loop on the 3-torus of the
SPS, i.e., on the level set of $(J^2, J_z, \Lambda)$, independent of the $J^2$- and
$J_z$-loops. We build it from flows in the EPS, where it is easier to evaluate,
and Eq.~\eqref{eq:EPS_SPS_loop} relates the result to the SPS action.

Since $J^2$ and $J_z$ generate rigid rotations, only the $\Lambda$-flow changes the
mutual angles of the triad $(\vb{L},\vb{S}_1,\vb{S}_2)$, so a loop containing a $\Lambda$-flow
segment is independent of the first two. Such a segment does not close by itself, and we complete
it with the flows of $J^2$, $L^2$, $S_1^2$, and $S_2^2$. 
All these COMs are in involution, so their flows commute and keep the system on the same level 
set, and they can therefore be applied one after the other.

This construction follows Ref.~\cite{Tanay2023}, where the fifth action of the non-OA 1.5PN system,
the analog of our $\mathcal{J}_3$, is built from the flow of $\vb{L}\cdot\vb{S}_{\text{eff}}$. 
Since $\Lambda = \vb{L}\cdot\vb{S}_0$ can be obtained from $\vb{L}\cdot\vb{S}_{\text{eff}}$ 
by the replacement $\sigma_a\to\Sigma_a$ (Sec.~\ref{sec:Notation}), the two flows have the same 
structure. From Eqs.~\eqref{eq:flow_def} and~\eqref{eq:SO3-PBs}, the $\Lambda$-flow of the physical 
vectors reads
\begin{align}
    \frac{d \vb{L}}{d \lambda_\Lambda} &= \vb{S}_0 \times \vb{L} , \label{eq:flow_L} \\
    \frac{d \vb{S}_a}{d \lambda_\Lambda} &= \Sigma_a\, \vb{L} \times \vb{S}_a , \label{eq:flow_SA}
\end{align}
and, by Eq.~\eqref{eq:fictitious_rotation}, each fictitious pair rotates in the same 
way as its physical vector, i.e., $(\vb{r}_L,\vb{p}_L)$ about $\vb{S}_0$, and 
$(\vb{r}_{S_a},\vb{p}_{S_a})$ about $\Sigma_a\vb{L}$. Apart from the replacement 
$\sigma_a\to\Sigma_a$, these flows are given by Eq.~(29) of Ref.~\cite{Tanay2023}, 
and the lengthy calculations of that reference simply carry over. 
We therefore only give the main steps and the final result.

The idea is to leave a point and come back to it using only these flows. 
This takes three steps
\begin{enumerate}
\item[(i)] Under the $\Lambda$-flow, the shape of the triad goes through one
full cycle after a flow amount $\Delta\lambda_\Lambda$, so that the mutual scalar
products and the triple product of $(\vb{L}, \vb{S}_1, \vb{S}_2)$ are back to
their initial values. The final triad is therefore a rigid rotation of the
initial one. Since $\Lambda$ commutes with all components of $\vb{J}$, the
$\Lambda$-flow leaves $\vb{J}$ fixed, and this rotation is about $\hat{\vb{J}}$,
by some angle $\Delta\Phi_L$.
\item[(ii)] The $J^2$-flow rotates all vectors about $\hat{\vb{J}}$ at angular
velocity $2J$. It brings the physical vectors back to their initial position
after the flow amount
\begin{equation} \label{eq:dlambda_J2_main}
    \Delta\lambda_{J^2} = -\frac{\Delta\Phi_L}{2J} .
\end{equation}
\item[(iii)] Each fictitious pair is still rotated about $\vb{V}$ by some angle
$\Delta\psi_V$. The $V^2$-flow turns it back at angular velocity $2V$ without
moving the physical vectors, after the flow amount
\begin{equation} \label{eq:dlambda_V2_main}
    \Delta\lambda_{V^2} = -\frac{\Delta\psi_V}{2V} .
\end{equation}
The loop is then closed in the EPS.
\end{enumerate}

The third action is the sum of the contributions of the five segments. 
Along the $\Lambda$-segment, the integrand of the action, computed with
\cref{eq:EPS_integrand} for $F=\Lambda$, is constant,
\begin{equation}
    \vb{L}\cdot\vb{S}_0 + \sum_a \Sigma_a\,\vb{S}_a\cdot\vb{L} = 2\Lambda ,
\end{equation}
and this segment contributes $2\Lambda\,\Delta\lambda_\Lambda$ to
$2\pi\mathcal{J}_3$.
The other generators $F\in\{J^2, L^2, S_1^2, S_2^2\}$ are also quadratic, and
each contributes $2F\,\Delta\lambda_F$ in the same way.\footnote{For a 
homogeneous function $F$ of degree $k$ in $(\vb{L},\vb{S}_1,\vb{S}_2)$, Euler's 
homogeneous function theorem gives $\sum_{\vb{V}} \vb{V}\cdot\partial F/\partial\vb{V} 
= kF$, with $k=2$ for all five generators.}
Hence, as in Eqs.~(27) and~(35) of Ref.~\cite{Tanay2023},
\begin{equation} \label{eq:action_3_final}
    \mathcal{J}_3 = \frac{1}{\pi} \left( \Lambda\, \Delta \lambda_{\Lambda} + J^2\, \Delta \lambda_{J^2}
    + L^2\, \Delta \lambda_{L^2} + S_1^2\, \Delta \lambda_{S_1^2} + S_2^2\, \Delta \lambda_{S_2^2} \right) .
\end{equation}

The $L^2$- and $S_a^2$-segments deserve a comment, since they leave the 
physical vectors fixed. Each of them rotates a fictitious pair about its 
physical vector, and therefore contributes only through the term $V\,d\psi_V$ 
of Eq.~\eqref{eq:EPS_SPS}. Their role is to undo the rotation of each pair 
about $\vb{V}$ accumulated along the $\Lambda$- and $J^2$-segments, so that 
$\psi_V$ returns to its initial value up to whole turns. Equation~\eqref{eq:action_3_final}, 
obtained in the EPS, therefore equals the action defined in the SPS by 
Eq.~\eqref{eq:action-defined}, up to $n_L L+n_{S_1}S_1+n_{S_2}S_2$ with integer 
$n_V$'s [Eq.~\eqref{eq:EPS_SPS_loop}]. 

The evaluation of $\mathcal{J}_3$ thus reduces to the computation of the 
five flow amounts, derived in App.~\ref{app:flow_amounts}. At fixed $J^2$, 
$\Lambda$, and Casimirs, the mutual scalar products of the triad are affine 
in the single variable $f = \vb{S}_1\cdot\vb{S}_2/(\Sigma_1-\Sigma_2)$, 
which oscillates between two roots $f_1 < f_2$ of a cubic polynomial. 
$\Delta\lambda_\Lambda$ is the period of this oscillation, given by 
a complete elliptic integral of the first kind [Eq.~\eqref{eq:delta_lambda_Lambda_result}]. 
The other four amounts follow from the angles by which $\vb{L}$ and the 
fictitious pairs rotate during one period, and involve complete elliptic 
integrals of the third kind [Eqs.~\eqref{eq:delta_Phi_L_def}--\eqref{eq:BDS_def}].

These expressions degenerate for $m_1 = m_2$. In this limit, $\Sigma_1 = \Sigma_2$, and both spins
precess about $\vb{L}$ at the same rate under the $\Lambda$-flow [Eq.~\eqref{eq:flow_SA}], so that
$\vb{S}_1\cdot\vb{S}_2$ is constant along it. The variable $f$ used in
App.~\ref{app:flow_amounts} [Eq.~\eqref{eq:f_def}] is then undefined, and the leading coefficient
of its cubic [Eq.~\eqref{eq:f_cubic}] vanishes. The system remains integrable, but this limit 
requires a different parametrization of the mutual angles, which we do not pursue here.

\section{Frequencies and angle variables}
\label{sec:frequencies_angles}

\subsection{Computing the frequencies}
\label{subsec:computing_frequencies}

In action-angle variables $(\mathcal{J}_i,\theta_i)$, the equations of motion 
read $\dot{\mathcal{J}}_i = 0$ and $\dot\theta_i = \omega_i \equiv \partial\bar H/\partial\mathcal{J}_i$,
so that the motion on each torus is characterized by three frequencies, instead of 
five for the non-OA 1.5PN system of Ref.~\cite{Tanay2023}. By Eq.~\eqref{eq:hamiltonian_dependent}, 
$\bar H$ depends on the COMs $\vb{C} = (J^2, J_z, \Lambda)$ only through $J^2$ and $\Lambda$, 
with the Casimirs $L$, $S_1$, and $S_2$ as fixed parameters. The frequencies can be computed from 
the chain rule
\begin{equation} \label{eq:freq_chain_rule}
    \omega_i = \sum_{j=1}^{3} \frac{\partial \bar{H}}{\partial C_j}\, \frac{\partial C_j}{\partial \mathcal{J}_i} .
\end{equation}
The first factor in the sum, $\partial \bar{H} / \partial C_j$, is evaluated by taking the 
partial derivatives of the Hamiltonian  in Eq.~\eqref{eq:hamiltonian_dependent}, yielding
\begin{equation}\label{eq:dH_dC}
    \frac{\partial \bar{H}}{\partial J^2} = \frac{1}{4 c^2 D^3} , \qquad
    \frac{\partial \bar{H}}{\partial J_z} = 0 , \qquad
    \frac{\partial \bar{H}}{\partial \Lambda} = \frac{3}{2 c^2 D^3} \left( 1 - \frac{\nu \Lambda}{L^2} \right) . 
\end{equation}
The second factor requires the inverse of the Jacobian
$\mathcal{M}_{ij} = \partial\mathcal{J}_i/\partial C_j$. Since $\mathcal{J}_1 = J = (J^2)^{1/2}$,
$\mathcal{J}_2 = J_z$, and among the COMs, $\mathcal{J}_3$ depends only on $J^2$ and $\Lambda$  
[Eq.~\eqref{eq:action_3_final}], $\mathcal{M}$ is lower triangular

\begin{equation} \label{eq:jacobian_matrix}
    \mathcal{M} = \frac{\partial \boldsymbol{\mathcal{J}}}{\partial \vb{C}} = \begin{pmatrix}
        \dfrac{1}{2J} & 0 & 0 \\
        0 & 1 & 0 \\
        \dfrac{\partial \mathcal{J}_3}{\partial J^2} & 0 & \dfrac{\partial \mathcal{J}_3}{\partial \Lambda}
    \end{pmatrix} , \qquad
    \det\mathcal{M} = \frac{1}{2J}\,\frac{\partial \mathcal{J}_3}{\partial \Lambda} .
\end{equation}
It is invertible whenever $J \neq 0$ and $\partial\mathcal{J}_3/\partial\Lambda$ is finite and non-zero. Its inverse is then
\begin{equation} \label{eq:inverse_jacobian_matrix}
    \mathcal{M}^{-1} = \frac{\partial \vb{C}}{\partial \boldsymbol{\mathcal{J}}} = \begin{pmatrix}
        2J & 0 & 0 \\
        0 & 1 & 0 \\
        -2J\,\dfrac{\partial \mathcal{J}_3/\partial J^2}{\partial \mathcal{J}_3/\partial \Lambda} & 0 &
        \left(\dfrac{\partial \mathcal{J}_3}{\partial \Lambda}\right)^{-1}
    \end{pmatrix} ,
\end{equation}
so that Eq.~\eqref{eq:freq_chain_rule} gives the three frequencies of the system,
\begin{subequations}
    \begin{align}
    \omega_1 &= 2J\left( \frac{\partial \bar{H}}{\partial J^2} - \frac{\partial \bar{H}}{\partial \Lambda}\,
    \frac{\partial \mathcal{J}_3 / \partial J^2}{\partial \mathcal{J}_3 / \partial \Lambda} \right) ,
    \label{eq:omega_1} \\
    \omega_2 &= 0 , \label{eq:omega_2} \\
    \omega_3 &= \frac{\partial \bar{H}}{\partial \Lambda}
    \left( \frac{\partial \mathcal{J}_3}{\partial \Lambda} \right)^{-1} .
    \label{eq:omega_3}
\end{align}
\end{subequations}
The derivatives of $\mathcal{J}_3$ can be obtained without explicitly differentiating the elliptic
integrals of App.~\ref{app:flow_amounts}. We instead use the structure of the loop of
Sec.~\ref{sec:third_action}: each of its segments is generated by a function
$F \in \{\Lambda, J^2, L^2, S_1^2, S_2^2\}$ over a flow amount $\Delta\lambda_F$. When $J^2$ and
$\Lambda$ are varied independently, at fixed $J_z$ and Casimirs, the loop moves to a neighboring
torus, and each segment contributes $\Delta\lambda_F\,\delta F$ to the first-order variation of the
action (see Appendix~\ref{app:dJ3_derivative} for a proof for the case when $F = \Lambda$ is varied)
\begin{equation} \label{eq:J3_variation}
    \delta(2\pi\mathcal{J}_3) = \sum_{F} \Delta\lambda_{F}\, \delta F .
\end{equation}
At fixed Casimirs, $\delta(L^2) = \delta(S_a^2) = 0$, so that only the $\Lambda$- and $J^2$-segments
contribute. Since $\mathcal{J}_3$ is a function of $J^2$ and
   $\Lambda$, we also have
   $\delta\mathcal{J}_3 = (\partial\mathcal{J}_3/\partial J^2)\,\delta(J^2)
   + (\partial\mathcal{J}_3/\partial\Lambda)\,\delta\Lambda$. Identifying the coefficients of
   $\delta(J^2)$ and $\delta\Lambda$ in Eq.~\eqref{eq:J3_variation} gives
\begin{equation} \label{eq:dJ3_derivatives}
    \frac{\partial\mathcal{J}_3}{\partial\Lambda}\bigg|_{J^2} = \frac{\Delta\lambda_\Lambda}{2\pi}, \qquad
    \frac{\partial\mathcal{J}_3}{\partial J^2}\bigg|_{\Lambda} = \frac{\Delta\lambda_{J^2}}{2\pi}.
\end{equation}
Inserting
Eq.~\eqref{eq:dJ3_derivatives} into Eqs.~\eqref{eq:omega_1} and~\eqref{eq:omega_3}, with
$\Delta\lambda_{J^2} = -\Delta\Phi_L/(2J)$ [Eq.~\eqref{eq:dlambda_J2_main}], gives
\begin{equation} \label{eq:frequencies_explicit}
    \omega_1 = 2J\,\frac{\partial\bar{H}}{\partial J^2} + \frac{\Delta\Phi_L}{\Delta\lambda_\Lambda}\,\frac{\partial\bar{H}}{\partial\Lambda},
    \qquad
    \omega_3 = \frac{2\pi}{\Delta\lambda_\Lambda}\,\frac{\partial\bar{H}}{\partial\Lambda},
\end{equation}
with $\partial\bar{H}/\partial J^2$ and $\partial\bar{H}/\partial\Lambda$ given by
Eq.~\eqref{eq:dH_dC}, and $\Delta\lambda_\Lambda$ and $\Delta\Phi_L$ by
Eqs.~\eqref{eq:delta_lambda_Lambda_result} and~\eqref{eq:delta_Phi_L_def}.

These frequencies have a direct physical meaning. Since $\bar{H} = \bar{H}(J^2,\Lambda)$, the time
derivative of any function $g$ of the physical variables is
\begin{equation} \label{eq:eom_two_flows}
    \frac{dg}{dt} =\{g,\bar{H}\}= \frac{\partial\bar{H}}{\partial J^2}\,\{g, J^2\}
    + \frac{\partial\bar{H}}{\partial\Lambda}\,\{g, \Lambda\},
\end{equation}
with coefficients $\partial\bar{H}/\partial J^2$ and $\partial\bar{H}/\partial\Lambda$
[Eq.~\eqref{eq:dH_dC}] that are constant along the motion, since they depend only on the COMs $J^2$
and $\Lambda$. For a mutual angle $g$ of the triad, $\{g, J^2\} = 0$, since $J^2$ generates rigid
rotations, and Eq.~\eqref{eq:eom_two_flows} reduces to
\begin{equation} \label{eq:mutual_angle_time}
    \frac{dg}{dt} = \frac{\partial\bar{H}}{\partial\Lambda}\,\{g,\Lambda\}
    = \frac{\partial\bar{H}}{\partial\Lambda}\,\frac{dg}{d\lambda_\Lambda} .
\end{equation}
Time and the $\Lambda$-flow parameter are thus proportional,
$\lambda_\Lambda = (\partial\bar{H}/\partial\Lambda)\,t$. Since $g$ returns to its initial value
after $\Delta\lambda_\Lambda$, it is periodic in time with period
$T_3 = \Delta\lambda_\Lambda/(\partial\bar{H}/\partial\Lambda) = 2\pi/\omega_3$, so that $\omega_3$
is the nutation frequency.

To interpret $\omega_1$, consider one nutation period $T_3$. 
By Eq.~\eqref{eq:eom_two_flows}, the motion combines the $\Lambda$-flow, 
with $\lambda_\Lambda = (\partial\bar{H}/\partial\Lambda)\,t$,
and the $J^2$-flow, with $\lambda_{J^2} = (\partial\bar{H}/\partial J^2)\,t$. 
Since these two flows commute, their effects over $T_3$ add 
(App.~\ref{sec:LC_of_COM}). During $T_3$, $\lambda_\Lambda$ increases by
$\Delta\lambda_\Lambda$, so that the $\Lambda$-flow restores the mutual 
angles and rotates the configuration about $\hat{\vb{J}}$ by $\Delta\Phi_L$, 
as we saw earlier in Sec.~\ref{sec:third_action}. Meanwhile, $\lambda_{J^2}$ 
increases by $(\partial\bar{H}/\partial J^2)\,T_3$, and the $J^2$-flow, a 
rotation about $\hat{\vb{J}}$ at angular velocity $2J$, rotates the configuration by 
$2J\,(\partial\bar{H}/\partial J^2)\,T_3$. After each nutation period, 
the three angular momenta are therefore rotated about 
$\vb{J}$ by $\Delta\Phi_L + 2J(\partial\bar{H}/\partial J^2)\,T_3$. 
Dividing by $T_3$ gives $\omega_1$ of Eq.~\eqref{eq:frequencies_explicit},
which is thus the precession frequency about $\vb{J}$ averaged over one nutation 
period. Finally, $\omega_2 = 0$ [Eq.~\eqref{eq:omega_2}], since $\bar{H}$ does not
depend on $J_z$, which reflects that $\vb{J}$ is fixed in space.
The same decomposition, a uniform precession about the total angular
momentum combined with a periodic nutation, holds for the analogous
planet-satellite problem~\cite{boue:hal-00335321}.

Once we account for the scaled (reduced) length and time variables employed in  
Ref.~\cite{colin-tanay}, our $\omega_3$ and $\omega_1$ coincide with their
nutation and precession frequencies $\omega_{\rm nut}$ and $\omega_{\rm prec}$, 
respectively [Eqs.~(5.1) and~(5.2)], obtained  from their time-domain solution of the
same OA EOMs.\footnote{The nutation variable of Ref.~\cite{colin-tanay} is
$\cos\kappa_1 = \hat{\vb{L}}\cdot\hat{\vb{S}}_1$. Since $\vb{L}\cdot\vb{S}_1 = \Sigma_2(f-\Delta_1)$
[Eq.~\eqref{eq:dot_products}], $\cos\kappa_1$ is an affine function of $f$. The
roots $x_- < x_+ < x_3$ of the cubic of Ref.~\cite{colin-tanay} are therefore the images of
$f_1 < f_2 < f_3$, and its elliptic modulus, denoted $\beta$ there, equals $k$. Its constant of
motion $\vb{l}\cdot\vb{s}_0/l^2$, denoted $\lambda$ there and written in its rescaled variables,
equals $\nu\Lambda/L^2$, so that $1-\lambda$ is proportional to $\partial\bar{H}/\partial\Lambda$.
In $\omega_{\rm prec}$, the term $\beta_{3L}/(c^2 d^3)$ equals $2mJ\,\partial\bar{H}/\partial J^2$, 
and the terms involving $\Pi$ equal $m\,\Delta\Phi_L/T_3$. Note also that $\Sigma_1$ and $\Sigma_2$ 
in Ref.~\cite{colin-tanay} denote constants of motion, not the mass ratios of $\vb{S}_0$.}

Since the frequencies are known in closed form, one can look for resonances,
i.e., $\vb{n} \cdot \boldsymbol{\omega} = 0$ for some nonzero
$\vb{n} \in \mathbb{Z}^3$. Since $\omega_2 = 0$, this involves only the ratio
\begin{equation} \label{eq:freq_ratio}
    \frac{\omega_1}{\omega_3} = \frac{\Delta\Phi_L}{2\pi}
    + \frac{J\,K(k)}{3\pi\,\Omega_\Lambda\,(1-\nu\Lambda/L^2)} ,
\end{equation}
where $\Delta\Phi_L$ is the rotation of $\vb L$ about $\hat{\vb J}$ during one
nutation cycle.
The spin-orbit resonances of Ref.~\cite{schnittman}, at
which $f_1 = f_2$, are not resonances in this sense~\cite{gerosa_private}, 
as pointed out by Correia~\cite{correia:hal-02319812}, who identified them
with Cassini states.
At fixed masses, spins, and $L$, the condition $f_1 = f_2$ is a single equation on
$(J,\Lambda)$, so the spin-orbit resonances form a one-parameter family,
the boundary of the region in Fig.~\ref{fig:JLambda}. The ratio
\eqref{eq:freq_ratio} varies continuously along this family, so
$f_1 = f_2$ does not make it rational. The condition instead sets to zero
the nutation amplitude $f_2-f_1$ in Eq.~\eqref{eq:f_solution_app}, so that
the shape of the triad is frozen and the triad precesses rigidly about
$\vb J$.

\subsection{The angle variables and action-angle-based solution}
\label{subsec:angle_variables}
In terms of action-angle variables, the motion is simple: the actions are constant 
and the angles advance linearly, $\theta_i(t) = \theta_i(0) + \omega_i 
t$~\cite{jose1998classical}. To use this solution, we need the map between 
the angles and the configuration $(\vb{L},\vb{S}_1,\vb{S}_2)$ on a given torus.
Section~VI~B of Ref.~\cite{Tanay2023} constructs the map from the angles
to the configuration, by flowing from a fiducial point by amounts set by 
the angles. We follow this construction, and also derive the inverse map, 
which was not given there. It expresses the angles of any configuration 
explicitly in terms of $(\vb{L},\vb{S}_1,\vb{S}_2)$ [Eq.~\eqref{eq:angles_explicit}] 
and gives them a physical interpretation.

Fixed values of $\vb{C}$ and of the Casimirs define an invariant 3-torus. 
We locate a configuration on it by three geometric quantities. The first is the 
shape of the triad. It depends only on the mutual scalar products, which are affine
in $f$, as it can be seen from Eq.~\eqref{eq:dot_products}. Thus the shape of the 
triad is fixed by $f\in[f_1,f_2]$ together with the handedness of the triad,
$s \equiv \operatorname{sgn}[\vb{L}\cdot(\vb{S}_1\times\vb{S}_2)]$. 
The second is the azimuth $\phi_J$ of $\hat{\vb{J}}$ about $\hat{\vb{z}}$.
The third is the orientation of the triad about $\hat{\vb{J}}$, given by the azimuth 
$\Phi_L$ of $\vb{L}$ about $\hat{\vb{J}}$, measured counterclockwise from the line of 
nodes $\hat{\vb{z}}\times\hat{\vb{J}}$. The capital letter distinguishes it from the 
canonical azimuth $\phi_L$ of $\vb{L}$ about $\hat{\vb{z}}$ (Sec.~\ref{sec:Hamiltonian}).

We assign $\boldsymbol{\theta} = \vb{0}$ to the fiducial configuration $f = f_1$, $\Phi_L = 0$,
$\phi_J = 0$, a turning point of the nutation. Any configuration on the torus is reached from it by
flowing under $\Lambda$, $J^2$, and $J_z$ by some amounts $\lambda_\Lambda$, $\lambda_{J^2}$, and
$\lambda_{J_z}$, in any order since these flows commute. Each flow changes only some of the coordinates 
$(f,s,\phi_J,\Phi_L)$, as summarized in Table~\ref{tab:flows}. In particular, the $J_z$-flow does not 
evolve $\Phi_L$, since the line of nodes rotates with the triad. The $\Lambda$-flow changes $\Phi_L$ 
as it deforms the triad (App.~\ref{app:flow_J2}). The flow amounts can therefore be read off one at a time:
$\lambda_\Lambda$ from $f$ and $s$, then $\lambda_{J_z}$ from $\phi_J$, and finally $\lambda_{J^2}$
from what remains of $\Phi_L$.

\begin{table}[h]
\caption{Effect of each flow on the coordinates $(f,s,\phi_J,\Phi_L)$ of a configuration. A bullet
marks the coordinates changed by the flow.}
\label{tab:flows}
\begin{ruledtabular}
\begin{tabular}{llccc}
Flow & Action on the configuration & $f$, $s$ & $\phi_J$ & $\Phi_L$ \\ \hline
$J_z$ & rigid rotation about $\hat{\vb{z}}$ (including $\hat{\vb{J}}$) & & $\bullet$ & \\
$J^2$ & rigid rotation about the fixed $\hat{\vb{J}}$ & & & $\bullet$ \\
$\Lambda$ & deformation of the triad, $\hat{\vb{J}}$ fixed & $\bullet$ & & $\bullet$ \\
\end{tabular}
\end{ruledtabular}
\end{table}

Under the $\Lambda$-flow (App.~\ref{app:flow_Lambda}), $f$ increases from $f_1$
to $f_2$ over the first half of the nutation cycle and comes back to $f_1$ over
the second half. Since $df/d\lambda_\Lambda = \vb{L}\cdot(\vb{S}_1\times\vb{S}_2)$,
the sign $s$ tells us which half a configuration belongs to, so $f$ and $s$
together fix $\lambda_\Lambda$. At the same time, the azimuth of $\vb{L}$ about
$\hat{\vb{J}}$ goes from $0$ to $\Phi_L^\Lambda(\lambda_\Lambda)$, given in
Eq.~\eqref{eq:Phi_L_solution_app}, and reaches
$\Phi_L^\Lambda(\Delta\lambda_\Lambda) = \Delta\Phi_L$ after one period, the
rotation angle of step~(i) in Sec.~\ref{sec:third_action}.

The two other flows are rigid rotations, at unit angular velocity for the $J_z$-flow
and at angular velocity $2J$ for the $J^2$-flow. Only the $J_z$-flow
moves $\hat{\vb{J}}$, so $\lambda_{J_z} = \phi_J$. The azimuth $\Phi_L$, on the
other hand, receives $\Phi_L^\Lambda(\lambda_\Lambda)$ from the $\Lambda$-flow
and $2J\lambda_{J^2}$ from the $J^2$-flow. The flow amounts that lead from the fiducial
configuration to $(f,s,\phi_J,\Phi_L)$ are therefore
\begin{equation} \label{eq:lambda_J}
    \lambda_\Lambda = \frac{U(f,s)}{\Omega_\Lambda} , \qquad
    \lambda_{J_z} = \phi_J , \qquad
    \lambda_{J^2} = \frac{\Phi_L - \Phi_L^{\Lambda}(\lambda_\Lambda)}{2J} ,
\end{equation}
where the first relation is Eq.~\eqref{eq:lambda_Lambda}, with $U$ and
$\Omega_\Lambda$ defined in App.~\ref{app:flow_Lambda}.

It remains to convert these flow amounts into angles. The flow of an action $\mathcal{J}_j$ shifts
$\theta_j$ only, since $d\theta_i/d\lambda_{\mathcal{J}_j} = \{\theta_i,\mathcal{J}_j\} = \delta_{ij}$.
The configuration with angles $\boldsymbol{\theta}$ is therefore reached from the fiducial one by
flowing under each $\mathcal{J}_i$ by $\theta_i$, in any order (Sec.~39~E of
Ref.~\cite{arnold2013mathematical}). Since $\mathcal{J}_i$ is a function of the COMs, flowing under
$\mathcal{J}_i$ by $\theta_i$ is equivalent to flowing under each $C_j$ by
$(\partial\mathcal{J}_i/\partial C_j)\,\theta_i$ (App.~\ref{sec:LC_of_COM}). Collecting the
contributions of the three actions, the flow amount under $C_j$ is
$\lambda_{C_j} = \sum_i (\partial\mathcal{J}_i/\partial C_j)\,\theta_i$\footnote{The trajectory 
depends on the actions only through $\bar H(C)$, since the Jacobian $\pd\mathcal J_i/\pd C_j$ 
cancels in $\lambda_{C_j}=(\pd\bar H/\pd C_j)\,t$. The actions are required for the physical 
frequencies and for canonical perturbation theory.}, and with the help of the
Jacobian~\eqref{eq:jacobian_matrix} we get
\begin{equation} \label{eq:lambda_theta}
    \lambda_{J^2} = \frac{\theta_1}{2J} + \frac{\partial\mathcal{J}_3}{\partial J^2}\,\theta_3 , \qquad
    \lambda_{J_z} = \theta_2 , \qquad
    \lambda_\Lambda = \frac{\partial\mathcal{J}_3}{\partial\Lambda}\,\theta_3 .
\end{equation}

Inverting these relations with Eqs.~\eqref{eq:dlambda_J2_main}
and~\eqref{eq:dJ3_derivatives}, and inserting the flow amounts of
Eq.~\eqref{eq:lambda_J}, we obtain the angles of any configuration on the torus,
\begin{subequations} \label{eq:angles_explicit}
\begin{align}
    \theta_1 &= \Phi_L - \left[ \Phi_L^{\Lambda}\big(\lambda_\Lambda(f,s)\big)
      - \frac{\Delta\Phi_L}{\Delta\lambda_\Lambda}\,\lambda_\Lambda(f,s) \right] , \\
    \theta_2 &= \phi_J , \\
    \theta_3 &= 2\pi\,\frac{\lambda_\Lambda(f,s)}{\Delta\lambda_\Lambda} ,
\end{align}
\end{subequations}
with $\lambda_\Lambda(f,s)$ and $\Phi_L^\Lambda$ given by
Eqs.~\eqref{eq:lambda_Lambda} and~\eqref{eq:Phi_L_solution_app}.

These angles also have a physical interpretation. Only $\theta_2$ is a 
geometric angle and it coincides with the azimuth $\phi_J$ of $\hat{\vb{J}}$, 
which is constant since $\omega_2=0$. The angle $\theta_3$ is the nutation phase, 
with $\theta_3/2\pi$ being the fraction of the nutation cycle completed since 
$f = f_1$. The angle $\theta_1$ differs from the azimuth $\Phi_L$ by the bracket, 
a periodic function of the triad shape, with period $\Delta\lambda_\Lambda$ in 
$\lambda_\Lambda$. For the motion that starts at the fiducial configuration, 
\begin{equation} \label{eq:phiL_motion}
    \Phi_L(t) = \omega_1 t + \left[ \Phi_L^{\Lambda}(\lambda_\Lambda) - \frac{\Delta\Phi_L}{\Delta\lambda_\Lambda}\,\lambda_\Lambda \right] ,
\end{equation}
which is the sum of a uniform precession at frequency $\omega_1$ and of an oscillation
with the nutation period. The angle $\theta_1 = \omega_1 t$ is thus the
precession of $\vb{L}$ about $\hat{\vb{J}}$ once the nutation is removed.

On a regular torus, the map $(\vb{L},\vb{S}_1,\vb{S}_2)\to\boldsymbol{\theta}$ of  
Eq.~\eqref{eq:angles_explicit} is singular only when $\vb{L} \parallel \pm \hat{\vb{J}}$, 
where $\Phi_L$ is undefined.

\section{Violation of integrability conditions and separatrices}
\label{sec:separatrix}

\subsection{Separatrices}

Informally, a separatrix is a level set of the COMs that separates the 
phase space into two regions with qualitatively different kinds of 
trajectories~\cite{ferraz2023chaotic}. The level set of the pendulum 
through its unstable equilibrium separates libration from rotation,
and the parabolic orbit under Newtonian gravitation separates bound from 
unbound orbits. In these simple systems, with one degree of freedom or one 
independent frequency, the separatrix is easy to find. 
The goal of this section is to propose a criterion that finds it in general. 
We also include level sets where the nature of the motion changes abruptly, 
without separating the phase space, for example, an isolated point of
the $(C_1, C_2)$ plane. We call all of them separatrix-like. These level 
sets are not tori.

Traditionally, separatrices are associated with hyperbolic fixed points or
with a vanishing frequency~\cite{lichtenberg2013regular, strogatz2024nonlinear}.
A hyperbolic fixed point, however, is a single point of a level set, and its
analysis only describes the motion close to it. A frequency is defined on the
whole level set, but it follows from the actions, which are themselves singular
there. We look instead for a criterion on the level set as a whole, written
directly in terms of the COMs. Our starting point is that a breakdown of the
Liouville--Arnold integrability conditions can lead to separatrix-like level
sets, where the topology of the level sets changes (see Sec.~50 of
Ref.~\cite{arnold2013mathematical}). In the next subsection, we give a
criterion that implies such a breakdown.

\subsection{A mathematical criterion}
\label{sec:math_criterion}

Consider an integrable Hamiltonian system with $2n$ phase-space variables
$(\vb{p},\vb{q})$ and $n$ independent COMs $\vb{C}$ in involution, whose level
sets are compact and connected in some region of phase space. There, the system
has action-angle variables $(\boldsymbol{\mathcal{J}},\boldsymbol{\theta})$. We
propose the following criterion: if
\begin{equation} \label{eq:separatrix_criterion}
   \lim_{\vb{C}\to\vb{C}_*} \det\left(\frac{\partial\vb{C}}{\partial\boldsymbol{\mathcal{J}}}\right) = 0 ,
\end{equation}
then integrability breaks down on the level set $\vb{C}=\vb{C}_*$
(App.~\ref{sec:separatrix_justification}).
We take the limit because the actions need not be defined on $\vb{C}_*$ itself.
In practice,  upon obtaining $\det(\partial\vb{C}/\partial\boldsymbol{\mathcal{J}})$ 
as an explicit function of $\vb{C}$,  $\vb{C}_*$ is found by setting it to zero. 

Since independence of COMs is equivalent to $dC_1 \wedge dC_2 \wedge \cdots \wedge dC_n \neq  0 $,
we confine ourselves to the COMs $\vb{C}$ whose differentials do not diverge at finite 
values of $(\vb{p}, \vb{q})$, for example $\vb{C} = (J^2, J_z, \Lambda)$ for the OA 
spinning BBH system. In general, action variables do not qualify.\footnote{The 
differential of the radial action variable of a Kepler orbit with Newtonian gravitation 
diverges when the energy corresponds to the parabolic trajectory.} That is why we do 
not set $\vb{C}=\boldsymbol{\mathcal{J}}$ in the criterion  $\det(\partial\vb{C}/\partial
\boldsymbol{\mathcal{J}}) = 0$; doing so yields an unrealizable condition $1 = 0$ anyway.

Vanishing frequencies, which are usually associated with separatrices, imply
the criterion. Indeed, by the chain rule of Eq.~\eqref{eq:freq_chain_rule},
\begin{equation}
    \omega_i = \sum_j \frac{\partial C_j}{\partial\mathcal{J}_i}\,\frac{\partial\bar{H}}{\partial C_j} .
\end{equation}
If all the frequencies go to zero in the limit $\vb{C}\to\vb{C}_*$, while the
$\partial\bar{H}/\partial C_j$ remain finite and not all zero, then the matrix
$\partial C_j/\partial\mathcal{J}_i$ becomes singular, which is
Eq.~\eqref{eq:separatrix_criterion}.
Finally, the converse of the criterion does not hold, as we will see for the
pendulum in Sec.~\ref{sec:pendulum}.
\subsection{Applications to various systems}

\subsubsection{Pendulum}\label{sec:pendulum}

For a pendulum of mass $m$ and length $l$ in a uniform gravitational field $g$,
with $q$ the angle from the downward vertical and $p_q$ its conjugate
momentum, the Hamiltonian is
\begin{equation} \label{eq:pendulum_H}
    H = \frac{p_q^2}{2 m l^2} - m g l \cos q .
\end{equation}
For a libration orbit, $|H| < mgl$, the action variable
$\mathcal{J} = (2\pi)^{-1}\oint p_q \, dq$ is~\cite{lichtenberg2013regular}
\begin{equation} \label{eq:pendulum_action}
    \mathcal{J} = \frac{8}{\pi}\, m l \sqrt{g l}\,
    \Big[ E(\kappa) - (1-\kappa^2)\, K(\kappa) \Big] , \qquad
    2\kappa^2 \equiv 1 + \frac{H}{m g l} ,
\end{equation}
where $K$ is the complete elliptic integral of the first kind defined in
Eq.~\eqref{eq:elliptic_first_defs}, and
\begin{equation} \label{eq:elliptic_second_def}
    E(k) = \int_0^{\pi/2} \sqrt{1-k^2\sin^2\vartheta}\; d\vartheta
\end{equation}
is the complete elliptic integral of the second kind. With $\vb{C} = H$, the
criterion reduces to the vanishing of the frequency
$\omega = dH/d\mathcal{J} = \pi\sqrt{g/l}/[2K(\kappa)]$, which tends to zero as
$\kappa\to1$, i.e., $H \to m g l$. This is the level set through the upright
unstable equilibrium, the separatrix between libration and rotation.

The integrability conditions also break down at the stable equilibrium,
$H = -m g l$, where the level set shrinks from a closed curve to a point. Yet
the frequency does not vanish there, since $\omega \to \sqrt{g/l}$.
This is no cause for worry since 
in Sec.~\ref{sec:math_criterion}, we claimed  that Eq.~\eqref{eq:separatrix_criterion} 
implies a breakdown of integrability conditions, and not the converse. 
Near this point, the action shrinks to zero together with the level set, 
so $H + mgl$ and $\mathcal{J}$ vanish together and their ratio $\omega$ 
stays finite. At the separatrix, on the contrary, $d\mathcal{J}/dH$ diverges. 
The criterion therefore captures the breakdowns where the actions become singular 
with respect to the COMs, which typically correspond to unstable equilibria.

\subsubsection{Orbits under Newtonian gravitation}

For bound orbits under Newtonian gravitation, with total mass $m$ and reduced mass $\mu$, the 
Hamiltonian depends on the radial action $\mathcal{J}_r = (2\pi)^{-1}\oint p_r\,dr$ and on $L$ only 
through their sum~\cite{goldstein2011classical},
\begin{equation} \label{eq:kepler_H}
    H = -\frac{m^2\mu^3}{2(\mathcal{J}_r + L)^2} .
\end{equation}
The radial and azimuthal frequencies are thus degenerate, $\pd H/\pd\mathcal{J}_r = \pd H/\pd L$,
which is why bound orbits close. With $\vb{C} = (H, L^2, L_z)$ and $\boldsymbol{\mathcal{J}} = 
(\mathcal{J}_r, L, L_z)$, the criterion reduces to the vanishing of the radial frequency
$\pd H/\pd\mathcal{J}_r = m^2\mu^3/(\mathcal{J}_r + L)^3 = (-2H)^{3/2}/(m\mu^{3/2})$, which 
tends to zero as $H \to 0^-$. This is the parabolic orbit, which separates the bound elliptic 
orbits from the unbound hyperbolic ones. There, all the frequencies vanish while $\pd H/\pd\vb{C}$ 
remains finite, as discussed at the end of Sec.~\ref{sec:math_criterion}.

\subsubsection{Orbit-averaged spinning BBH system} \label{sec:separatrix_OA_BBH}

We now apply the criterion to the OA spinning BBH system. Recall that, at fixed
$(J,\Lambda)$, the shape of the triad is determined by $f$, defined in
Eq.~\eqref{eq:dot_products} by $\vb{S}_1\cdot\vb{S}_2 = (\Sigma_1-\Sigma_2)f$.
Along the $\Lambda$-flow, $f$ oscillates between the two smallest roots, $f_1$
and $f_2$, of the cubic $P$ of Eq.~\eqref{eq:f_cubic}, whose roots are
$f_1 \le f_2 \le f_3$, as
\begin{equation}\label{eq:sol_f}
    f(\lambda_\Lambda) = f_1 + (f_2-f_1)\,\sn^2(\Omega_\Lambda\lambda_\Lambda,k) ,
    \qquad
    \Omega_\Lambda = \tfrac12\sqrt{a_3(f_3-f_1)} , \qquad
    k^2 = \frac{f_2-f_1}{f_3-f_1} ,
\end{equation}
where $\sn$ is the Jacobi elliptic sine.
These roots also govern the physical motion. Since $\bar{H} = \bar{H}(J^2,\Lambda)$
[Eq.~\eqref{eq:hamiltonian_dependent}] and $f$, a scalar product, is invariant under
the rotations generated by $\vb{J}$, so that $\{f, J^2\} = 0$, Eq.~\eqref{eq:eom_two_flows} gives
\begin{equation} \label{eq:f_time}
    \frac{df}{dt} = \frac{\pd\bar{H}}{\pd\Lambda}\,\frac{df}{d\lambda_\Lambda} .
\end{equation}
On a level set, $\pd\bar{H}/\pd\Lambda$ is constant, so the time evolution of $f$
is the $\Lambda$-flow of Eq.~\eqref{eq:sol_f} with
$\lambda_\Lambda = (\pd\bar{H}/\pd\Lambda)\,t$. In real time too, $f$ therefore oscillates
between the turning points $f_1$ and $f_2$, with the nutation frequency $\omega_3$ of
Eq.~\eqref{eq:frequencies_explicit}.

With $\vb{C} = (J^2, J_z, \Lambda)$ and
$\boldsymbol{\mathcal{J}} = (J, J_z, \mathcal{J}_3)$, the criterion reduces,
by Eqs.~\eqref{eq:inverse_jacobian_matrix}, \eqref{eq:dJ3_derivatives},
and~\eqref{eq:delta_lambda_Lambda_result}, to
\begin{equation} \label{eq:criterion_OA}
    \det\left(\frac{\pd\vb{C}}{\pd\boldsymbol{\mathcal{J}}}\right)
    = \frac{2 \pi\,J\,\Omega_\Lambda}{K(k)} \to 0 .
\end{equation}
Since $K(k) \ge \pi/2$, this happens when one of the following
conditions holds: (i) $K(k)$ diverges, (ii) $\Omega_\Lambda = 0$, or (iii) $J = 0$.
We discuss them in turn.

Condition (i) means $k \to 1$, i.e., by Eq.~\eqref{eq:sol_f}, that the two
largest roots coalesce, $f_1 < f_2 = f_3$. Since $a_3 > 0$, condition (ii) means
$f_3 = f_1$, i.e., that all three roots coincide, $f_1 = f_2 = f_3$. In both cases,
the nutation frequency $\omega_3 = \pi\Omega_\Lambda(\pd\bar{H}/\pd\Lambda)/K(k)$ of
Eq.~\eqref{eq:frequencies_explicit} vanishes, i.e., the nutation period $2\pi/\omega_3$
diverges. In case (i), the divergence is logarithmic, since
$K(k) \simeq \ln\big(4/\sqrt{1-k^2}\big)$ as $k \to 1$, and it is the divergence of the
precession period $\tau$ found in Ref.~\cite{gerosa2015updown}, which we call the nutation period.

Conditions (i) and (ii) correspond to a multiple root $f_\ast$ of $P$, double in 
case (i) and triple in case (ii). For any multiple root, the configuration with 
$f=f_\ast$ is an equilibrium of the shape of the triad. Since $P(f_\ast)=0$, 
Eq.~\eqref{eq:f_cubic} gives $df/d\lambda_\Lambda = 0$ at $f_\ast$, and 
differentiating it gives $d^2f/d\lambda_\Lambda^2 = P'(f)/2$, which also 
vanishes at $f_\ast$, so $f$ remains equal to $f_\ast$. Its stability follows 
from the continuity of the roots of $P$ with respect to $(J,\Lambda)$. 
A small perturbation of the configuration moves it to a nearby level set, on which $f$ 
oscillates between the two smallest roots of the perturbed cubic. If 
$f_\ast = f_1 = f_2 < f_3$, these roots are both close to $f_\ast$, and $f$ 
stays close to $f_\ast$, so the equilibrium is stable. These equilibria are the
spin-orbit resonances~\cite{schnittman, gerosa2015multi}, or Cassini states~\cite{correia:hal-02319812}, which the criterion 
does not detect, since $K(0)=\pi/2$ keeps the determinant finite. 
If $f_\ast = f_2 = f_3$, the two smallest roots of the perturbed cubic
are instead close to $f_1$ and $f_\ast$. 
However small the perturbation, $f$ then oscillates over almost the whole interval $[f_1, f_\ast]$, 
and thus moves away from $f_\ast$ by about $f_\ast - f_1$, so the equilibrium is unstable. 

We now apply these results to the four aligned configurations,  
in which $\vb{S}_1$ and $\vb{S}_2$ are each parallel or antiparallel to
$\vb{L}$~\cite{gerosa2015updown, PhysRevD.101.124037}.
We first show that each of them corresponds to a double root of $P$. 
Since the three vectors are collinear,
all the cross products in the $\Lambda$-flow, Eqs.~\eqref{eq:flow_L}
and~\eqref{eq:flow_SA}, vanish, so an aligned configuration is a fixed point of the flow.
Then $f$ is constant, $df/d\lambda_\Lambda = d^2f/d\lambda_\Lambda^2 = 0$, and
Eq.~\eqref{eq:f_cubic} gives $P(f_\ast) = P'(f_\ast) = 0$: the value $f_\ast$ of $f$ at an
aligned configuration is a double root of $P$. Which kind of double root it is follows
from the third root, $f_{\rm rd}$, given by the sum of the roots,
$2f_\ast + f_{\rm rd} = -a_2/a_3$. For the up-down configuration, with $\vb{S}_1$ along
$\vb{L}$ and $\vb{S}_2$ opposite to it, $\vb{S}_1\cdot\vb{S}_2 = -S_1S_2$, so that, by
Eq.~\eqref{eq:f_def}, $f_\ast = -S_1S_2/(\Sigma_1-\Sigma_2)$, on the level set
\begin{equation} \label{eq:levelset_aligned}
    (J,\Lambda) = \big(|L+ S_1- S_2|,\; L(\Sigma_1 S_1-\Sigma_2 S_2)\big) .
\end{equation}
Inserting Eq.~\eqref{eq:levelset_aligned} into Eqs.~\eqref{eq:Delta_def}
and~\eqref{eq:a_def}, we find
\begin{equation} \label{eq:up_down_range}
    f_{\rm rd} - f_\ast = \frac{m_1-m_2}{2m}(L-L_+)(L-L_-) , \qquad
    L_\pm = \frac{m_1 m_2}{m_1^2-m_2^2}\big(\sqrt{\Sigma_1 S_1} \pm \sqrt{\Sigma_2 S_2}\big)^2 ,
\end{equation}
with $0 \le L_- < L_+$ for $m_1>m_2$. For $L_- < L < L_+$, the third root lies below $f_\ast$, so
that $f_{\rm rd} = f_1 < f_\ast = f_2 = f_3$: condition (i) holds, and the up-down
configuration is unstable. The other configurations of this level set, with
$f_1 \le f < f_\ast$, are not aligned, but approach the up-down configuration as
$\lambda_\Lambda \to \pm\infty$, in agreement with the divergence of $K(k)$. At
$L = L_\pm$, $f_{\rm rd} = f_\ast$ is a triple root: condition (ii) holds, and the
up-down configuration changes stability. For $L < L_-$ or $L > L_+$,
$f_\ast = f_1 = f_2 < f_{\rm rd} = f_3$, and it is stable. The same computation for the
up-up, down-down, and down-up configurations gives $f_{\rm rd} > f_\ast$ for all $L$, so
their double root is always $f_1 = f_2 < f_3$, and they are stable, in agreement with
Refs.~\cite{gerosa2015updown, PhysRevD.101.124037}.

At Newtonian order, $L = \mu\sqrt{m\,a(1-e^2)}$. With
$S_a = \chi_a m_a^2$ and $q = m_2/m_1$, Eq.~\eqref{eq:up_down_range} becomes
$L_\pm = \mu m\big(\sqrt{\chi_1}\pm\sqrt{q\chi_2}\big)^2/(1-q)$, and the up-down
configuration of an eccentric binary is unstable for
\begin{equation} \label{eq:p_ud}
    p_{\rm ud-} < a(1-e^2) < p_{\rm ud+} , \qquad
    p_{\rm ud\pm} \equiv \frac{L_\pm^2}{\mu^2 m}
    = \frac{\big(\sqrt{\chi_1} \pm \sqrt{q\chi_2}\big)^4}{(1-q)^2}\, m ,
\end{equation}
and changes stability at $a(1-e^2) = p_{\rm ud\pm}$. At fixed eccentricity, the unstable
range of the semi-major axis is $p_{\rm ud-}/(1-e^2) < a < p_{\rm ud+}/(1-e^2)$.
For circular orbits, $e = 0$ and
$a = r$, and Eq.~\eqref{eq:p_ud} reduces to $r_{\rm ud-} < r < r_{\rm ud+}$, with
\begin{equation} \label{eq:r_ud}
    r_{\rm ud\pm} = \frac{\big(\sqrt{\chi_1} \pm \sqrt{q\chi_2}\big)^4}{(1-q)^2}\, m ,
\end{equation}
which is the instability range found in Refs.~\cite{gerosa2015updown, PhysRevD.101.124037}.
In practice, $r_{\text{ud}-}$ lies below the domain of validity of the PN approximation, 
and only $r_{\text{ud}+}$ is relevant. 

\begin{figure*}[t]
    \centering
    \includegraphics[width=0.95\textwidth]{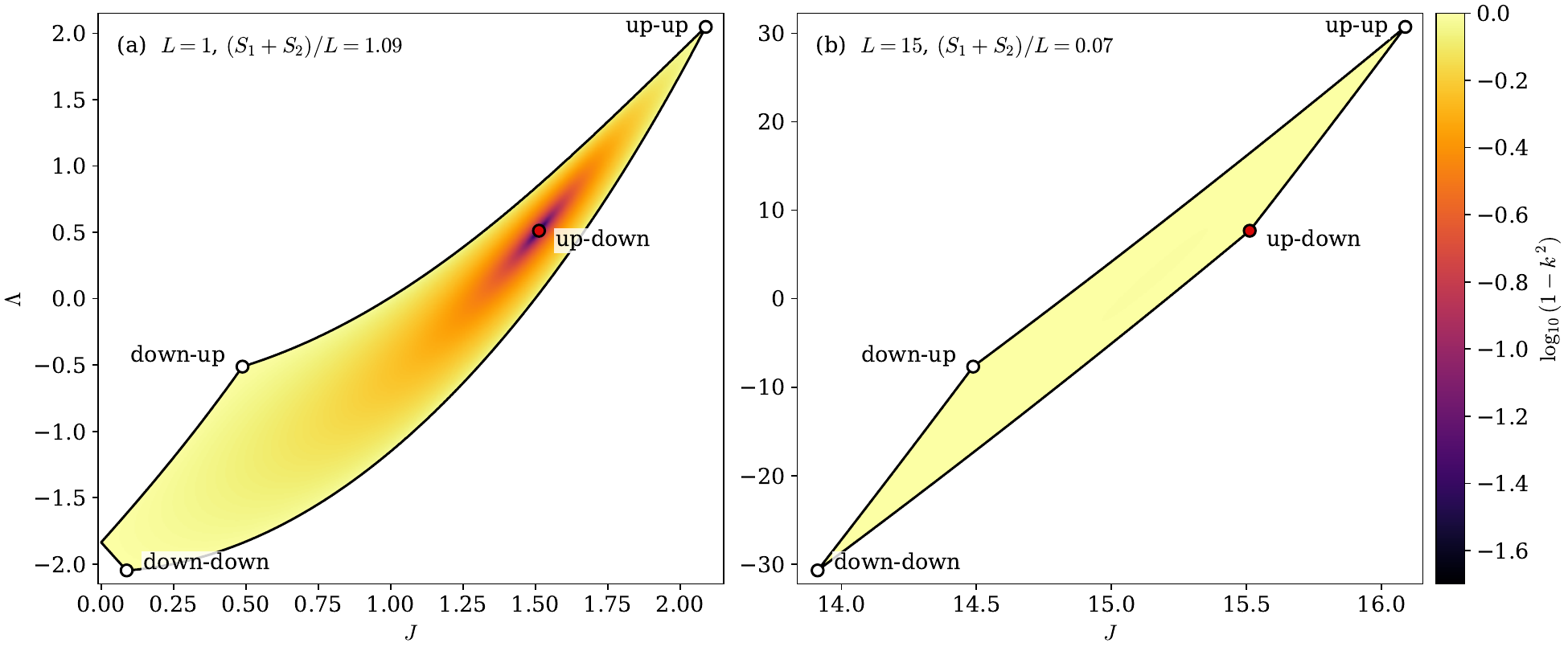}
    \caption{
    Physical region of the $(J,\Lambda)$ plane for $m_1 = 1$, $m_2 = 0.6$, 
    and $\chi_1 = \chi_2 = 0.8$, for which $(L_-, L_+) = (0.061, 3.78)$. 
    The color gives $\log_{10}(1-k^2)$, with $1-k^2 = (f_3-f_2)/(f_3-f_1)$. 
    The boundary corresponds to the spin-orbit resonances, $f_1 = f_2$. 
    The white points are the stable aligned configurations, which 
    are corners of the region, and the red point is the up-down configuration. 
    (a) $L = 1$ and $a(1-e^2) \simeq 2.8m$.  The up-down configuration is 
    unstable and lies inside the region, where $1-k^2$ vanishes. 
    The additional corner at $J = 0$ is not an aligned configuration but 
    corresponds to $\vb{L}+\vb{S}_1+\vb{S}_2 = 0$, which requires $L \le S_1 + S_2$. 
    (b) $L = 15$ and $a(1-e^2) \simeq 625m$. The up-down configuration is 
    stable and is a corner of the region, $J=0$ is not reached, and $1-k^2 \geq 0.98$ everywhere.}
    \label{fig:JLambda}
\end{figure*}

Condition (iii), $J = 0$, is of a different nature. It requires $\vb{L}+\vb{S}_1+\vb{S}_2 = 0$, 
meaning that the three angular momenta form a closed triangle. It can happen only if $|S_1 - S_2| \le L \le S_1 + S_2$.
This level set is the corner at $J=0$ of the region in Fig.~\ref{fig:JLambda}(a).
Squaring $\vb{S}_1+\vb{S}_2 = -\vb{L}$ gives $\vb{S}_1\cdot\vb{S}_2 = (L^2 - S_1^2 - S_2^2)/2$, 
and similarly for the other two dot products. The shape of the triangle is therefore fixed by the 
magnitudes alone, $f$ takes a single value, so that $f_1=f_2$, and the shape is stable.
The determinant vanishes for a different reason. The gradient of $J^2$ is proportional to the vector 
$\vb J$, and thus vanishes at $J=0$, so the COMs are no longer independent. Accordingly, the level 
set consists of all rotations of a single triangle, which is not a torus. The criterion therefore 
detects a degeneracy in the orientation of the triad, not an instability of its shape. 
Moreover, the condition $L \le S_1 + S_2$ corresponds to separations of a few $m$, outside the 
domain of validity of the PN approximation. 

Figure~\ref{fig:JLambda} shows the physical region of the $(J,\Lambda)$ plane, 
that is, the pairs $(J,\Lambda)$ reached by at least one configuration of the 
triad, at fixed masses and magnitudes. We compute it by reconstructing, at each 
$(J,\Lambda)$ and $f$, the dot products of $\vb L$, $\vb S_1$, and $\vb S_2$, 
and requiring their Gram matrix to be positive semidefinite, so that a triad 
with these dot products exists. Up to the rescaling $\chi_{\rm eff} \equiv c\Lambda/(L m^2)$, 
this is the $(J, \chi_{\rm eff})$ plane of Ref.~\cite{gerosa2015multi}, with the spin-orbit 
resonances, $f_1 = f_2$, on its boundary. What is new here is the color, which 
shows $1-k^2$. Since $1-k^2$ vanishes when $f_2=f_3$, it measures how
close each level set is to condition (i). For $L_- < L < L_+$, the up-down 
configuration lies inside the physical region, and $1-k^2$ reaches its minimum, zero, there. 
For $L>L_+$, the up-down configuration is a corner of the region, and the 
instability disappears. Far above the threshold, as in panel (b), 
$1-k^2$ stays close to unity everywhere, so no level set is near condition (i). 

For $L_- < L < L_+$, the motion on the level set of the up-down configuration is not
periodic. Every other configuration on it tends to the up-down configuration, in both the 
infinite past and the infinite future, as on the separatrix of the pendulum. Unlike the 
separatrix of the pendulum, however, this level set is a single point of the $(J,\Lambda)$ 
plane and does not split it into two regions. It is therefore separatrix-like, in the sense 
of Sec.~\ref{sec:separatrix}. At $L=L_\pm$, the triple root marks the change of stability 
of this configuration. The  other level set detected by the criterion, $J=0$ (as 
per~\cref{eq:criterion_OA}), is stable and reflects a degeneracy in the orientation of the triad.

\begin{figure}
  \includegraphics[width=0.6\columnwidth]{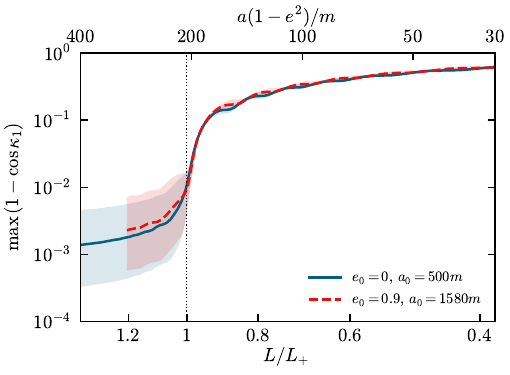}
\caption{\label{fig:updown}
Onset of the up-down instability during the inspiral, for two ensembles of
$3000$ BBHs ($q=0.8$, $\chi_1=\chi_2=0.8$) starting close to the up-down
configuration, with Gaussian misalignments of width $3^\circ$, on a circular
orbit with $a_0 = 500m$ (solid) and an eccentric orbit with $e_0 = 0.9$ and
$a_0 = 1580m$ (dashed), evolved with spin-independent radiation reaction. The
maximum of $1-\cos\kappa_1$ is taken over about three precession periods at
fixed orbit. Lines show the median over the ensemble and bands the 16--84\%
range. The bottom axis gives $L/L_+$, the top axis $a(1-e^2)/m$, and the
vertical line marks $L=L_+$. 
The two curves collapse at $L = L_+$, which verifies that the onset of the
instability is set by $a(1-e^2)$, in Eq.~\eqref{eq:p_ud}.}
\end{figure}

\paragraph{Onset of the up-down instability during the inspiral.}
For circular orbits, the onset of the up-down instability during 
the inspiral was studied numerically in Refs.~\cite{gerosa2015updown, 
PhysRevD.101.124037, PhysRevD.93.124074}. We test here the eccentric 
threshold of Eq.~\eqref{eq:p_ud}, which predicts that the onset 
is set by $L$, i.e., by $a(1-e^2)$, and not by $a$ alone. We evolve 
two ensembles of $N=3000$ BBHs, with $q=0.8$ and $\chi_1=\chi_2=0.8$, 
starting close to the up-down configuration, with misalignments of 
$\hat{\vb S}_1$ from $\hat{\vb L}$ and of $\hat{\vb S}_2$ from $-\hat{\vb L}$ 
drawn from a Gaussian distribution of width $3^\circ$. The first 
ensemble starts on a circular orbit at $a_0=500m$, and the second on
an eccentric orbit with $e_0=0.9$ at $a_0=1580 m$. Both start above 
the threshold and are evolved down to $a=15m$, with the OA precession, 
while Peters’ equations~\eqref{eq:peters}~\cite{PhysRev.136.B1224} 
drive the inspiral. Above the threshold, the spins oscillate around 
the up-down configuration. To measure the amplitude of this nutation 
independently of its phase, we interrupt the inspiral at 150 stages, 
equally spaced in $\log \,a$. At each stage, a copy of the ensemble 
is evolved with the conservative OA dynamics only, over about three 
precession periods at fixed orbit, and we record for each member 
the largest value of $1-\cos\kappa_1$. The inspiral then continues 
from the state before this conservative evolution. 

Figure~\ref{fig:updown} shows this amplitude as a function of $L/L_+$. For both ensembles, it remains 
at the level of the initial spread as long as $L>L_+$, then grows by two orders of magnitude
between $L\simeq1.1L_+$ and $L\simeq0.9L_+$, as the spins move away from the up-down configuration. 
The two curves coincide, although the eccentric ensemble reaches the threshold at $e\simeq0.55$ and $a\simeq295m$
and the circular one at $a = r_{\text{ud}+}= 206m$. The onset is thus set by $a(1-e^2)$, as predicted, whereas a 
threshold on $a$ alone would place it much later in the inspiral of eccentric binaries. 

In summary, the determinant in the
criterion of Eq.~\eqref{eq:separatrix_criterion}
vanishes at the up-down configuration  for
$L_- < L < L_+$, which coincides with the instability range found in
Refs.~\cite{gerosa2015updown, PhysRevD.101.124037}, and it accounts for the
logarithmic divergence of the nutation period, called precession period in
Refs.~\cite{gerosa2015multi, gerosa2015updown}. Combined with the root structure
of $P$, it identifies the instability with the coalescence of the two largest
roots, $f_2 = f_3$, and its threshold with the triple root, $f_1 = f_2 = f_3$.
Since the criterion depends on the orbit only through $L$, this result extends
to eccentric orbits, for which the instability is set by $a(1-e^2)$ rather than
by $a$, consistent with Ref.~\cite{PhysRevD.103.084025} and as confirmed 
numerically in Fig.~\ref{fig:updown}.

\subsubsection{Non-orbit-averaged spinning BBH system at 1.5PN order} \label{sec:separatrix_nonOA_BBH}
Finally, we apply the criterion of Eq.~\eqref{eq:separatrix_criterion} to the
non-OA spinning BBH system. Unlike for the 2PN OA system, its action variables
are known only at 1.5PN order~\cite{Tanay2020, Tanay2023}. Following
Ref.~\cite{Tanay2023}, we take $\vb{C} = (J^2, J_z, L^2, H, \vb{S}_{\ty{eff}} \cdot \vb{L})$, 
i.e., the smooth versions of the COMs of Ref.~\cite{Tanay2023}, and
$\boldsymbol{\mathcal{J}} = (J, J_z, L, \mathcal{J}_4, \mathcal{J}_5)$, where
$\mathcal{J}_4$ is the radial action of Ref.~\cite{Tanay2020} and
$\mathcal{J}_5$ is the spin action of Ref.~\cite{Tanay2023}. Since
$\mathcal{J}_4$ depends only on $(H, L^2, \vb{S}_{\ty{eff}} \cdot \vb{L})$ and
$\mathcal{J}_5$ only on $(J^2, L^2, \vb{S}_{\ty{eff}} \cdot \vb{L})$, the Jacobian
$\pd\boldsymbol{\mathcal{J}}/\pd\vb{C}$, given in Eq.~(55) of
Ref.~\cite{Tanay2023} with $(J, L)$ in place of $(J^2, L^2)$, is block triangular, and the criterion
factorizes,
\begin{equation} \label{eq:criterion_nonOA}
    \det\left(\frac{\pd\vb{C}}{\pd\boldsymbol{\mathcal{J}}}\right)
    = 4JL\left(\frac{\pd\mathcal{J}_4}{\pd H}\right)^{-1}
      \left(\frac{\pd\mathcal{J}_5}{\pd(\vb{S}_{\ty{eff}} \cdot \vb{L})}\right)^{-1} \to 0 .
\end{equation}
The factor $4JL$ comes from $\pd J^2/\pd J = 2J$ and $\pd L^2/\pd L = 2L$.
Apart from $J = 0$ and $L = 0$, the criterion is met as soon as one of the two
derivatives diverges. For $L = 0$, the bodies move along a straight line, so the
orbital plane is undefined, as is the orientation of the triad for $J = 0$.

The first factor is the radial frequency,
$\omega_4 = (\pd\mathcal{J}_4/\pd H)^{-1}$. From Eq.~(38) of
Ref.~\cite{Tanay2020},
\begin{equation} \label{eq:dJ4_dH}
    \frac{\pd\mathcal{J}_4}{\pd H} = \frac{m \mu^{3/2}}{(-2H)^{3/2}}
    - \frac{m}{c^2}\, \frac{\mu^{1/2}(\nu - 15)}{2\sqrt{32}\,\sqrt{-H}} ,
\end{equation}
which diverges as $H \to 0^-$. This is the PN counterpart of the Newtonian
parabolic orbit. The radial period diverges, and the level set $H = 0$
separates the bound orbits from the unbound ones.

The second factor is the counterpart of Eq.~\eqref{eq:criterion_OA}. As
discussed in Sec.~\ref{sec:third_action}, $\mathcal{J}_5$ depends on
$\vb{S}_{\ty{eff}} \cdot \vb{L}$ in the same way as $\mathcal{J}_3$ depends on
$\Lambda$, and the corresponding cubic is obtained from Eq.~\eqref{eq:f_cubic}
by replacing $\Sigma_a$ with $\sigma_a$. The analysis of
Sec.~\ref{sec:separatrix_OA_BBH} then carries over. For $J \neq 0$, the
criterion is met when the two largest roots of this cubic coalesce, which again
 happens at the up-down configuration. Repeating the computation of
Eq.~\eqref{eq:up_down_range} with $\sigma_a$, we find that this configuration
is unstable for $L'_- < L < L'_+$, with
\begin{equation} \label{eq:up_down_range_nonOA}
    L'_\pm = \frac{4 m_1 m_2}{3(m_1^2-m_2^2)} \Big(\sqrt{\sigma_1S_1} \pm \sqrt{\sigma_2S_2}\Big)^2 .
\end{equation}
For the system of Fig.~\ref{fig:JLambda}, the absence of the 2PN spin-spin
coupling increases the separation at which the instability starts by about
$37\%$ compared with the 2PN OA system. The same criterion thus captures both
the bound-unbound transition of the orbit and the instability of the spins.

\paragraph{Remark.}
For all the systems of this section, the criterion selects the separatrix-like
level sets, together with the level set $J = 0$ of the spinning BBH system,
which is stable but where the orientation of the triad is undefined. It is a
sufficient condition for the breakdown of integrability, but not a necessary
one, since it misses the stable equilibrium of the pendulum and the spin-orbit
resonances. Its main limitation is that it requires the actions as functions of
the COMs, which are in general hard to obtain. 
Another point to note is that 
for the systems considered here, these level sets could 
also have been found by other means, by elementary calculations for the pendulum 
and the Newtonian two-body problem, and from the stability analysis of
Refs.~\cite{gerosa2015updown, PhysRevD.101.124037} for the spinning BBH system.
The merit of the criterion is to provide a single condition covering all these
cases, justified by the breakdown of the Liouville–Arnold integrability
conditions (App.~\ref{sec:separatrix_justification}).
Also, it gives the entire $\vb{C}$ level set in terms of $\vb{C}$
without needing to explore the individual equilibrium points on this 
level set, where the integrability conditions actually break 
down.\footnote{In general, the integrability conditions break down 
only at select points of the level set given by this criterion.}

\section{Hamiltonian mechanics and isotropic spins}
\label{sec:isotropic_spins}

\subsection{The background}  

The evolution of an ensemble of spin directions during the inspiral has been
studied numerically before. Schnittman~\cite{schnittman} derived OA equations 
for the precession, which contain the spin-orbit and spin-spin couplings, 
and evolved the orbit with Peters' equations \cite{PhysRev.136.B1224}, which 
change only the magnitude of $\vb{L}$. Bogdanovi\'{c} et al.~\cite{Bogdanovic_2007} 
integrated these equations for ensembles of circular BBHs with identical masses, 
spin magnitudes and initial relative separation and found that an initially 
isotropic distribution of spin directions remains isotropic during the inspiral. 
Gerosa et al.~\cite{Gerosa:2013laa} discussed how this result is maintained
during the inspiral, but its origin has not been explained. 
Lousto et al.~\cite{Lousto:2009ka} evolved similar ensembles with the non-OA PN 
equations of motion, with conservative terms up to 3PN and the spin-dependent RR 
force of Ref.~\cite{Buonanno:2005xu}, which tilts $\hat{\vb{L}}$. They found 
instead a small bias of the spins toward an anti-alignment with $\hat{\vb{L}}$. 
Correia~\cite{correia:hal-02319812} showed that the Cassini states move
during the inspiral, which modifies the dynamics around them, and argued
that this is why the final distribution of the spins may differ from the
initial one.

An ensemble of BBHs is isotropic if $(\hat{\vb L}, \hat{\vb S}_1, \hat{\vb S}_2)$
is uniformly distributed on $S^2 \times S^2 \times S^2$.
We show that the preservation of isotropy follows from the Hamiltonian
structure of the OA precession, provided that all the members share 
the same parameters and that RR does not act on the 
directions of the angular momenta. This covers the conservative regime 
and spin-independent RR, for circular as well as eccentric 
orbits. The equations of Ref.~\cite{schnittman} satisfy both conditions, 
explaining the result of Ref.~\cite{Bogdanovic_2007}. The RR 
force used in Ref.~\cite{Lousto:2009ka} violates both conditions, 
since the energy and angular momentum fluxes depend on the spin orientations, 
and part of the radiated angular momentum is not along $\hat{\vb L}$. 

Our derivation is close to the proof of Liouville’s volume theorem, 
and follows the presentation of Ref.~\cite{hand1998analytical}. 
We derive the result in Sec.~\ref{sec:isotropy-liouville} and 
test it numerically in Sec.~\ref{sec:isotropy-numerics}.

\subsection{Preservation of isotropy}  \label{sec:isotropy-liouville}

In this subsection, the member BBHs of our ensemble evolve under either 
just the OA EOMs~\eqref{eq:OA-eom} in the conservative regime where 
$L$ and $D$ are constant, or together with spin-independent RR,
where $L$ and $D$ decrease. Spin-dependent RR
is discussed at the end of this subsection and in  Sec.~\ref{sec:isotropy-numerics}. 
The derivation has four steps. We first show that all members share the same 
parameters at all times, and therefore follow the same equations of motion
for the directions of their angular momenta. We then choose coordinates 
in which an isotropic distribution corresponds to a uniform density. 
Next, we show that the flow generated by these equations preserves volume 
in these coordinates, whether $L$ and $D$ are constant or decreasing. 
Finally, our extension of Liouville's theorem implies that the density is 
constant along each trajectory, so an initially uniform density remains uniform.

\paragraph{Common parameters.}
Consider an ensemble of spinning BBHs whose members share the same masses $m_a$, spin 
magnitudes $S_a$, initial semi-major axis $a_0$ and eccentricity 
$e_0$, but have different orientations of $\vb L$, $\vb S_1$, and 
$\vb S_2$. In the conservative regime, $a$ and $e$ are constants. 
With leading-order RR, they evolve according to Peters'
equations~\cite{PhysRev.136.B1224}
\begin{subequations}\label{eq:peters}
    \begin{align}
    \frac{da}{dt} &= -\frac{64}{5c^5}\frac{\nu m^3}{a^3(1-e^2)^{7/2}}
    \left(1+\frac{73}{24}e^2+\frac{37}{96}e^4\right) , \\
    \frac{de}{dt}  &= -\frac{304}{15c^5}\frac{\nu m^3 e}{a^4(1-e^2)^{5/2}}
    \left(1+\frac{121}{304}e^2\right) . 
\end{align}
\end{subequations}

 These depend solely on $a$, $e$ and the masses, thus $a(t)$ and $e(t)$ 
 remain common for all members, as do the values of the orbital angular
 momentum and the semi-minor axis, since
\begin{equation} \label{eq:L_magnitude}
    L(t) = \mu \sqrt{m\, a(t)\,[1-e(t)^2]}, \qquad D(t) = a(t) \sqrt{1-e(t)^2}.
\end{equation}
The same holds for any spin-independent corrections to RR. 
At leading order, RR does not act on the spins, and the averaged 
flux of angular momentum is along $\hat{\vb L}$~\cite{PhysRev.136.B1224}. 
It therefore only adds to $d \vb L / dt $ a term along $\hat {\vb L}$, which 
changes only the magnitude of $\vb L$, while the precession term of Eq.~\eqref{eq:dldt_avg}
is orthogonal to $\vb L$ and changes only its direction. The spins thus still obey 
Eqs.~\eqref{eq:OA-eom}, and $\hat{\vb L}$ obeys 
a slightly modified form of Eq.~\eqref{eq:dldt_avg}
\begin{equation}\label{eq:eom_Lhat}
    \frac{d\hat{\vb L}}{dt} = \frac{1}{c^2 D^3}
    \left\{ 2\,\vb S_{\rm eff} - \frac{3\nu}{2}\frac{(\hat{\vb L}\cdot\vb S_0)}{L}\,\vb S_0 \right\}
    \times \hat{\vb L}\,,
\end{equation}
with the common $L(t)$ and $D(t)$ of Eq.~\eqref{eq:L_magnitude}.

\paragraph{Coordinates and density.}
In the dissipative regime, the phase space 
$S^2(\vb L) \times S^2(\vb S_1) \times S^2(\vb S_2)$ changes continuously 
with time as the radius $L(t)$ of the first sphere shrinks due to GW emission. 
In both regimes, we therefore work with 
$S^2(\hat{\vb L}) \times S^2(\vb S_1) \times S^2(\vb S_2)$, 
whose spheres have constant radii. On each sphere, we use cylindrical coordinates 
defined by the azimuth around a fixed axis $\hat z$ and the component along it, 
$(\phi_L, \hat L_z)$ for $\hat{\vb L}$ and $(\phi_{S_a}, S_{az})$ for the spins. 
For a vector of magnitude $V$, $d\phi_V\, dV_z = V\, d\Omega_V$, with $d\Omega_V$ 
the solid angle, so a distribution uniform in $(\phi_V, V_z)$ is isotropic.\footnote{
This is specific to cylindrical coordinates. In spherical coordinates, 
$d\Omega = \sin\theta\, d\theta\, d\phi$, and a distribution uniform in
$(\theta, \phi)$ accumulates members near the poles.}
We describe the ensemble by its density
$\rho(\phi_L, \hat L_z, \phi_{S_1}, S_{1z}, \phi_{S_2}, S_{2z}, t)$, 
defined by
\begin{equation}
    dN = \rho\, d\phi_L\, d\hat L_z\, d\phi_{S_1}\, dS_{1z}\, d\phi_{S_2}\, dS_{2z} ,
\end{equation}
where $dN$ is the number of members in a cell of phase space at time $t$. The ensemble is
isotropic at time $t$ if $\rho$ does not depend on the coordinates. Our aim is to show that if
this property holds at $t=0$, it holds at all later times.

\paragraph{Equations of motion.}
In these coordinates, Eqs.~\eqref{eq:OA-eom} for the spins and
Eq.~\eqref{eq:eom_Lhat} for $\hat{\vb L}$ read
\begin{equation} \label{eq:dissipation_EOM}
\begin{aligned}
   \frac{d\phi_L}{dt} &= \frac{1}{L}\frac{\pd \bar H}{\pd \hat L_z} , &
   \frac{d\hat L_z}{dt} &= -\frac{1}{L}\frac{\pd \bar H}{\pd \phi_L} , 
   \\
   \frac{d\phi_{S_a}}{dt} &= \frac{\pd \bar H}{\pd S_{az}} , &
   \frac{dS_{az}}{dt} &= -\frac{\pd \bar H}{\pd \phi_{S_a}} ,
\end{aligned}
\end{equation}
where $\bar{H}$ is seen as a function of the six coordinates and of the
parameters $L$, $D$, $S_a$, and $m_a$, which are held fixed in the partial
derivatives. The magnitude $L$ enters at two places, through $\bar{H}$ and 
through the prefactor $1/L$ in Eqs.~\eqref{eq:dissipation_EOM}. 
Together with Eqs.~\eqref{eq:peters} and \eqref{eq:L_magnitude}, which 
give $L(t)$ and $D(t)$, these equations form the complete system of evolution for 
$(\vb{S}_1, \vb{S}_2, \vb{L})$ in both regimes. RR therefore enters only through 
these two functions of time. In the conservative regime, where $L$ is constant, 
Eqs.~\eqref{eq:dissipation_EOM} are Hamilton's equations with canonical
pairs $(\phi_L, L_z)$ and $(\phi_{S_a}, S_{az})$. When $L$ varies, 
the evolution is no longer Hamiltonian with $\bar{H}$ and the PBs 
of Eq.~\eqref{eq:SO3-PBs} since, with these PBs, any Hamiltonian 
implies $dL/dt = 0$ ($L^2$ is a Casimir), which does not hold with RR. This 
is why we cannot apply Liouville's theorem directly. Our derivation, however, 
relies only on the conservation of phase-space volume, which still holds, as 
we now show.

\paragraph{Volume preservation.}
Since all members share the same parameters, their velocity in phase space
depends only on their position and on time, and defines a velocity field
$\vb v = (\dot\phi_L, \dot{\hat L}_z, \dot\phi_{S_1}, \dot S_{1z}, \dot\phi_{S_2}, \dot S_{2z})$.
In these coordinates, the divergence takes its Cartesian form, and
Eqs.~\eqref{eq:dissipation_EOM} give
\begin{equation} \label{eq:divergence}
  \nabla\cdot\vb v
  = \frac{1}{L}\left(\frac{\pd^2\bar H}{\pd\phi_L\,\pd\hat L_z}
     - \frac{\pd^2\bar H}{\pd\hat L_z\,\pd\phi_L}\right)
  + \sum_a\left(\frac{\pd^2\bar H}{\pd\phi_{S_a}\pd S_{az}}
     - \frac{\pd^2\bar H}{\pd S_{az}\pd\phi_{S_a}}\right) = 0 .
\end{equation}
For each pair of coordinates, the two mixed derivatives of $\bar H$ cancel. 
For the pair $(\phi_L, \hat L_z)$, this requires $1/L$ to factor out, which holds
because $L$ depends only on time and not on the coordinates. 
Since a small volume $\delta V$ carried by the flow changes as 
$d(\delta V)/dt = (\nabla\cdot\vb v)\,\delta V$, the flow preserves volume.

\paragraph{Extension of Liouville's theorem.}
The ensembles satisfy the following two conditions: 
(i) the members of the ensemble are neither created nor destroyed, 
and (ii) all members share the same parameters ($m_a, S_a, L(t), D(t)$) 
at any given time thereby resulting in a single-valued velocity field 
on the phase space. The consequence is that $\rho$ obeys the continuity
equation $\pd_t\rho + \nabla\cdot(\rho\,\vb v) = 0$, which with
Eq.~\eqref{eq:divergence}, reduces to
\begin{equation} \label{eq:liouville_equation}
    \frac{d\rho}{dt} = \frac{\pd\rho}{\pd t} + \vb v\cdot\nabla\rho = 0 ,
\end{equation}
where $d/dt$ is the material or Lagrangian 
derivative along the trajectory of a member. 
This extends Liouville's theorem to our non-Hamiltonian flow. 
The density is constant along each trajectory, so an initially 
uniform density remains uniform. An initially isotropic ensemble 
thus remains isotropic, in the conservative regime and with 
spin-independent RR, for any eccentricity. The
circular ensembles of Ref.~\cite{Bogdanovic_2007} correspond to $e=0$ 
with leading-order RR.

\paragraph{Conditions.}
The argument relies on two conditions. First, all members share the same
parameters, $m_a$, $S_a$, $L(t)$, and $D(t)$. Otherwise, the velocity of a
member would depend on its parameters as well as on its position, thus making 
the velocity field multi-valued at  a given point in the phase space,
leading to non-applicability of the continuity equation.
Second, in the evolution considered here, RR acts only 
on $L$ and $D$, and the directions of $\vb L$ and $\vb S_a$ evolve under
Eqs.~\eqref{eq:dissipation_EOM}.

Spin-dependent RR (see, e.g., Eq.~(17) of Ref.~\cite{Gerosa:2013laa} and 
Refs.~\cite{PhysRevD.58.124001,Gergely:1999pd,Gergely:2000jz,Gergely:2002fd}) 
violates both conditions. Its spin-orbit and spin-spin contributions to the 
fluxes depend on $\hat{\vb L}\cdot\vb S_a$, so members with different orientations 
acquire different $L(t)$ and $D(t)$. The radiated angular momentum is also no 
longer along $\hat{\vb L}$, so RR tilts $\hat{\vb L}$~\cite{kidder1995} and adds 
a term to Eq.~\eqref{eq:eom_Lhat}. Isotropy is then no longer protected. 
The violation is suppressed by the relative PN order of these terms, 
and we show in Sec.~\ref{sec:isotropy-numerics} that it lowers the 
mean of the effective spin 
$\chi_{\rm eff} = c\Lambda/(L m^2) = c\,\hat{\vb L}\cdot\vb S_0/m^2$.

When the masses, the initial semi-major axis and eccentricity, or the spin
magnitudes vary between members, the first condition fails, but the argument
applies to each subset of members with identical parameters. Within each subset, 
if the initial directions of the angular momenta are isotropic and independent of 
these parameters, the subset starts isotropic and remains isotropic, and so does 
the whole ensemble. If some parameters are correlated with the directions, 
however, the subsets may not be isotropic at the start, even if the whole ensemble 
is, and isotropy can then be lost. Both cases are tested in Sec.~\ref{sec:isotropy-numerics}.

\subsection{Numerical check}  \label{sec:isotropy-numerics}
We evolve four ensembles of $N=10^4$ BBHs with $\chi_1=\chi_2=0.8$ and 
isotropic initial directions, all starting at $a_0=780m$. They differ 
in the mass ratio and initial eccentricity of their members, given in 
Table~\ref{tab:ensembles}. Each ensemble is evolved either in the conservative 
regime over one inspiral time, or down to $a=15m$, with the OA precession 
and RR. Spin-independent RR is given by Eqs.~\eqref{eq:peters}. Spin-dependent 
RR adds the LO spin-orbit and spin-spin contributions to the fluxes and 
the torque on $\hat{\vb{L}}$~\cite{PhysRevD.58.124001,Gergely:1999pd,Gergely:2000jz,Gergely:2002fd}. 
For the isotropic ensemble, the RR runs start from the final state of the conservative
run, so that the members have first precessed over many cycles and RR acts on a generic 
state of the precessing ensemble, and not on the sampled initial directions.

\begin{table}[t]
\small
\caption{Mass ratio $q$ and initial eccentricity $e_0$ of the members of each
ensemble. In the correlated ensembles, they depend on the hemisphere of
$\hat{\vb S}_1$ relative to $\hat{\vb L}$.}
\label{tab:ensembles}
\begin{ruledtabular}
\begin{tabular}{lll}
Ensemble & $q$ & $e_0$ \\
\hline
Isotropic      & $0.8$ & $0.6$ \\
$q$ correlated & $0.8$ if $\cos\kappa_1>0$, $0.4$ otherwise & $0.6$ \\
$q$ random     & $0.8$ or $0.4$, at random & $0.6$ \\
$L$ correlated & $0.8$ & $0.6$ if $\cos\kappa_1>0$, $0.2$ otherwise \\
\end{tabular}
\end{ruledtabular}
\end{table}

\begin{table}[t]
\small
\caption{
All numbers are in units of $10^{-3}$.
Mean cosines of the final angles and mean change of the effective spin,
$\Delta\chi_{\rm eff}$, for ensembles of $N = 10^4$ BBHs with isotropic 
initial directions. The last row gives the theoretical standard deviations 
for mean cosines. For each binary, the change of $\chi_{\rm eff}$, between 
initial and final state, vanishes to numerical precision ($\sim 10^{-10}$) 
except with spin-dependent RR, where we give $\langle \Delta \chi_{\text{eff}}
\rangle$ along with its error. Rows 1, 2, 6 and 7 indicate isotropy preservation 
(all mean cosines lie within $2\sigma$ of zero). Rows 3, 4 and 5 indicate isotropy 
breaking: row 5 at about $10\sigma$, row 4 marginally at about $3\sigma$, and row 
3 through $\langle\Delta\chi_{\rm eff}\rangle$ only. All of these are in line with 
the theoretical predictions. Rows 8 and 9, however, show no evidence of isotropy
breaking, although the theory no longer predicts isotropy preservation there.
}
\label{tab:isotropy}
\begin{ruledtabular}
\begin{tabular}{llrrrrr}
Ensemble & Evolution & $\langle\cos\kappa_1\rangle$ & $\langle\cos\kappa_2\rangle$
  & $\langle\cos\gamma\rangle$ & $\langle\cos\Delta\Phi\rangle$
  & $\langle\Delta\chi_{\rm eff}\rangle$ \\
\hline
Isotropic      & Conservative    & $-0.7$  & $0.8$  & $2.5$  & $10.6$  & $0$ \\
               & Spin-indep.\ RR & $3.4$   & $-4.3$ & $-6.6$ & $-12.0$ & $0$ \\
               & Spin-dep.\ RR   & $-2.7$  & $-4.6$ & $-4.0$ & $-1.6$  & $-2.86\pm0.02$ \\
$q$ correlated & Conservative    & $-3.3$  & $3.9$  & $19.2$ & $20.3$  & $0$ \\
               & Spin-indep.\ RR & $-26.9$ & $30.9$ & $60.8$ & $69.1$  & $0$ \\
$q$ random     & Conservative    & $-1.0$  & $1.1$  & $0.1$  & $7.4$   & $0$ \\
               & Spin-indep.\ RR & $0.0$   & $0.1$  & $-1.7$ & $1.7$   & $0$ \\
$L$ correlated & Conservative    & $-1.8$  & $2.3$  & $9.0$  & $11.1$  & $0$ \\
               & Spin-indep.\ RR & $-0.5$  & $0.6$  & $-2.2$ & $-0.9$  & $0$ \\
\hline
$\sigma$      &                 & $5.8$   & $5.8$  & $5.8$  & $7.1$   & \\
\end{tabular}
\end{ruledtabular}
\end{table}

Table~\ref{tab:isotropy} gives the mean cosines of the final angles, and the 
mean change of the effective spin $\chi_{\rm eff} = c\,\hat{\vb L}\cdot\vb S_0/m^2$, 
computed binary by binary. The theoretical mean of the cosines of $\kappa_1, \kappa_2, \gamma$ 
and $\Delta\Phi$ vanishes for an isotropic distribution. For an isotropic ensemble, 
$\cos\kappa_1$, $\cos\kappa_2$, and $\cos\gamma$ are uniform on $[-1,1]$, with variance $1/3$, 
and $\Delta\Phi$ is uniform on $[0,2\pi)$, so $\cos\Delta\Phi$ has variance $1/2$. The last 
row gives the resulting theoretical standard deviation of each mean as a point estimate, 
$\sigma = \sqrt{1/(3N)}$ and $\sqrt{1/(2N)}$, respectively; see Chapter 31 of 
Ref.~\cite{Riley_Hobson_Bence_2006}.

When both conditions of Sec.~\ref{sec:isotropy-liouville} hold, all mean cosines
are compatible with zero, within $2 \sigma$. This is the case for the isotropic 
ensemble, in the conservative regime and with spin-independent RR, and for the ensemble 
with random $q$, where for the latter, the argument applies to each of its two subsets 
(Rows 1, 2, 6, and 7 of Table~\ref{tab:isotropy}). The correlation between $q$ and the 
directions of the angular momenta breaks isotropy (Rows 4 and 5). This is clear in the 
spin-independent RR case, at about $10 \sigma$, where the spins tend to align with each other, 
and marginal, at about $ 3 \sigma$, in the conservative case. The correlation between $L$ 
and the directions of the angular momenta does not exhibit a breaking of isotropy, although 
isotropy is no longer protected by our theoretical derivation (Rows 8 and 9). 

With spin-dependent RR, the mean cosines are also compatible with zero, but 
isotropy is broken. Since $\chi_{\rm eff}$ is linear in $\cos\kappa_1$ and 
$\cos\kappa_2$, it has a zero mean in an isotropic ensemble whose masses 
are independent of the directions (Row 3). It is conserved for each binary by the
OA precession and spin-independent RR, as confirmed by Table~\ref{tab:isotropy},
but not by the torque on $\hat{\vb{L}}$ caused by the spin-dependent RR. Its mean
then changes by $(-2.86\pm0.02)\times10^{-3}$, and by $(-2.87\pm0.02)\times10^{-3}$
for a second realization of the ensemble with different random initial directions.  
Here the error ($0.02$) for the sample mean ($-2.86$) was computed by dividing the 
sample standard deviation by $\sqrt{N}$~\cite{Riley_Hobson_Bence_2006}.
Since the ensemble starts isotropic and the final mean of $\chi_{\rm eff}$ is nonzero,  
the final ensemble is not isotropic anymore. The sign agrees with the bias 
toward anti-alignment found by Lousto et al.~\cite{Lousto:2009ka}.

\section{Effective spin of binary neutron stars}
\label{sec:chi_eff_BNS}

\subsection{Hamiltonian and integrability}

For a neutron star (NS), the spin-induced quadrupole moment differs from
its Kerr value. It is parametrized by a constant $C_{Qa}$ for each body,
with $C_{Qa}=1$ for a BH and $C_{Qa}>1$ for a
NS~\cite{Poisson1998quadrupole}. Only the self-spin terms of $H_{\rm SS}$
change, each being multiplied by $C_{Qa}$~\cite{Schafer2024}. Averaging
them with Eqs.~\eqref{eq:how_to_average} and dropping the Casimir terms
gives
\begin{equation} \label{eq:H_bar_BNS}
\bar H^{\rm BNS} = \bar H - \frac{3\nu}{4c^2D^3}\sum_a \left(C_{Qa}-1\right)
\Sigma_a^2\,(\hat{\vb L}\cdot\vb S_a)^2 .
\end{equation}
The arguments of Sec.~\ref{sec:derivation_OA} apply equally to BNSs, so
$\bar H^{\rm BNS}$ generates the OA BNS EOMs through
Eq.~\eqref{eq:LP-eom}, with the PBs~\eqref{eq:SO3-PBs}. Since
$\bar H^{\rm BNS}$ is invariant under rotations,
$(J^2, J_z, \bar H^{\rm BNS})$ are three COMs in involution, and the OA BNS
system is integrable, by the same argument as for
BBHs~\cite{boue:hal-00335321, correia:hal-02319812}. These results hold for
any $C_{Qa}$, and therefore also for NS-BH binaries.

$\Lambda$, and therefore $\chi_{\rm eff} = c\Lambda/(m^2L) = (m_1\boldsymbol{\chi}_1
+ m_2\boldsymbol{\chi}_2)\cdot\hat{\vb L}/m$, is no longer 
conserved~\cite{PhysRevD.103.084024, PhysRevD.108.043018}. 
Since $\{\Lambda,\bar H\}=0$ and
$\{\vb L\cdot\vb S_1,\vb L\cdot\vb S_2\} = \vb L\cdot(\vb S_1\times\vb S_2)$,
only the last term of Eq.~\eqref{eq:H_bar_BNS} contributes, and
\begin{equation} \label{eq:Lambda_dot_BNS}
\frac{d\Lambda}{dt} = \frac{3}{2c^2D^3L^2}
\Big[\left(C_{Q1}-1\right)\Sigma_1\,\vb L\cdot\vb S_1
- \left(C_{Q2}-1\right)\Sigma_2\,\vb L\cdot\vb S_2\Big]\,
\vb L\cdot(\vb S_1\times\vb S_2) .
\end{equation}
This vanishes for BBHs, and for BNSs when the triad is coplanar. In
general, $\chi_{\rm eff}$ oscillates on the precession timescale.

\subsection{Conserved effective spin}

As in Sec.~\ref{sec:Integrability_AAV}, we can rewrite 
$\bar H^{\rm BNS}$ of Eq.~\eqref{eq:H_bar_BNS} in terms of $J^2$, 
the Casimirs, and the projections $\vb L\cdot\vb S_a$ only,
\begin{equation} \label{eq:H_bar_BNS_K}
4c^2D^3\bar H^{\rm BNS} = J^2 - L^2 - S_1^2 - S_2^2 + 4\mathcal{K} ,
\end{equation}
with
\begin{equation} \label{eq:K_BNS}
\mathcal{K} \equiv \frac{3}{2}\Lambda - \frac{3\nu}{4L^2}\Big[\Lambda^2
+ \sum_a \left(C_{Qa}-1\right)\Sigma_a^2(\vb L\cdot\vb S_a)^2\Big] .
\end{equation}
Since $\bar H^{\rm BNS}$ and $J^2$ are conserved, so is $\mathcal{K}$. 
For BBHs, $\mathcal{K} = 3\Lambda/2 - 3\nu\Lambda^2/(4L^2)$, and we recover
Eq.~\eqref{eq:hamiltonian_dependent}. 
For BNSs, the additional terms in $(\vb L\cdot\vb S_a)^2$ make
$\mathcal{K}$ no longer a function of $\Lambda$ alone, and in general no
linear combination of $\vb L\cdot\vb S_1$ and $\vb L\cdot\vb S_2$ is
conserved.

To obtain an effective spin that reduces to $\chi_{\rm eff}$, we invert the
BBH relation, i.e., we define $\Lambda^{\rm BNS}$ by
$\mathcal{K} = 3\Lambda^{\rm BNS}/2 - 3\nu(\Lambda^{\rm BNS})^2/(4L^2)$.
Of the two roots of this quadratic equation, we keep the one that reduces
to $\Lambda$ for $C_{Q1}=C_{Q2}=1$,
\begin{equation} \label{eq:Lambda_BNS}
\Lambda^{\rm BNS} = \frac{L^2}{\nu}\left(1-\sqrt{1-\frac{4\nu\mathcal{K}}{3L^2}}\right) ,
\end{equation}
the other one, which exceeds $L^2/\nu$, lies outside the PN regime. 
In terms of the spins, with $l \equiv cL/(\mu m)$, this gives
\begin{equation} \label{eq:chi_eff_BNS}
\chi_{\rm eff}^{\rm BNS} \equiv \frac{c\,\Lambda^{\rm BNS}}{m^2L}
= l - \sqrt{\left(l-\chi_{\rm eff}\right)^2
+ \sum_a \left(C_{Qa}-1\right)\frac{m_a^2}{m^2}\,
(\hat{\vb L}\cdot\boldsymbol{\chi}_a)^2} .
\end{equation}

$\chi_{\rm eff}^{\rm BNS}$ is thus not an independent COM, but a function
of $\bar H^{\rm BNS}$ and $J^2$. It is exact in $C_{Qa}-1$, holds for
eccentric orbits through $l = c\sqrt{a(1-e^2)/m}$ at Newtonian order,
and satisfies $\chi_{\rm eff}^{\rm BNS}\le\chi_{\rm eff}$, with equality
when the spins of the bodies with $C_{Qa}>1$ lie in the orbital plane.
Unlike $\chi_{\rm eff}$ for BBHs~\cite{Racine2008}, it depends on $l$, 
and is therefore not conserved even under spin-independent RR. 
Since the OA BNS system has only three independent COMs, the conserved 
quantity of Eq.~(4.14) of Ref.~\cite{PhysRevD.108.043018}, obtained 
there for circular orbits as a constant of integration, is a function 
of $J^2$ and $\chi_{\rm eff}^{\rm BNS}$.

\subsection{Toward action-angle variables}

For the action variables, $\mathcal{K}$ is the natural third COM. Since
$\Lambda^{\rm BNS}$ is a function of $\mathcal{K}$, both generate the same
loops, but $\mathcal{K}$ is a polynomial in $\vb L\cdot\vb S_a$ and
$\bar H^{\rm BNS}$ is affine in it, with
$\partial\bar H^{\rm BNS}/\partial\mathcal{K} = 1/(c^2D^3)$. As in
Eq.~\eqref{eq:mutual_angle_time}, the mutual angles of the triad therefore
evolve along the $\mathcal{K}$-flow with
$\lambda_{\mathcal{K}} = t/(c^2D^3)$, and the nutation period is
$c^2D^3\Delta\lambda_{\mathcal{K}}$, where $\Delta\lambda_{\mathcal{K}}$ is
the period of the $\mathcal{K}$-flow.
We do not compute the BNS actions here. The construction of
Sec.~\ref{sec:third_action} only uses flows generated by COMs and the
homogeneity of these COMs. Since $\mathcal{K}$, like $\Lambda$, is
homogeneous of degree two, the third action would take the form of
Eq.~\eqref{eq:action_3_final}, with $\Lambda\,\Delta\lambda_\Lambda$
replaced by $\mathcal{K}\,\Delta\lambda_{\mathcal{K}}$. The difficulty lies
in the flow amounts. For BBHs, fixing $J^2$ and $\Lambda$ makes all the
scalar products of the triad affine in the single variable $f$. For BNSs,
$\mathcal{K}$ is quadratic in $\vb L\cdot\vb S_a$, the scalar products are
no longer affine in a single variable, and the elliptic integrals of
App.~\ref{app:flow_amounts} no longer apply. We leave the computation of
the flow amounts, and hence of the BNS actions, frequencies, and angles, to
future work.

\section{Concluding remarks}\label{sec:conclusions}

\subsection{Scope of the present work}

Most analytical results obtained so far for the OA spinning BBH system rely
on the conservation of the projected effective spin~\cite{Racine2008,
Kesden2015, gerosa2015multi, Klein2017b, gerosa2015updown,
PhysRevD.101.124037}. In this work, we studied this system from the point
of view of Hamiltonian mechanics. 
Our main results are the following:

\begin{itemize}

\item \textbf{Hamiltonian nature:} 
    In Sec.~\ref{sec:Hamiltonian}, we confirmed that the 2PN OA system is
    Hamiltonian~\cite{correia:hal-02319812}, with the SO(3) Poisson brackets
    of $\vb{L}$, $\vb{S}_1$, and $\vb{S}_2$. We derived its Hamiltonian both
    by orbit averaging the full 2PN Hamiltonian, which requires the
    conservation of $\vb{J}$ for $\vb{L}$, and from an ansatz built from the
    OA EOMs alone.

\item \textbf{Integrability and action-angle variables:} 
    In Secs.~\ref{sec:Integrability_AAV} and~\ref{sec:frequencies_angles}, we
    showed that the OA system is integrable, as noted in
    Ref.~\cite{correia:hal-02319812}, and computed its action
    variables, frequencies, and angle variables, which give an
    action-angle-based solution. The actions were obtained with the
    fictitious variables of Ref.~\cite{Tanay2023}, a second application of
    this method. We also obtained four further results. First, we related
    the OA COMs to the COMs of the full non-OA system at 2PN. Second, we
    showed that the actions computed in the EPS and in the SPS differ only
    by integer multiples of the Casimirs. These two results clarify points
    discussed in Refs.~\cite{Tanay2020} and~\cite{Tanay2023}, respectively.
    Third, we computed the frequencies without differentiating the elliptic
    integrals. Fourth, these frequencies confirm that the spin-orbit
    resonances are not resonances in the usual sense, as argued in
    Ref.~\cite{correia:hal-02319812}.

    \item \textbf{Breakdown of integrability and separatrices:} In
    Sec.~\ref{sec:separatrix}, we proposed a criterion for the breakdown of
    the integrability conditions, based on the Jacobian between the COMs
    and the actions. It applies to any integrable Hamiltonian system. For
    the pendulum and the Kepler problem, it recovers the standard
    separatrices. For the OA BBH system, it recovers the up-down
    instability, the logarithmic divergence of its nutation period, and the
    thresholds $L_\pm$~\cite{gerosa2015updown}. It also gives these thresholds
    explicitly for eccentric orbits, for which the onset is set by
    $a(1-e^2)$, consistent with the linear analysis of
    Ref.~\cite{PhysRevD.103.084025}, and we verified this numerically.
    For the non-OA 1.5PN system, it gives both the separatrix between bound 
    and unbound orbits and the up-down instability.

    \item \textbf{Ensembles of binaries:} 
    In Sec.~\ref{sec:isotropic_spins}, we showed that the preservation of 
    isotropy observed in Ref.~\cite{Bogdanovic_2007} follows from the 
    Hamiltonian nature of the OA system. Our derivation predicts that 
    isotropy is also preserved for eccentric orbits, and that it can be lost 
    when the parameters of the members are correlated with the directions 
    of their angular momenta, or when RR depends on the spins. We verified 
    these predictions numerically.

    \item \textbf{Binary neutron stars:} 
    In Sec.~\ref{sec:chi_eff_BNS}, we used the conserved OA Hamiltonian of BNSs
    to generalize $\chi_{\rm eff}$ to BNSs. The result is exact in the
    quadrupole parameters and holds for eccentric orbits. We also identified
    $\mathcal{K}$ as the natural third COM for their action-angle variables.

\end{itemize}

\subsection{Future perspectives}   \label{sec:future_work}

For OA BNSs, the next step is to compute the action-angle variables from
the flow of $\mathcal{K}$ (Sec.~\ref{sec:chi_eff_BNS}). Alternatively,
since the quadrupole terms are of relative 0.5PN order with respect to the
leading spin-orbit coupling, approximate BNS action-angle variables can be
constructed from the BBH ones with canonical perturbation
theory~\cite{jose1998classical}. The criterion of Sec.~\ref{sec:separatrix}
would then give their separatrices, complementing the linear stability
analysis of Ref.~\cite{PhysRevD.103.084025}. Since
$\chi_{\rm eff}^{\rm BNS}$ depends on $L$, it is not conserved even under
spin-independent RR, unlike $\chi_{\rm eff}$ for BBHs, and its evolution
during the inspiral also remains to be studied.
Finally, since our criterion is sufficient but not necessary
(Sec.~\ref{sec:separatrix_nonOA_BBH}), whether it captures all unstable
equilibria remains an open question.

\begin{acknowledgments}

We would like to thank Laura Bernard and José Tomás Gálvez Ghersi 
for motivating us to pursue this work. We thank Leo Stein, Samuel Lisi, 
Guillaume Faye, Paul Ramond,  Soichiro Isoyama, Thibault Damour, Alan Weinstein, and Davide Gerosa for fruitful discussions. S.T. was supported by the 
PSL Postdoctoral  Fellowship and UTEC Internal Research Grant
during the course of this work.

\end{acknowledgments}

\appendix

\section{Loop integrals and the poles of the azimuthal coordinates}
\label{app:invariant_action_stokes_rotation}

The azimuth $\phi_V$ of $\vb{V}\in\{\vb{L},\vb{S}_1,\vb{S}_2\}$ about $\hat{\vb{z}}$ is undefined at
the poles $\vb{V}\parallel\pm\hat{\vb{z}}$. In this appendix, we evaluate the integrals of
Eq.~\eqref{eq:first_action_expansion} for any orientation of the $z$-axis, derive the
relation~\eqref{eq:EPS_SPS} between the EPS and SPS integrands, and show that the resulting additive
constants do not affect the dynamics. We first fix the sign convention of Stokes’ theorem.

\subsection{Sign convention with Stokes' theorem}
\label{app:cap_stokes_convention}

With the $x$ and $y$ axes pointing right and up, let a loop in the
$(x,y)$ plane enclose a region of area $A>0$.
Since $\oint d(xy)=0$, we have
\begin{equation} \label{eq:stokes_sign_convention}
    \oint y\, dx = -\oint x\, dy =
    \begin{cases}
        -A, & \text{loop traversed counterclockwise},\\
        +A, & \text{loop traversed clockwise}.
    \end{cases}
\end{equation}

\subsection{\texorpdfstring{$J^2$}{J2}-loop for any position of the poles}

\begin{figure*}[t]
    \centering
    \includegraphics[width=\textwidth]{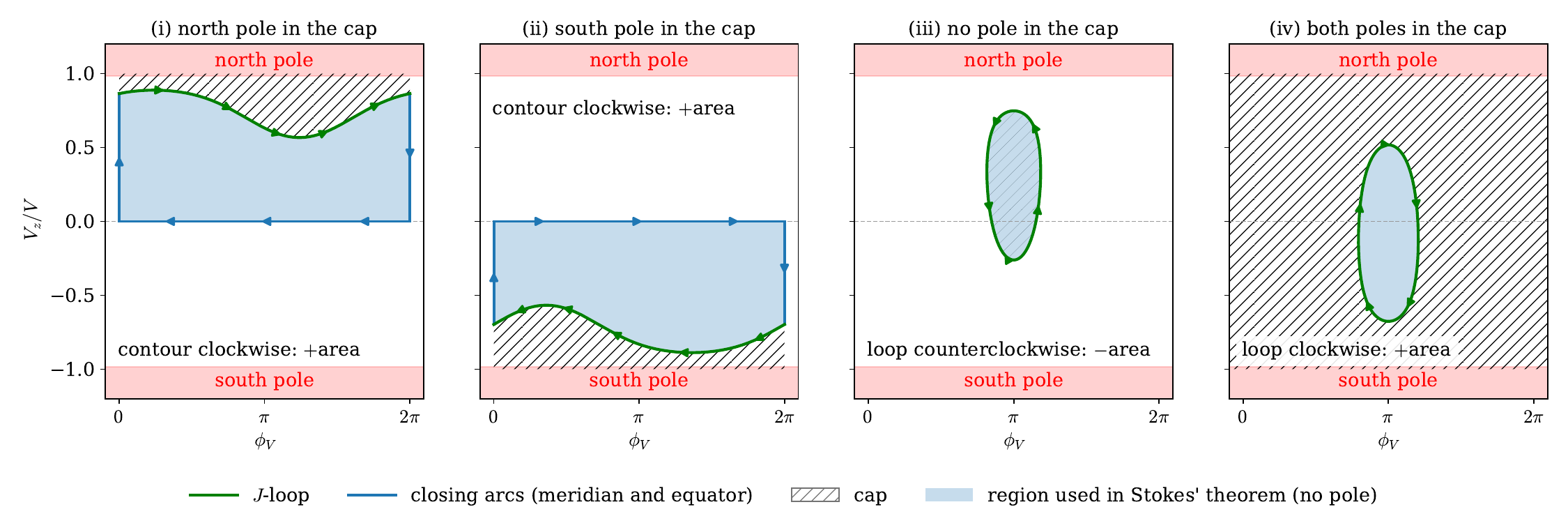}
    \caption{The $J^2$-loop (green) in the $(\phi_V,V_z)$ plane, for the four positions of the poles
    relative to the cap (hatched). The north and south poles are the top and bottom edges, and the
    dashed line is the equator. The shaded region contains no pole and is the one to which Stokes'
    theorem is applied. In (i) and (ii), the loop is closed by a meridian arc and the equator
    (blue), as in Fig.~\ref{fig:stokes_theorem}(c); in (iii) and (iv), the loop is itself a closed
    contour.}
    \label{fig:loop_cases}
\end{figure*}

To evaluate $\oint V_z\,d\phi_V$ along the $J^2$-loop, we use Stokes' theorem in the $(\phi_V,V_z)$
plane. $\oint V_z\,d\phi_V$ equals the area enclosed by the contour, counted positively for a
clockwise contour and negatively for a counterclockwise one, and this area is $1/V$ times the
corresponding area on the sphere. Stokes' theorem requires the integrand to be defined on the
contour and everywhere inside it, so the enclosed region must contain no pole. In the
$(\phi_V,V_z)$ plane, $V_z$ ranges from $-V$ to $V$, so that the north and south poles become 
the top and bottom edges, $V_z=\pm V$ (Fig.~\ref{fig:loop_cases}). Seen from outside the 
sphere, the loop runs counterclockwise around the cap, of area $2\pi V(V-\vb{V}\cdot\hat{\vb{J}})$,
and its image in the $(\phi_V,V_z)$ plane runs in the same sense.
Recall that we gave a specific definition of cap in Sec.~\ref{sec:first_two_actions}.
Four cases arise, depending on
whether the north and south poles lie inside the cap ($n_N=1$ and $n_S=1$, respectively):
\begin{enumerate}
\item[(i)] North pole only ($n_N=1$, $n_S=0$) [Fig.~\ref{fig:loop_cases}(i)].
This is the case of Sec.~\ref{sec:first_two_actions}, where closing the
loop with a meridian arc and the equator gives
$\oint V_z\,d\phi_V = 2\pi\,\vb{V}\cdot\hat{\vb{J}}$
[Eq.~\eqref{eq:loop_integral}].

\item[(ii)] South pole only ($n_N=0$, $n_S=1$) [Fig.~\ref{fig:loop_cases}(ii)].
Since the cap contains the south pole but not the north pole,
$\hat J_z<0$, and the loop goes once around the $z$-axis with $\phi_V$
decreasing by $2\pi$. Closing it with a meridian arc and the equator, as
in case~(i), and using Eq.~\eqref{eq:stokes_sign_convention}, we again
obtain $\oint V_z\,d\phi_V = 2\pi\,\vb{V}\cdot\hat{\vb{J}}$.

\item[(iii)] No pole ($n_N=n_S=0$) [Fig.~\ref{fig:loop_cases}(iii)]. The loop does not go around
the $z$-axis, so the net change of $\phi_V$ along it is zero, instead of $\pm2\pi$ in cases~(i)
and~(ii). In the $(\phi_V,V_z)$ plane, the loop is an
oval that touches neither the top nor the bottom edge, and the cap is the inside of the oval. The
oval thus runs counterclockwise in the $(\phi_V, V_z)$ plane,  and
$\oint V_z\,d\phi_V = -2\pi V(V-\vb{V}\cdot\hat{\vb{J}})/V = -2\pi\,(V-\vb{V}\cdot\hat{\vb{J}})$.
This is negative as per Eq.~\eqref{eq:stokes_sign_convention}.
\item[(iv)] Both poles ($n_N=n_S=1$) [Fig.~\ref{fig:loop_cases}(iv)]. The loop is again an oval,
but the cap now touches both the top and the bottom edge, so it lies outside the oval. The inside
of the oval is the complementary cap, centered on $-\hat{\vb{J}}$, which contains no pole and has
area $4\pi V^2 - 2\pi V(V-\vb{V}\cdot\hat{\vb{J}})$. Seen from outside the sphere, a loop running
counterclockwise around the cap runs clockwise around the complementary cap. The oval thus runs
clockwise, and
$\oint V_z\,d\phi_V = [4\pi V^2 - 2\pi V(V-\vb{V}\cdot\hat{\vb{J}})]/V
= 2\pi\,(\vb{V}\cdot\hat{\vb{J}} + V)>0$.
\end{enumerate}
The four cases combine into
\begin{equation} \label{eq:loop_general}
    \oint_{\text{loop}} V_z\,d\phi_V = 2\pi\left[\vb{V}\cdot\hat{\vb{J}} + (n_N+n_S-1)\,V\right] .
\end{equation}
The integral is therefore invariant under rotations of the coordinate axes, as long as the
number of poles
inside the cap remains the same. When a pole lies exactly on the loop, $\phi_V$ is undefined at one
point of it. This happens only when a pole crosses the loop, i.e., in the transition between two
cases, where $n_N$ or $n_S$ changes by one and the integral jumps by $2\pi V$. Such orientations of
$\hat{\vb{z}}$ form a set of measure zero and are excluded. The same holds for any loop on the invariant
torus: a deformation that crosses a pole shifts the contribution of $\vb{V}$ to the action by $\pm V$.

\subsection{EPS and SPS integrands}

Let $\theta_V$ and $\phi_V$ be the polar and azimuthal angles of $\vb{V}$ about $\hat{\vb{z}}$, so
that $V_z = V\cos\theta_V$, and let
\begin{equation}
    \hat{\vb{e}}_\theta = (\cos\theta_V\cos\phi_V,\ \cos\theta_V\sin\phi_V,\ -\sin\theta_V) , \qquad
    \hat{\vb{e}}_\phi = (-\sin\phi_V,\ \cos\phi_V,\ 0)
\end{equation}
be the spherical unit vectors at $\hat{\vb{V}}$: $\hat{\vb{e}}_\theta$ is tangent to the meridian and
points toward the south pole, and $\hat{\vb{e}}_\theta\times\hat{\vb{e}}_\phi = \hat{\vb{V}}$. Since
$\vb{r}_V\perp\vb{V}$, we define $\psi_V$, the angle of $\vb{r}_V$ about $\vb{V}$ such that
\begin{equation} \label{eq:psi_def}
    \hat{\vb{r}}_V = \cos\psi_V\,\hat{\vb{e}}_\theta + \sin\psi_V\,\hat{\vb{e}}_\phi .
\end{equation}
We compute in two ways the angle $\hat{\vb{V}}\cdot(\hat{\vb{r}}_V\times d\hat{\vb{r}}_V)$ by which
$\hat{\vb{r}}_V$ turns about $\hat{\vb{V}}$ along a flow of the form~\eqref{eq:fictitious_rotation}.

\emph{From the flow.} Along the flow,
$d\hat{\vb{r}}_V = (\partial F/\partial\vb{V})\times\hat{\vb{r}}_V\,d\lambda_F$, so that
$\hat{\vb{r}}_V\times d\hat{\vb{r}}_V = [\partial F/\partial\vb{V}
- \hat{\vb{r}}_V\,(\hat{\vb{r}}_V\cdot\partial F/\partial\vb{V})]\,d\lambda_F$. Since
$\hat{\vb{r}}_V\perp\hat{\vb{V}}$, only the first term survives the projection on $\hat{\vb{V}}$, and
Eq.~\eqref{eq:pdr_rotation} gives
\begin{equation} \label{eq:pdr_turn}
    \vb{p}_V\cdot d\vb{r}_V = \vb{V}\cdot\frac{\partial F}{\partial\vb{V}}\,d\lambda_F
    = V\,\hat{\vb{V}}\cdot(\hat{\vb{r}}_V\times d\hat{\vb{r}}_V) .
\end{equation}

\emph{From the frame.} Differentiating Eq.~\eqref{eq:psi_def} and using
$\hat{\vb{V}}\cdot(\hat{\vb{r}}_V\times d\hat{\vb{r}}_V) = (\hat{\vb{V}}\times\hat{\vb{r}}_V)\cdot d\hat{\vb{r}}_V$,
with $\hat{\vb{V}}\times\hat{\vb{r}}_V = -\sin\psi_V\,\hat{\vb{e}}_\theta + \cos\psi_V\,\hat{\vb{e}}_\phi$,
we obtain
\begin{equation}
    \hat{\vb{V}}\cdot(\hat{\vb{r}}_V\times d\hat{\vb{r}}_V)
    = d\psi_V + \cos^2\psi_V\,\hat{\vb{e}}_\phi\cdot d\hat{\vb{e}}_\theta
    - \sin^2\psi_V\,\hat{\vb{e}}_\theta\cdot d\hat{\vb{e}}_\phi
    = d\psi_V + \hat{\vb{e}}_\phi\cdot d\hat{\vb{e}}_\theta ,
\end{equation}
where we used $\hat{\vb{e}}_\theta\cdot d\hat{\vb{e}}_\theta = \hat{\vb{e}}_\phi\cdot d\hat{\vb{e}}_\phi = 0$
and $\hat{\vb{e}}_\theta\cdot d\hat{\vb{e}}_\phi = -\hat{\vb{e}}_\phi\cdot d\hat{\vb{e}}_\theta$. The
first term is the rotation of $\hat{\vb{r}}_V$ relative to the frame, and the second is the rotation of the
frame itself about $\hat{\vb{V}}$. From the components of $\hat{\vb{e}}_\theta$ and
$\hat{\vb{e}}_\phi$, $\hat{\vb{e}}_\phi\cdot d\hat{\vb{e}}_\theta = \cos\theta_V\,d\phi_V$, we have
\begin{equation} \label{eq:turn_frame}
    \hat{\vb{V}}\cdot(\hat{\vb{r}}_V\times d\hat{\vb{r}}_V) = d\psi_V + \cos\theta_V\,d\phi_V .
\end{equation}

Inserting Eq.~\eqref{eq:turn_frame} into Eq.~\eqref{eq:pdr_turn}, we finally have
Eq.~\eqref{eq:EPS_SPS},
\begin{equation}
    \vb{p}_V\cdot d\vb{r}_V = V_z\,d\phi_V + V\,d\psi_V .
\end{equation}
The first term is the SPS integrand, due to the motion of $\vb{V}$ and the second is due to the rotation
of the pair about $\vb{V}$, which the SPS integrand does not see.
Like $\phi_V$, the angle $\psi_V$ is undefined at the poles.

\subsection{Additive constants}

By Eqs.~\eqref{eq:loop_general} and~\eqref{eq:EPS_SPS_loop}, the actions are defined up to constants
$n_L L+n_{S_1}S_1+n_{S_2}S_2$ with integer $n_V$'s. These are combinations of the Casimirs, which are
fixed parameters on the SPS. The integers can jump only where a loop crosses a pole, and are constant 
on the tori between such crossings. On these tori, the actions are
thus shifted by constants, which leaves the angle variables and the frequencies unchanged, since
both involve only derivatives with respect to the COMs at fixed Casimirs (see Proposition~11.3 of
Ref.~\cite{Fasano2006}).

\section{Derivative of the third action with respect to \texorpdfstring{$\Lambda$}{Lambda}}
\label{app:dJ3_derivative}

We derive Eqs.~\eqref{eq:dJ3_derivatives}. Denote the canonical pairs of the EPS
collectively by $(q^i, p_i)$, with a sum over repeated $i$ implied. We vary $\Lambda$ at fixed $J^2$,
$J_z$, and Casimirs. Each value of $\Lambda$ labels a torus, and on it the loop of
Sec.~\ref{sec:third_action}, which we parametrize by $\tau \in [0, 2\pi)$. The loop is then
described by $q^i(\tau, \Lambda)$ and $p_i(\tau, \Lambda)$. Since the
interval of integration in
\begin{equation}
    2\pi\mathcal{J}_3 = \oint p_i\, dq^i
    = \int_0^{2\pi} p_i\, \frac{\partial q^i}{\partial\tau}\, d\tau
\end{equation}
does not depend on $\Lambda$, the derivative goes under the integral sign
\begin{equation} \label{eq:app_dJ_raw}
    2\pi\, \frac{\partial\mathcal{J}_3}{\partial\Lambda}
    = \int_0^{2\pi} \left( \frac{\partial p_i}{\partial\Lambda}\, \frac{\partial q^i}{\partial\tau}
    + p_i\, \frac{\partial^2 q^i}{\partial\Lambda\,\partial\tau} \right) d\tau .
\end{equation}
Integrating the second term by parts, with the boundary term vanishing because the loop is closed
for every $\Lambda$, gives
\begin{equation} \label{eq:app_dJ}
    2\pi\, \frac{\partial\mathcal{J}_3}{\partial\Lambda}
    = \int_0^{2\pi} \left( \frac{\partial p_i}{\partial\Lambda}\, \frac{\partial q^i}{\partial\tau}
    - \frac{\partial q^i}{\partial\Lambda}\, \frac{\partial p_i}{\partial\tau} \right) d\tau .
\end{equation}

On the segment of the loop generated by $F \in \{\Lambda, J^2, L^2, S_1^2, S_2^2\}$,
Eq.~\eqref{eq:flow_def} gives
$\partial q^i/\partial\tau = (\partial F/\partial p_i)\, d\lambda_F/d\tau$ and
$\partial p_i/\partial\tau = -(\partial F/\partial q^i)\, d\lambda_F/d\tau$, and 
by the chain rule the integrand of Eq.~\eqref{eq:app_dJ} becomes
\begin{equation}
    \left( \frac{\partial F}{\partial p_i}\, \frac{\partial p_i}{\partial\Lambda}
    + \frac{\partial F}{\partial q^i}\, \frac{\partial q^i}{\partial\Lambda} \right)
    \frac{d\lambda_F}{d\tau}
    = \frac{\partial F\big(q(\tau,\Lambda), p(\tau,\Lambda)\big)}{\partial\Lambda}\,
    \frac{d\lambda_F}{d\tau} .
\end{equation}
Along the loop, $F$ equals its constant value on the torus. Hence the derivative on the right-hand
side is $1$ for $F = \Lambda$, and $0$ for the other four, which are held fixed. Only the
$\Lambda$-segment contributes, and integrating $d\lambda_\Lambda/d\tau$ over it gives
$\Delta\lambda_\Lambda$, so that
$\partial\mathcal{J}_3/\partial\Lambda|_{J^2} = \Delta\lambda_\Lambda/(2\pi)$. Varying instead 
$J^2$ at fixed $\Lambda$, only the $J^2$-segment contributes, which gives $\partial\mathcal{J}_3/
\partial J^2|_{\Lambda} = \Delta\lambda_{J^2}/(2\pi)$.

\section{The \texorpdfstring{$\Lambda$}{Lambda}-flow and the flow amounts}
\label{app:flow_amounts}

In this appendix, we give the explicit solution of the flow generated by $\Lambda$ and 
the five flow amounts $\Delta\lambda_\Lambda$, $\Delta\lambda_{J^2}$, $\Delta\lambda_{L^2}$,
$\Delta\lambda_{S_1^2}$, and $\Delta\lambda_{S_2^2}$ that close the loop of
Sec.~\ref{sec:third_action} and enter the third action through Eq.~\eqref{eq:action_3_final}. 
The $\Lambda$-flow, Eqs.~\eqref{eq:flow_L} and~\eqref{eq:flow_SA}, coincides with the flow 
generated by $\vb{L}\cdot\vb{S}_{\text{eff}}$ in the non-OA 1.5PN system of Ref.~\cite{Tanay2023} 
under the replacement $\sigma_a\to\Sigma_a$, i.e., $\vb{S}_{\text{eff}}\to\vb{S}_0$ and
$\vb{L}\cdot\vb{S}_{\text{eff}}\to\Lambda$. The derivations of that reference therefore carry over
directly, and we only adapt the results to the 2PN OA problem. We treat in turn the mutual angles of
the triad, which yield $\Delta\lambda_\Lambda$ (App.~\ref{app:flow_Lambda}), the orientation of
$\vb{L}$ about $\hat{\vb{J}}$, which yields $\Delta\lambda_{J^2}$ (App.~\ref{app:flow_J2}), and the
rotation of the fictitious pairs about their physical vectors, which yields $\Delta\lambda_{L^2}$,
$\Delta\lambda_{S_1^2}$, and $\Delta\lambda_{S_2^2}$ (App.~\ref{app:flow_L2}).

\subsection{Mutual angles and \texorpdfstring{$\Delta\lambda_\Lambda$}{Delta lambda Lambda}}
\label{app:flow_Lambda}

Along the $\Lambda$-flow, the Casimirs and the COMs $J^2$ and $\Lambda$ are fixed. The three scalar
products of the triad are then related by
\begin{subequations} \label{eq:dot_constraints}
\begin{align}
    \Sigma_1\,\vb{L}\cdot\vb{S}_1 + \Sigma_2\,\vb{L}\cdot\vb{S}_2 &= \Lambda , \\
    \vb{L}\cdot\vb{S}_1 + \vb{L}\cdot\vb{S}_2 + \vb{S}_1\cdot\vb{S}_2
    &= \tfrac12\left(J^2 - L^2 - S_1^2 - S_2^2\right) ,
\end{align}
\end{subequations}
so that only one of them is free. We choose
\begin{equation} \label{eq:f_def}
    f \equiv \frac{\vb{S}_1\cdot\vb{S}_2}{\Sigma_1-\Sigma_2} ,
\end{equation}
in terms of which Eqs.~\eqref{eq:dot_constraints} give
\begin{subequations} \label{eq:dot_products}
\begin{align}
    \vb{S}_1\cdot\vb{S}_2 &= (\Sigma_1-\Sigma_2)\,f , \\
    \vb{L}\cdot\vb{S}_1 &= \Sigma_2\,(f-\Delta_1) , \\
    \vb{L}\cdot\vb{S}_2 &= \Sigma_1\,(\Delta_2-f) ,
\end{align}
\end{subequations}
with the constants
\begin{equation} \label{eq:Delta_def}
    \Delta_a = \frac{1}{\Sigma_1-\Sigma_2}
    \left[\tfrac12\left(J^2-L^2-S_1^2-S_2^2\right) - \frac{\Lambda}{\Sigma_b}\right] ,
    \qquad (a,b)\in\{(1,2),(2,1)\} .
\end{equation}

Using Eq.~\eqref{eq:flow_SA}, we obtain
\begin{equation} \label{eq:df_dlambda}
    \frac{df}{d\lambda_\Lambda} = \vb{L}\cdot(\vb{S}_1\times\vb{S}_2) .
\end{equation}
The square of the triple product is fixed by the magnitudes and the scalar products,
\begin{equation} \label{eq:triple_identity}
    \big[\vb{L}\cdot(\vb{S}_1\times\vb{S}_2)\big]^2
    = L^2 S_1^2 S_2^2 + 2(\vb{L}\cdot\vb{S}_1)(\vb{L}\cdot\vb{S}_2)(\vb{S}_1\cdot\vb{S}_2)
    - L^2(\vb{S}_1\cdot\vb{S}_2)^2 - S_1^2(\vb{L}\cdot\vb{S}_2)^2 - S_2^2(\vb{L}\cdot\vb{S}_1)^2 .
\end{equation}
Inserting Eq.~\eqref{eq:dot_products}, we obtain a cubic polynomial in $f$,
\begin{equation} \label{eq:f_cubic}
    \left(\frac{df}{d\lambda_\Lambda}\right)^2 = P(f) \equiv a_3 f^3 + a_2 f^2 + a_1 f + a_0 ,
\end{equation}
with
\begin{subequations} \label{eq:a_def}
\begin{align}
    a_3 &= 2\Sigma_1\Sigma_2(\Sigma_2-\Sigma_1) , \\
    a_2 &= 2\Sigma_1\Sigma_2(\Sigma_1-\Sigma_2)(\Delta_1+\Delta_2) - L^2(\Sigma_1-\Sigma_2)^2
           - \Sigma_1^2 S_1^2 - \Sigma_2^2 S_2^2 , \\
    a_1 &= 2\left[\Sigma_1^2 S_1^2\Delta_2 + \Sigma_2^2 S_2^2\Delta_1
           + \Sigma_1\Sigma_2(\Sigma_2-\Sigma_1)\Delta_1\Delta_2\right] , \\
    a_0 &= L^2 S_1^2 S_2^2 - \Sigma_1^2 S_1^2\Delta_2^2 - \Sigma_2^2 S_2^2\Delta_1^2 .
\end{align}
\end{subequations}
For $m_1 > m_2$, $a_3 > 0$, and $P$ has three real roots $f_1 < f_2 < f_3$. Since $P(f) \ge 0$
along the flow, $f$ oscillates between the turning points $f_1$ and $f_2$, where the triple product
vanishes and the triad is coplanar.

We measure $\lambda_\Lambda$ from the fiducial configuration of Sec.~\ref{subsec:angle_variables},
a turning point with $f = f_1$. The solution of Eq.~\eqref{eq:f_cubic} then reads
\begin{equation} \label{eq:f_solution_app}
    f(\lambda_\Lambda) = f_1 + (f_2-f_1)\,\sn^2(\Omega_\Lambda\lambda_\Lambda,k) ,
\end{equation}
with
\begin{equation} \label{eq:Omega_k_def}
    \Omega_\Lambda \equiv \tfrac12\sqrt{a_3(f_3-f_1)} , \qquad
    k^2 \equiv \frac{f_2-f_1}{f_3-f_1} .
\end{equation}
In Eq.~\eqref{eq:f_solution_app}, $\sn$ is the Jacobi elliptic sine and $\am$ the Jacobi amplitude,
\begin{equation} \label{eq:sn_def}
    \sn(u,k) = \sin\am(u,k) , \qquad
    \am\big(F(\varphi,k),k\big) = \varphi ,
\end{equation}
where $\am$ inverts, with respect to $\varphi$, the incomplete elliptic integral of the first kind
\begin{equation} \label{eq:elliptic_first_defs}
    F(\varphi,k) = \int_0^{\varphi}\frac{d\vartheta}{\sqrt{1-k^2\sin^2\vartheta}} , \qquad
    K(k) = F(\pi/2,k) ,
\end{equation}
and $K$ is the complete elliptic integral of the first kind. Since $\sn^2(u,k)$ has period $2K(k)$, the
mutual angles return to their initial values after
\begin{equation} \label{eq:delta_lambda_Lambda_result}
    \Delta\lambda_\Lambda = \frac{2K(k)}{\Omega_\Lambda} .
\end{equation}
By Eq.~\eqref{eq:df_dlambda}, $f$ increases during the first half of this period and decreases
during the second, according to the sign $s$ of $\vb{L}\cdot(\vb{S}_1\times\vb{S}_2)$. Inverting
Eq.~\eqref{eq:f_solution_app}, a configuration with given $f$ and $s$ is reached after
\begin{equation} \label{eq:lambda_Lambda}
    \lambda_\Lambda = \frac{U(f,s)}{\Omega_\Lambda} , \qquad
    U(f,s) \equiv \begin{cases} F\big(\varphi(f), k\big) , & s \ge 0 , \\
    2K(k) - F\big(\varphi(f), k\big) , & s < 0 , \end{cases}
\end{equation}
with $\varphi(f) \equiv \arcsin\sqrt{(f-f_1)/(f_2-f_1)}$.

\subsection{Orientation of \texorpdfstring{$\vb{L}$}{L} and \texorpdfstring{$\Delta\lambda_{J^2}$}{Delta lambda J2}}
\label{app:flow_J2}

After a $\Lambda$-flow by $\Delta\lambda_\Lambda$, the triad recovers its shape but has rotated about
$\hat{\vb{J}}$. To obtain this rotation, we follow the azimuth $\Phi_L$ of $\vb{L}$ about
$\hat{\vb{J}}$, measured as in Sec.~\ref{subsec:angle_variables}. From the fiducial configuration,
where $f = f_1$ and $\Phi_L = 0$, a $\Lambda$-flow by $\lambda_\Lambda$ brings it
to~\cite{Tanay2023}
\begin{equation} \label{eq:Phi_L_solution_app}
    \Phi_L^\Lambda(\lambda_\Lambda) = \frac{1}{\Omega_\Lambda}\sum_{i=1}^{2}\frac{B_{iL}}{D_{iL}}\,
    \Pi\big(\alpha_{iL}^2, \am(\Omega_\Lambda\lambda_\Lambda,k), k\big) ,
\end{equation}
with
\begin{subequations} \label{eq:BD_def}
\begin{align}
    B_{1,2\,L} &= \frac12\Big\{J\big[\Lambda + L^2(\Sigma_1+\Sigma_2)\big] \\ 
  &\qquad  \pm L\big[\Lambda + L^2(\Sigma_1+\Sigma_2) + \Sigma_1 S_1^2 + \Sigma_2 S_2^2
    + (\Sigma_1+\Sigma_2)(\Delta_2\Sigma_1-\Delta_1\Sigma_2)\big]\Big\} , \\
    D_{1,2\,L} &= L(L \pm J) + \Delta_2\Sigma_1 - \Delta_1\Sigma_2 - (\Sigma_1-\Sigma_2)\,f_1 , \qquad
    \alpha_{iL}^2 = \frac{(\Sigma_1-\Sigma_2)(f_2-f_1)}{D_{iL}} ,
\end{align}
\end{subequations}
where the upper (lower) signs refer to $i = 1$ ($i = 2$), and where
\begin{equation} \label{eq:elliptic_third_defs}
    \Pi(n,\varphi,k) = \int_0^{\varphi}\frac{d\vartheta}{(1-n\sin^2\vartheta)\sqrt{1-k^2\sin^2\vartheta}} , \qquad
    \Pi(n,k) = \Pi(n,\pi/2,k)
\end{equation}
are the incomplete and complete elliptic integrals of the third kind. 

Since $\vb{J}$ is fixed, $f$, $s$, and $\Phi_L$ determine the triad entirely. 
Indeed, the scalar products [Eq.~\eqref{eq:dot_products}], and hence 
$\vb{J}\cdot\vb{L} = L^2 + \vb{L}\cdot\vb{S}_1 + \vb{L}\cdot\vb{S}_2$, are affine in $f$,
and $s$ fixes the handedness. 

After one period
$\Delta\lambda_\Lambda$, $\Omega_\Lambda\lambda_\Lambda$ has increased by $2K$ and $\am$ by $\pi$, so
that $\vb{L}$ has rotated about $\hat{\vb{J}}$ by
\begin{equation} \label{eq:delta_Phi_L_def}
    \Delta\Phi_L = \Phi_L^\Lambda(\Delta\lambda_\Lambda)
    = \frac{2}{\Omega_\Lambda}\sum_{i=1}^{2}\frac{B_{iL}}{D_{iL}}\,\Pi(\alpha_{iL}^2,k) .
\end{equation}
Since the $J^2$-flow rotates all vectors rigidly about $\hat{\vb{J}}$ by an angle $2J\lambda_{J^2}$,
the flow amount that brings them back to their initial orientation is
\begin{equation} \label{eq:delta_lambda_J2_result}
    \Delta\lambda_{J^2} = -\frac{\Delta\Phi_L}{2J} .
\end{equation}

\subsection{Rotation of the fictitious pairs and \texorpdfstring{$\Delta\lambda_{V^2}$}{Delta lambda V2}}
\label{app:flow_L2}

After the $\Lambda$- and $J^2$-segments, the physical vectors are back to their initial values, but
each fictitious pair $(\vb{r}_V,\vb{p}_V)$ has rotated about its physical vector $\vb{V}$ by an angle
$\Delta\psi_V$, with $\psi_V$ defined as in Eq.~\eqref{eq:psi_def}. Since the $V^2$-flow leaves the
physical vectors fixed and rotates the pair about $\vb{V}$ by an angle $2V\lambda_{V^2}$
[Eq.~\eqref{eq:fictitious_rotation}], the flow amounts that undo these rotations are
\begin{equation} \label{eq:delta_lambda_V2_result}
    \Delta\lambda_{L^2} = -\frac{\Delta\psi_L}{2L} , \qquad
    \Delta\lambda_{S_1^2} = -\frac{\Delta\psi_{S_1}}{2S_1} , \qquad
    \Delta\lambda_{S_2^2} = -\frac{\Delta\psi_{S_2}}{2S_2} ,
\end{equation}
with~\cite{Tanay2023}
\begin{subequations} \label{eq:delta_psi_def}
\begin{align}
    \Delta\psi_L &= \frac{2}{\Omega_\Lambda}\left[\frac{B_{1L}}{D_{1L}}\,\Pi(\alpha_{1L}^2,k)
    - \frac{B_{2L}}{D_{2L}}\,\Pi(\alpha_{2L}^2,k)\right] - L(\Sigma_1+\Sigma_2)\,\Delta\lambda_\Lambda , \\
    \Delta\psi_{S_a} &= \frac{2}{\Omega_\Lambda}\left[\frac{B_{1S_a}}{D_{1S_a}}\,\Pi(\alpha_{1S_a}^2,k)
    - \frac{B_{2S_a}}{D_{2S_a}}\,\Pi(\alpha_{2S_a}^2,k)\right]
    - S_a(\Sigma_a-\Sigma_b)\,\Delta\lambda_\Lambda , \quad b\neq a ,
\end{align}
\end{subequations}
where $B_{iL}$, $D_{iL}$, and $\alpha_{iL}^2$ are given by Eq.~\eqref{eq:BD_def}, and
\begin{subequations} \label{eq:BDS_def}
\begin{align}
    B_{1,2\,S_1} &= \frac12\Big\{J\big[S_1^2(\Sigma_1-2\Sigma_2) - \Delta_1\Sigma_2(\Sigma_1-\Sigma_2)\big]\nonumber\\&\qquad
    \pm S_1\big[L^2\Sigma_1 - J^2\Sigma_2 + S_1^2(\Sigma_1-\Sigma_2)
    + \Delta_2\Sigma_1^2 + \Delta_1\Sigma_2^2 - 2\Delta_1\Sigma_1\Sigma_2\big]\Big\} , \\
    D_{1,2\,S_1} &= S_1(S_1 \pm J) - \Delta_1\Sigma_2 + \Sigma_1 f_1 , \\
    \alpha_{iS_1}^2 &= -\frac{\Sigma_1(f_2-f_1)}{D_{iS_1}} ,
\end{align}
\end{subequations}
with the same sign convention as in Eq.~\eqref{eq:BD_def}. The coefficients $B_{iS_2}$, $D_{iS_2}$,
and $\alpha_{iS_2}^2$ follow from $B_{iS_1}$, $D_{iS_1}$, and $\alpha_{iS_1}^2$ by exchanging the
two bodies, $S_1\leftrightarrow S_2$ and $\Sigma_1\leftrightarrow\Sigma_2$, at fixed $i$. Since
$\Sigma_1-\Sigma_2$ then changes sign, so do $f$ and $\Delta_a$
[Eqs.~\eqref{eq:f_def} and~\eqref{eq:Delta_def}], and the exchange must be supplemented by
$(f_1,f_2)\to(-f_1,-f_2)$ and $(\Delta_1,\Delta_2)\to(-\Delta_2,-\Delta_1)$.

The angles $\Delta\psi_V$ are defined modulo $2\pi$. A shift by $2\pi$ changes $\mathcal{J}_3$ by
$\pm V$, as discussed in Sec.~\ref{sec:third_action}, which does not affect the dynamics since $L$,
$S_1$, and $S_2$ are Casimirs.

\section{Flow generated by a function of commuting constants of motion}    \label{sec:LC_of_COM}

Consider $n$ COMs $C_1,\dots,C_n$ in involution, i.e., such that $\{C_i,C_j\}=0$ for all
$i,j\in\{1,\dots,n\}$. Under the flow of $C_i$, a phase-space function $g$ evolves as
$dg/d\lambda_{C_i} = \mathcal{L}_i\,g$, with $\mathcal{L}_i \equiv \{\,\cdot\,,C_i\}$. The Jacobi
identity gives
\begin{equation} \label{eq:commuting_operators}
    \mathcal{L}_i\mathcal{L}_j\,g - \mathcal{L}_j\mathcal{L}_i\,g = \{g,\{C_j,C_i\}\} = 0 ,
\end{equation}
so the operators $\mathcal{L}_i$ commute. For constant coefficients $c_i$, the flow of
$C = \sum_i c_i C_i$ by $\lambda_C$ is then
\begin{equation} \label{eq:exp_LC}
    g(\lambda_C) = \exp\Big(\lambda_C\sum_{i} c_i\mathcal{L}_i\Big)\, g(0)
    = \prod_{i}\exp\left(c_i\lambda_C\,\mathcal{L}_i\right) g(0) ,
\end{equation}
where $g(0)$ is the initial value of $g$ and each exponential is a power series 
in its operator. The second equality holds because the operators $\mathcal{L}_i$ 
commute, so that the exponential of their sum factorizes into a product of exponentials. 
The factor $\exp(c_i\lambda_C\,\mathcal{L}_i)$ is the flow of $C_i$ by $c_i\lambda_C$. 
Flowing under $C$ by $\lambda_C$ is thus the same as flowing under each $C_i$ by 
$c_i\lambda_C$, in any order. See Sec.~5.3.4 of Ref.~\cite{jose1998classical} or 
Sec.~9.6 of Ref.~\cite{goldstein2011classical} for more details. 

This extends to any function $G(C_1,\dots,C_n)$. Indeed,
$\{g,G\} = \sum_i (\partial G/\partial C_i)\{g,C_i\}$, and the coefficients $\partial G/\partial C_i$
do not change along the flow. Flowing under $G$ by $\lambda_G$ is therefore the same as flowing under
each $C_i$ by $(\partial G/\partial C_i)\,\lambda_G$. We use this with $G = \bar{H}$ in
Sec.~\ref{subsec:computing_frequencies} and with $G = \mathcal{J}_1, \mathcal{J}_2, \mathcal{J}_3$ in
Sec.~\ref{subsec:angle_variables}.

\section{Justification of  the criterion for the breakdown of integrability conditions}     \label{sec:separatrix_justification}

We first briefly review the Liouville–Arnold theorem, and then justify the criterion.

\subsection{Background on the Liouville-Arnold theorem}
Consider a Hamiltonian system with $n$ degrees of freedom, on a $2n$-dimensional 
symplectic manifold $\mathcal{S}$ with symplectic form $\sum_{i=1}^n dp_i 
\wedge dq_i$
and canonical coordinates
$(p_i, q_i)$.
Now assume the following integrability conditions:
\begin{enumerate}
    \item There exist $n$ functions $\mathbf{C} = (C_1, \ldots, C_n)$ in mutual involution,
    $\pb{C_i}{C_j} = 0$, whose level set $\mathcal{T}$, defined by $\vb{C} = \text{const}$, is compact
    and connected.
    \item The $C_i$'s are independent on $\mathcal{T}$. 
    This is equivalent to the linear independence of their differentials, 
    $dC_1 \wedge dC_2 \wedge \cdots \wedge dC_n \neq 0$.
\end{enumerate}

Then as per the Liouville-Arnold theorem~\cite{arnold2013mathematical, Fasano2006}, $\mathcal{T}$ 
is diffeomorphic to a torus, and in the neighborhood of $\mathcal{T}$, there exist action-angle 
coordinates $(\mathcal{J}_i, \theta_i)$, i.e.,
\begin{align}     \label{eq:arnold_symplectic}
    \sum_i dp_i \wedge dq_i  = \sum_i d\mathcal{J}_i\wedge d\theta_i.
\end{align}
For an integrable Hamiltonian system, the Hamiltonian is one of the
$C_i$'s, or a function of them. The $C_i$'s then commute with it
and are therefore COMs.

\paragraph{Preferred sets of COMs.} As discussed in
Sec.~\ref{sec:math_criterion}, we use COMs whose differentials remain
finite at finite values of $(p_i, q_i)$, such as $(J^2, J_z, \Lambda)$ for
the OA spinning BBH system, and denote them by $\vb{C}$.

\subsection{Justification of the criterion}

We proceed in two steps. We first show that the condition of \cref{eq:separatrix_criterion} 
does not depend on the choice of action-angle variables, and then that it implies the 
breakdown of at least one integrability condition.

\textit{Step 1. The condition of Eq.~\eqref{eq:separatrix_criterion} cannot be removed by a change of action-angle variables.} Starting from a point on the level set $\vb{C} = \vb{C}_*$ 
satisfying Eq.~\eqref{eq:separatrix_criterion}, we define a curve $\gamma: [0, 1] \rightarrow 
\mathcal{S}$ with $\gamma(0)$ on  this level set and $\gamma(p)$, $p \in (0, 1]$ off the 
level set, such that for points $\gamma(p)$ with $p \in (0, 1]$ the integrability 
conditions hold. Implicit is  the assumption that such a curve can indeed be defined.

The level set of $\mathbf{C}$ through  $\gamma(p)$ for each $p \in (0,1]$ is then a torus 
$\mathcal{T}_p$ with action-angle variables, and $\det(\partial \mathbf{C}/\partial 
\boldsymbol{\mathcal{J}})$ varies smoothly with $p$. We also denote by $\mathcal{T}^*$ 
the level set $\vb{C} = \vb{C}_*$ through $\gamma(0)$. Hence
\begin{align}
    \lim_{p \rightarrow 0}  \vb{C}(\gamma(p)) = \vb{C}_* ,   \\
      \lim_{p \rightarrow 0}  \mathcal{T}_p = \mathcal{T}^*.
\end{align}

Action-angle variables are not unique, so one may ask whether the vanishing determinant 
in Eq.~\eqref{eq:separatrix_criterion} is merely an artifact of a poor choice of actions. 
Suppose that there exist other action-angle variables $(\tilde{\boldsymbol{\mathcal{J}}}, 
\tilde{\boldsymbol{\theta}})$ on the tori $\mathcal{T}_p$ for which $\det(\partial \mathbf{C}/
\partial \tilde{\boldsymbol{\mathcal{J}}})$ remains finite and non-zero as $p \to 0$. Then, 
by the chain rule,
\begin{equation} \label{eq:chain_rule_det}
    \det\left(\frac{\partial \mathbf{C}}{\partial \boldsymbol{\mathcal{J}}}\right) = \det\left(\frac{\partial \mathbf{C}}{\partial \tilde{\boldsymbol{\mathcal{J}}}}\right) \det\left(\frac{\partial \tilde{\boldsymbol{\mathcal{J}}}}{\partial \boldsymbol{\mathcal{J}}}\right) ,
\end{equation}
so that Eq.~\eqref{eq:separatrix_criterion} would require
\begin{equation} \label{eq:action_det_zero}
    \lim_{p \to 0} \det\left(\frac{\partial \tilde{\boldsymbol{\mathcal{J}}}}{\partial \boldsymbol{\mathcal{J}}}\right) = 0 .
\end{equation}

We now show that this is impossible by contradiction. For $p > 0$, both sets of 
actions are invertible functions of $\mathbf{C}$ only. Hence $\tilde{\boldsymbol{\mathcal{J}}} 
= \tilde{\boldsymbol{\mathcal{J}}}(\boldsymbol{\mathcal{J}})$, and the Jacobian of the 
transformation $(\boldsymbol{\mathcal{J}}, \boldsymbol{\theta}) \to (\tilde{\boldsymbol{
\mathcal{J}}}, \tilde{\boldsymbol{\theta}})$ is lower block-triangular,
\begin{equation} \label{eq:matrix_M}
    \mathcal{M} =
    \begin{bmatrix}
        \dfrac{\partial \tilde{\boldsymbol{\mathcal{J}}}}{\partial \boldsymbol{\mathcal{J}}} & \mathbf{0} \\[8pt]
        \dfrac{\partial \tilde{\boldsymbol{\theta}}}{\partial \boldsymbol{\mathcal{J}}} & \dfrac{\partial \tilde{\boldsymbol{\theta}}}{\partial \boldsymbol{\theta}}
    \end{bmatrix} .
\end{equation}
Since the transformation is canonical, its Jacobian has unit determinant, and the 
block-triangular structure gives
\begin{align}
    1 = \det \mathcal{M}
    &= \det\left(\frac{\partial \tilde{\boldsymbol{\mathcal{J}}}}{\partial \boldsymbol{\mathcal{J}}}\right) \det\left(\frac{\partial \tilde{\boldsymbol{\theta}}}{\partial \boldsymbol{\theta}}\right) , \nonumber \\
    \Longrightarrow \quad
    \det\left(\frac{\partial \tilde{\boldsymbol{\theta}}}{\partial \boldsymbol{\theta}}\right)
    &= \left[\det\left(\frac{\partial \tilde{\boldsymbol{\mathcal{J}}}}{\partial \boldsymbol{\mathcal{J}}}\right)\right]^{-1} . \label{eq:det_canonical}
\end{align}
By Eq.~\eqref{eq:action_det_zero}, the angular Jacobian would then diverge as $p \to 0$.

The key point is that this divergence is in contradiction with our assumption that both 
$\boldsymbol{\theta}$ and $\tilde{\boldsymbol{\theta}}$ are angle coordinates on the same 
torus $\mathcal{T}_p$, each covering it once with range $[0, 2\pi)^n$. The total angular 
volume of the torus, computed in either set of angles, must therefore be $(2\pi)^n$. 
Changing variables from $\tilde{\boldsymbol{\theta}}$ to $\boldsymbol{\theta}$, and 
noting from Eq.~\eqref{eq:det_canonical} that $\det(\partial \tilde{\boldsymbol{\theta}}
/\partial \boldsymbol{\theta})$ depends on the actions only, and is hence constant on 
$\mathcal{T}_p$, we obtain
\begin{equation} \label{eq:torus_integral}
    (2\pi)^n = \oint_{\mathcal{T}_p} d^n \tilde{\boldsymbol{\theta}}
    = \oint_{\mathcal{T}_p} \left|\det\left(\frac{\partial \tilde{\boldsymbol{\theta}}}{\partial \boldsymbol{\theta}}\right)\right| d^n \boldsymbol{\theta}
    = \left|\det\left(\frac{\partial \tilde{\boldsymbol{\theta}}}{\partial \boldsymbol{\theta}}\right)\right| \oint_{\mathcal{T}_p} d^n \boldsymbol{\theta}
    = (2\pi)^n \left|\det\left(\frac{\partial \tilde{\boldsymbol{\theta}}}{\partial \boldsymbol{\theta}}\right)\right| ,
\end{equation}
for every $p > 0$. Hence $|\det(\partial \tilde{\boldsymbol{\theta}}/\partial 
\boldsymbol{\theta})| = 1$ on every torus $\mathcal{T}_p$, and the angular Jacobian 
cannot diverge as $p \to 0$. This contradicts Eqs.~\eqref{eq:action_det_zero} and 
\eqref{eq:det_canonical}. So our supposition that the vanishing of $\det(\partial 
\mathbf{C}/\partial \boldsymbol{\mathcal{J}})$ can be removed by a change of 
action-angle variables must be false.

As a by-product, Eqs.~\eqref{eq:det_canonical} and \eqref{eq:torus_integral} show that 
any two sets of action-angle variables on the same torus satisfy
\begin{equation} \label{eq:action_jacobian_unit}
    \left|\det\left(\frac{\partial \tilde{\boldsymbol{\mathcal{J}}}}{\partial \boldsymbol{\mathcal{J}}}\right)\right| = 1 .
\end{equation}

\textit{Step 2. Breakdown of the integrability conditions.}
Let $(\boldsymbol{\mathcal{J}}, \boldsymbol{\theta})$ be any set of action-angle 
variables on the tori $\mathcal{T}_p$. For all $p > 0$,
\begin{equation} \label{eq:wedge_relation}
    dC_1 \wedge \cdots \wedge dC_n = \det\left(\frac{\partial \mathbf{C}}{\partial 
    \boldsymbol{\mathcal{J}}}\right) d\mathcal{J}_1 \wedge \cdots \wedge d\mathcal{J}_n ,
\end{equation}
where both wedge products are expanded in the original canonical coordinates 
$(q_i, p_i)$. By Eq.~\eqref{eq:separatrix_criterion} and Step 1, the determinant 
tends to zero as $p \to 0$ for every such choice, so at least one of the following holds.

\textit{(i) $d\mathcal{J}_1 \wedge \cdots \wedge d\mathcal{J}_n$ does not diverge 
as $p \to 0$.} Then Eq.~\eqref{eq:wedge_relation} forces $dC_1 \wedge \cdots \wedge 
dC_n \to 0$. Since the COMs are smooth, $dC_1 \wedge \cdots \wedge dC_n = 0$ somewhere 
on $\mathcal{T}^*$, so the second integrability condition is violated.

\textit{(ii) $d\mathcal{J}_1 \wedge \cdots \wedge d\mathcal{J}_n$ diverges as $p \to 0$.} 
Then at least one $d\mathcal{J}_i$ diverges, so Eq.~\eqref{eq:arnold_symplectic} cannot 
hold on $\mathcal{T}^*$. Since, by the Liouville--Arnold theorem, Eq.~\eqref{eq:arnold_symplectic} 
is an implication of the integrability conditions, at least one of these conditions must 
fail on $\mathcal{T}^*$.

\bibliography{references}

\end{document}